\documentclass[twocolumn]{aastex63}

\usepackage{graphicx,xcolor,rotating}
\usepackage{lipsum}
\usepackage{gensymb}
\usepackage{placeins}
\usepackage{float}
\usepackage{hyperref}
\usepackage{threeparttablex}
\usepackage{longtable}
\usepackage{booktabs,multirow,tabularx} 
\usepackage{amsmath}  
\usepackage{cleveref}
\usepackage{graphicx} 
\usepackage{tabulary,capt-of}
\usepackage{booktabs}
\usepackage{rotating}
\usepackage{booktabs} 
\usepackage{tablefootnote}
\usepackage{natbib}
\makeatletter
\newcommand\footnoteref[1]{\protected@xdef\@thefnmark{\ref{#1}}\@footnotemark}
\makeatother
\usepackage{float}
\usepackage{lineno}

\usepackage{xcolor}
\usepackage[dvipsnames]{xcolor}
\definecolor{darkgreen}{rgb}{0, 0.4, 0}
\newcommand{\reviewchange}[1]{\textcolor{darkgreen}{#1}}

\shorttitle{NEP-NuSTAR Spectral Variability}
\shortauthors{Creech et al.}

\begin{document}

\title{PEARLS: NuSTAR and XMM-Newton Extragalactic Survey of the JWST North Ecliptic Pole
Time-Domain Field V: Unraveling X-ray Spectral Variability}


\author[0000-0002-9041-7437]{Samantha Creech}
\affiliation{Department of Physics \& Astronomy, The University of Utah, 115 South 1400 East, Salt Lake City, UT 84112, USA}
\email{s.creech@utah.edu}

\author[0000-0002-2115-1137]{Francesca Civano}
\affiliation{NASA Goddard Space Flight Center}

\author[0000-0001-9110-2245]{Daniel R. Wik}
\affiliation{Department of Physics \& Astronomy, The University of Utah, 115 South 1400 East, Salt Lake City, UT 84112, USA}

\author[0000-0002-2115-1137]{Núria Torres-Alba}
\affiliation{Department of Physics \& Astronomy, The University of Virginia}

\author[0000-0001-6564-0517]{Ross Silver}
\affiliation{NASA Goddard Space Flight Center}

\author[0000-0002-7791-3671]{Xiurui Zhao}
\affiliation{Cahill Center for Astrophysics, California Institute of Technology, 1216 East California Boulevard, Pasadena, CA 91125, USA}

\author[0000-0001-8211-3807]{Tonima Ananna}
\affiliation{Department of Physics and Astronomy, Wayne State University, Detroit, MI 48202, USA}

\author[0009-0007-0782-0721]{Gibson Bowling}
\affiliation{School of Earth and Space Exploration, Arizona State University,
Tempe, AZ 85287-1404, USA}

\author[0000-0001-9440-8872]{Norman A. Grogin}
\affiliation{Space Telescope Science Institute, 3700 San Martin Drive, Baltimore, MD 21218, USA}

\author[0000-0002-6610-2048]{Anton M. Koekemoer}
\affiliation{Space Telescope Science Institute, 3700 San Martin Drive,
Baltimore, MD 21218, USA}

\author[0000-0002-2203-7889]{W. Peter Maksym}
\affiliation{NASA Marshall Space Flight Center, Huntsville, AL 35812, USA}

\author[0000-0002-2115-1137]{Sylvia Mesicek}
\affiliation{Department of Physics \& Astronomy, The University of Utah, 115 South 1400 East, Salt Lake City, UT 84112, USA}

\author[0000-0002-6150-833X]{Rafael Ortiz III}
\affiliation{School of Earth and Space Exploration, Arizona State University,
Tempe, AZ 85287-1404, USA}

\author[0000-0002-5319-6620]{Payaswini Saikia}
\affiliation{Department of Astronomy, Yale University, PO Box 208101, New Haven, CT 06520-8101, USA}

\author[0000-0002-9895-5758]{S. P. Willner}
\affiliation{Center for Astrophysics \textbar\ Harvard \& Smithsonian, 60 Garden Street, Cambridge, MA, USA 02138}

\author[0000-0001-8156-6281]{Rogier A. Windhorst}
\affiliation{School of Earth and Space Exploration, Arizona State University,
Tempe, AZ 85287-1404, USA}



\defcitealias{Zhao24}{Z24}
\defcitealias{Silver25}{S26a}
\defcitealias{Creech25}{C25}
\defcitealias{Ortiz26}{O26}
\defcitealias{Silver26}{S26b}

\begin{abstract}

As one of the first deep time-domain surveys in the hard X-ray band, the NuSTAR campaign in the JWST North Ecliptic Pole Time Domain Field (NEP-TDF) offers a prime opportunity to measure variability in a faint population of Active Galactic Nuclei (AGN) in the medium-to-high redshift universe.
In the X-ray band, AGN variability is generally caused by changes in the brightness of the corona ($F_\text{int}$ variability) or changes in the column density of obscuring material near the SMBH ($N_{\mathrm{H}}$ variability). 
Broad-band spectroscopy ($\sim$0.5--24~keV) is required to reliably disentangle these two phenomena.
With simultaneous NuSTAR and XMM-Newton coverage---as well as 1.8 Ms of Chandra monitoring---the X-ray surveys in the NEP-TDF are ideally situated for this purpose.
This work presents the first spectral modeling of the 52 NuSTAR-identified sources discovered in the latter half of the campaign, as well as a spectroscopic variability study of the entire 112-source NuSTAR NEP-TDF catalog.
Seven variable candidates are identified.
We find that $\gtrsim 1000$ total counts split across many ($\gtrsim 10$) epochs are needed in order to detect variability.
Additionally, variable sources are compared to the survey sensitivity in order to inform population synthesis models of how variations can affect the detectability of faint AGN, which dominate the SMBH accretion history.
Lastly, the timescale of $N_{\mathrm{H}}$ variability is investigated as a probe for the obscurer location. Significant variability is slightly more common on $> 100$ day timescales, suggesting that the torus may have only slight dominance in driving $N_{\mathrm{H}}$ variability in faint AGN.
\end{abstract}


\section{INTRODUCTION}\label{sec:intro}


When supermassive black holes (SMBHs) undergo phases of accretion-driven growth, they become luminous over many wavelengths and are classified as Active Galactic Nuclei (AGN).
Within a light hour of the SMBH, a reservoir of hot electrons Comptonizes UV and optical emission from the accretion disk. 
This structure is called the X-ray corona, and it is the primary engine for X-ray emission in AGN \citep[e.g][]{Martocchia96-corona_location, Fabian09-corona}.

X-ray variability has been found to occur in AGN on many different timescales.
Some sources have been observed to vary over hours \citep[e.g.,][]{Elvis04-CO_event, Risaliti05-CO_event, Puccetti07-CO_event, Maiolino10-Comets_around_AGN, Sanfrutos13-Corona_size},
and others over months or years \citep[e.g.,][]{Lamer03-CO_event, Markowitz14-eclipse, Miniutti14-nh_var, TA23, Cox25, Pizzetti25}.
While variability in the overall brightness of the X-ray corona can occur due to stochastic changes in accretion processes \citep[e.g.,][]{Nandra01}, much variability in the X-ray band is instead driven by changes in the line-of-sight column density of obscuring material ($N_{\mathrm{H}}$).
This obscuring material is linked to structures near the SMBH, including clouds within the broad line region (BLR) and the dusty `torus,' which is composed of gas and molecular dust located $\sim 1$~pc from the SMBH.
While the exact geometry of the torus is not well understood, infrared analyses \citep[e.g.,][]{Nenkova08-clumpy1, Nenkova08-clumpy2} and $N_{\mathrm{H}}$ variability identified in the X-ray band \citep[e.g.,][]{Risaliti02, Risaliti05-CO_event, Bianchi05, Markowitz14-eclipse, Buchner19_clumpy, Laha20} point to a clumpy structure, which describes the torus as a non-uniform collection of dusty clouds orbiting the SMBH.
In addition to observational evidence, cosmological simulations \citep[e.g.,][]{Hopkins24,Bardati26-tori} support the clumpy torus scenario as well.

Due to the clumpy nature of the torus, dust clouds can pass in or out of the line of sight, causing $N_{\mathrm{H}}$ to change by orders of magnitude in relatively short periods of time.
These eclipses are sometimes referred to as changing-obscuration events, which form a subclass of changing-look AGN: objects that undergo dramatic variability in either $N_{\mathrm{H}}$ (changing-obscuration) or in the presence/absence of optical emission components such as broad absorption lines (changing-state) \citep[see, e.g,][and references therein]{Ricci23}.
Observing changing obscuration events allows powerful constraints to be placed on the geometry of the dust clumps, including shape, size, and distance from the X-ray corona \citep[e.g.,][]{Maiolino10-Comets_around_AGN, Markowitz14-eclipse}.
However, multi-epoch, broad-band X-ray data (e.g., $\sim 0.5$--24~keV) are required in order to reliably distinguish flux variability from changes in $N_{\mathrm{H}}$.
No broad-band X-ray observatory currently exists, necessitating simultaneous observations from hard \citep[e.g., NuSTAR,][]{Harrison13-NuSTAR} and soft (e.g., XMM-Newton, Chandra) observatories.
Monitoring campaigns are resource-intensive, which results in confirmed observations of eclipse events being exceptionally rare.

In the absence of eclipse events, torus geometry can be inferred from statistical analysis of large multi-epoch surveys. 
For example, a NuSTAR, Chandra, and XMM-Newton campaign of local obscured AGN yielded 31 variable candidates \citep{Zhao21_torus}.
In-depth spectroscopic analysis reveals $N_{\mathrm{H}}$ variability in ten of the candidates \citep{Pizzetti22,TA23,Pizzetti25,Sengupta25,TA25-Mrk477}, and a statistical analysis of the population probes the innermost geometry of the torus \citep{TA23,Pizzetti25,Gianolli26}.

In this work, we aim to supplement the local variability study with an investigation of the hard X-ray-selected sources in the JWST North Ecliptic Pole Time Domain Field \citep[NEP-TDF;][]{Windhorst23-NEP}.
The NEP-TDF is the subject of a multi-wavelength survey designed to detect and monitor variability in the medium and high-redshift universe.
This region has been surveyed across the electromagnetic spectrum, from the radio \citep[][]{Payaswini25-VLBA,Willmer23-MMT} to IR \citep[][]{Jansen18-JWST_NEP,Windhorst23-NEP,unWISE19} and visible \citep[][]{Willmer23-MMT,HC23,OBrien24}.
In the X-ray band, the NuSTAR and XMM-Newton campaigns in the NEP-TDF have been awarded 3.57 Ms and 288 ks, respectively, over five contiguous years and four observation cycles.
As a result, the NEP field hosts the deepest NuSTAR survey to date.
The multi-epoch, broad-band X-ray data available within this field data offers a unique opportunity to probe variability in the observationally faint AGN population.
This variability is of particular interest for population synthesis models.
Faint and low-luminosity AGN dominate the cosmic accretion history of SMBHs; as a result, the dynamics of accretion and obscuration displayed by these faint sources is vital in order to fully capture SMBH growth \citep[e.g.,][]{Comastri95-CXB,Gilli07-CXB,Ueda14-CXB,Annana19-Accretion_HistoryI,Gerolymatou25}.
These dynamics cannot be extrapolated from previous studies of bright sources alone.
In order to inform SMBH population synthesis models, multi-epoch campaigns must target faint AGN as well.

Using the first two cycles of NuSTAR observations of the NEP-TDF, \citet{Zhao24} (\citetalias{Zhao24} hereafter) discovered 60 sources, and spectral analysis of this 60-source catalog was performed by \citet{Creech25} (\citetalias{Creech25} hereafter).
\cite{Silver25} (\citetalias{Silver25} hereafter) used the following two cycles of NuSTAR observations to discovered an additional 52 NuSTAR-selected sources, which were not detected by \citetalias{Zhao24}.
The first spectroscopic analysis of these 52 new sources is presented here.

In this work, we investigate the spectroscopic variability of the 112 NuSTAR-selected sources from the \citetalias{Zhao24} and \citetalias{Silver25} catalogs.
The companion paper to this manuscript \citep[][\citetalias{Silver26} hereafter]{Silver26} presents a photometric variability analysis of the entire X-ray-selected NEP-TDF sample.
In Section \ref{sec:DR}, we describe the sample, observations, and data reduction. 
In Section \ref{sec:analysis}, we discuss the models and methodology used for X-ray spectral analysis, and Section \ref{sec:var} describes the methodology for assessing variability.
Section \ref{sec:discussion} discusses the results, and this work is summarized in Section \ref{sec:conclusion}.


Unless stated otherwise, all uncertainties reported here are at the $90\%$ level.
This study assumes a $\Lambda$CDM cosmology with {\it H$_{0}$} = 70 km s$^{-1}$ Mpc$^{-1}$, $\Omega_{M}$ = 0.27, and $\Omega_{\Lambda}$ = 0.73 \citep[e.g.,][]{LIGO17-cosmo,Hotokezaka19-cosmo,Freedman25-cosmo}.

\section{SAMPLE AND DATA REDUCTION}\label{sec:DR}
\tabletypesize{\footnotesize}
\begin{deluxetable*}{ccccc}
\tablecaption{X-ray Observations of the NEP-TDF Grouped by Epoch\label{tab:obs}}
\startdata
\\
 \textbf{Date} &\textbf{Epoch ID}& \textbf{NuSTAR IDs}\tablenotemark{\small a} & \textbf{XMM-Newton ID}  & \textbf{Chandra ID}    \\
 \hline
2018-04-13 & C1 & - & - & 20363 \\
2018-07-26 & C2 & - & - & 20362 \\
2018-07-31 & C3 & - & - & 20358 \\
2018-10-05 & C4 & - & - & 20359 \\
2018-11-13 & C5 & - & - & 20361, 21875 \\
2018-12-12 & C6 & - & - & 21959, 20360 \\
2019-10-02 & Nu1 & 60511001002, 60511002002, 60511003002 & - & - \\
2019-10-12 & C7 & - & - & 21650 \\
2020-01-04 & Nu2 & 60511004002, 60511005002, 60511006002 & - & - \\
2020-03-02 & Nu3 & 60511007002, 60511008001, 60511009001 & - & - \\
2020-04-07 & C8 & - & - & 21652, 23212 \\
2020-05-20 & C9 & - & - & 21654, 23253 \\
2020-08-08 & C10 & - & - & 21653, 23338 \\
2020-09-21 & C11 & - & - & 21651, 24655, 24656 \\
2020-10-13 & Nu4 & 60666003002, 60666002002, 60666001002 & 0870860101 & - \\
2021-01-09 & C12 & - & - & 23016 \\
2021-01-16 & Nu5 & 60666004002, 60666005002, 60666006002 & 0870860201 & - \\
2021-01-23 & C13 & - & - & 23014 \\
2021-03-29 & C14 & - & - & 23015, 23013 \\
2021-04-20 & C15 & - & - & 23017, 25010 \\
2021-06-14 & C16 & - & - & 25062, 22518, 22519 \\
2021-06-28 & C17 & - & - & 24933, 25082 \\
2021-08-31 & C18 & - & - & 23874, 24258, 24259 \\
2021-10-14 & Nu6 & 60666007002, 60666008002, 60666009002 & - & - \\
2021-11-18 & C19 & - & - & 23875, 24514 \\
2021-12-14 & C20 & - & - & 24515 \\
2022-01-22 & Nu7 & 60666010002, 60666011002, 60666012002 & 0870860401 & - \\
2022-02-27 & C21 & - & - & 23876 \\
2022-03-15 & C22 & - & - & 25974, 25975 \\
2022-05-19 & C23 & - & - & 23877 \\
2022-06-09 & C24 & - & - & 25976 \\
2022-08-26 & C25 & - & - & 23878 \\
2022-08-29 & Nu8 & 60666013002, 60666014002, 60666015002 & 0870860501 & - \\
2022-09-12 & C26 & - & - & 25979 \\
2022-09-19 & C27 & - & - & 25978 \\
2022-09-25 & C28 & - & - & 27451 \\
2022-11-11 & C29 & - & - & 23879 \\
2022-11-20 & C30 & - & - & 25981 \\
2022-11-29 & Nu9 & 60810001002, 60810002002, 60810003002 & 0913590101 & 25980, 27576 \\
2023-02-26 & Nu10 & 60810004002, 60810005002, 60810006002 & 0913590501 & - \\
2023-03-07 & C31 & - & - & 27725, 27726 \\
2023-03-15 & C32 & - & - & 23880, 27724, 27764 \\
2023-05-26 & Nu11 & 60810007002, 60810009002, 60810009004, 60810008002 & 0913590601 & - \\
2023-06-01 & C33 & - & - & 23881, 27876, 27877 \\
2023-06-17 & C34 & - & - & 27878 \\
2023-08-23 & Nu12 & 60910001002, 60910002002, 60910003002, 60910001004 & 0931420701 & - \\
2023-09-07 & C35 & - & - & 23882, 28880, 28881, 28882 \\
2023-11-19 & Nu13 & 60910004002, 60910005002, 60910006002 & 0931420101 & - \\
2023-12-07 & C36 & - & - & 23883, 29095, 29096, 29097, 29098 \\
2024-02-21 & Nu14 & 60910010002, 60910011002, 60910012002 & 0931420501 & - \\
2024-05-02 & Nu15 & 60910007002, 60910008002, 60910009002 & 0931420601 & - \\
\hline
\enddata
\tablecomments{ \\
$^{a}$ The third number in each NuSTAR ID denotes the observation cycle.\\\
}
\end{deluxetable*}

The X-ray surveys of the NEP-TDF were performed by the NuSTAR, XMM-Newton, and Chandra observatories, and they were designed with the time domain in mind.
In this work, we perform a spectroscopic variability analysis of the 112 NuSTAR-detected sources discovered by \citetalias{Zhao24} (60 sources) and \citetalias{Silver25} (52 additional sources).
Throughout the manuscript, we identify sources using their unique NuSTAR IDs from the total 112-source catalog (presented by \citetalias{Silver26}).
For comparison with the IDs published in the \citetalias{Zhao24} and \citetalias{Silver25} catalogs---as well as a summary of the AGN properties---see Table \ref{tab:agn_props}.

\citetalias{Creech25} details the spectral extraction procedure for the 60 sources discovered in the \citetalias{Zhao24} catalog. 
We replicate that procedure for the 52 new \citetalias{Silver25} sources using all available NuSTAR and XMM-Newton data.
Additionally, we perform the first spectroscopic analysis of Chandra data in the NEP-TDF, for the seventeen NuSTAR-selected sources that match with a Chandra detection.

Due to the faint nature of the sources analyzed in this work, grouping observations into epochs allows for improved statistics and tighter constraints on model fitting.
In general, observations $<5$ days apart are grouped into epochs.
On these short timescales, the probability of $N_{\mathrm{H}}$ variability is only $\sim 20\%$, and the average change is small ( $\Delta N_{\mathrm{H}} < 3 \times 10^{22}$~cm$^{-2}$ for a sample between $10^{23} < N_{\mathrm{H}} < 10^{24}$) \citep{TA25-Mrk477}.
Further, cosmological time dilation shortens the timescales of many sources.
So, this epoch grouping procedure balances the need for maximized source counts with the goal of detecting variability.
Table \ref{tab:obs} lists the observation IDs---grouped by epoch---from the NuSTAR, XMM-Newton, and Chandra campaigns of the NEP-TDF.
Epoch selection, data reduction, and spectral extraction steps are summarized below.

\subsection{NuSTAR}

The NuSTAR campaign of the NEP-TDF was designed such that each epoch contains three or four observations $<5$ days apart (Table \ref{tab:obs}) with varying degrees of overlap on the sky.
The fifteen NuSTAR epochs took place during NuSTAR Cycle 5 (PI: Civano, ID: 5192, three epochs), Cycle 6 (PI: Civano, ID: 6218, five epochs), Cycle 8 (PI: Civano, ID: 8180, four epochs), and Cycle 9 (PI: Civano, ID: 9267, four epochs). 
The NuSTAR data were reduced by \cite{Zhao21_cycle5} (Cycle 5), \citetalias{Zhao24} (Cycle 6), and \citetalias{Silver25} (Cycles 8 and 9). 
Additionally, \citetalias{Creech25} summarizes the procedures used by all three studies.

Source and background spectra from the \citetalias{Zhao24} catalog were extracted by \citetalias{Creech25}.
In this work, we follow the same procedure to extract spectra from the \citetalias{Silver25} catalog.
As discussed in Appendix \ref{sec:a_r}, region sizes are chosen to maximize the signal-to-noise of each source.
These regions were then passed on to \textsc{nuproducts} for extraction of source spectra, ARFs, and RMFs.
The same regions were also used to extract background spectra by following the \textsc{nubgdspec} routine from \textsc{nuskybgd}  \citep{Wik14-nuskybgd}.
Once source and background spectra were extracted, any observation with $< 1$ net photons were excluded from the analysis of that source.
Spectra were then binned to contain a minimum of three counts per bin.

\begin{figure*}
    \centering
    \includegraphics[width=0.8\textwidth]{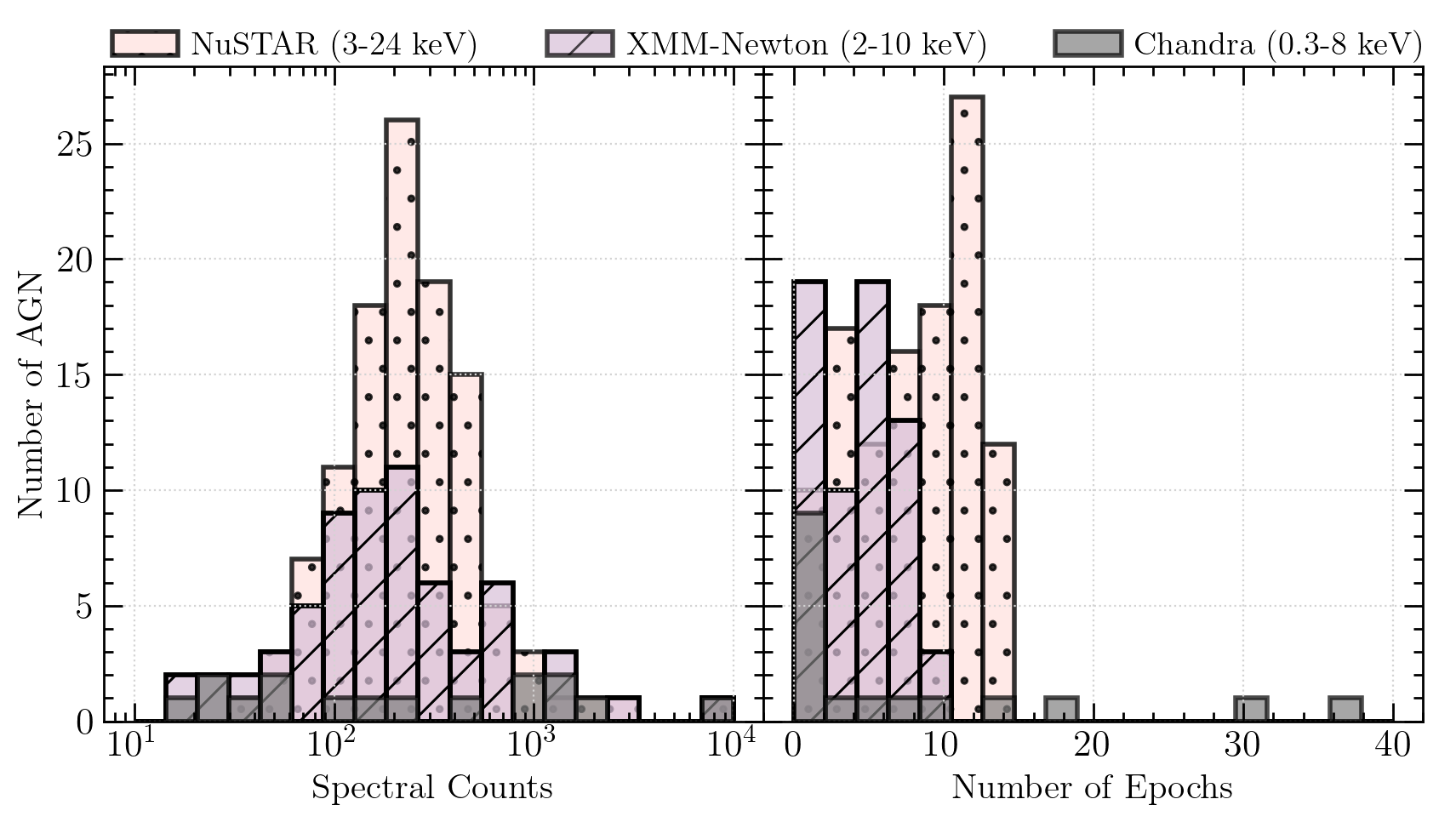}
    \caption{The net spectral counts (left) and number of epochs (right) of the 112 sources in the NEP-TDF NuSTAR-selected sample. The results are shown for NuSTAR (pink; dotted), XMM-Newton (purple; hatched), and Chandra (black; solid).} 
    \label{fig:cts_eps}
\end{figure*}

In order to improve statistics for model fitting, the spectra for each source were stacked using the \textsc{addspec} python routine\footnote{\url{https://github.com/JohannesBuchner/addspec.py}\label{ft:add}}, with FPMA and FPMB kept separate. 
First, all existing NuSTAR spectra were stacked in order to maximize signal to noise when obtaining non-variable model fits, in which model parameters are assumed to remain constant over time (Section \ref{sec:non_var}).
Next, spectra were added on an epoch-by-epoch basis.
Retaining separate spectra for each epoch allows parameter variability to be searched for.
Any epochs with $<5$ photons were excluded from the variability analysis.
The net spectral counts and number of NuSTAR epochs for the entire sample are shown in Figure \ref{fig:cts_eps}.

\subsection{XMM-Newton}

Each epoch in NuSTAR cycles 6, 8, and 9 has a corresponding XMM-Newton observation (with one exception; Obs. ID 0870860301, corresponding to NuSTAR epoch 6, was dominated by an exceptionally high particle background and therefore excluded from all analysis).
Data reduction was performed by \citetalias{Zhao24} (NuSTAR Cycle 6) and \citetalias{Silver25} (NuSTAR Cycles 8 and 9) using the XMM-Newton Science Analysis System (SAS; version 20.0.0) and following the procedures outlined by  \cite{Brunner08_LockmanHole}, \cite{Cappelluti09-XMM_COSMOS_PointCat}, and \cite{LaMassa09_XMM_Seyfert}.
This methodology is described in detail by \citetalias{Zhao24} and \citetalias{Silver25}, and \citetalias{Creech25} summarizes them as well.

Cross-matching between the NuSTAR and XMM-Newton catalogs was performed by \citetalias{Zhao24} and \citetalias{Silver25}.
From the NuSTAR catalogs, 64 sources have XMM-Newton counterparts (36 from \citetalias{Zhao24}, 28 from \citetalias{Silver25}).
Unlike NuSTAR, XMM-Newton's relatively small PSF (FWHM $5--6 \arcsec$) forgoes the need for SNR-maximized source regions.
Instead, source spectra were extracted from $15\arcsec$ regions centered around the XMM-Newton coordinates (using \textsc{evselect}).
Extraction regions for background spectra were determined using the \textsc{ebkreg} routine, and serendipitous XMM-Newton sources were excluded from the background extraction using $15\arcsec$ circular regions.
Next, RMFs and ARFs were generated using the \textsc{rmfgen} and \textsc{arfgen} tasks.
In order to correct for cross-calibration between NuSTAR and XMM-Newton instruments, the flags \textsc{applyabsfluxcorr=yes} (to correct the effective area of PN spectra in order to better match NuSTAR) and \textsc{applyxcaladjustment=yes} (to improve consistency between MOS and PN spectra by applying an energy-dependent correction function) were applied to the \textsc{arfgen} routine.
Finally, any spectrum with $<5$ counts were excluded from variability analysis, and the remaining spectra were grouped with a minimum of 3 counts per bin.

As with NuSTAR, all XMM-Newton spectra are added (using \textsc{addspec}\footref{ft:add})---keeping PN, M1, and M2 separate---in order to assess the non-variable scenario discussed in Section \ref{sec:non_var}. 
Because each epoch only has one XMM-Newton observation at most, all spectra are kept separate when searching for variability in Section \ref{sec:var_fit}.
Figure \ref{fig:cts_eps} shows the number of epochs with XMM-Newton data and the net spectral counts.

\subsection{Chandra}\label{sec:chandra}

\begin{table}
    \centering
    \caption{Chandra Cross-Match Information\label{tab:ch_obs}}
    \begin{tabular}{ccccc}
     \hline 
     \hline
     \\[-0.75em]
     \textbf{Nu ID} & \textbf{RA} &  \textbf{DEC} & $\textbf{N}_{\textbf{epochs}}$ & \textbf{Counts}\\
1 & 261.00566 & 66.00935 & - & 36.1 \\
3 & 260.69069 & 65.73808 & - & 58.2 \\
5 & 260.82356 & 65.94343 & 10 & 1148.1 \\
7 & 260.42646 & 65.92038 & 13 & 1665.2 \\
8 & 261.02091 & 65.84280 & - & 20.6 \\
9 & 260.80881 & 65.79616 & 37 & 17690.8 \\
13 & 260.58043 & 65.73816 & - & 188.6 \\
15 & 260.75284 & 65.91039 & - & 151.2 \\
19 & 260.77648 & 65.85526 & 6 & 792.0 \\
20 & 260.89658 & 65.81742 & 3 & 524.2 \\
21 & 260.43193 & 65.76753 & 31 & 7305.0 \\
23 & 260.67150 & 65.71161 & 17 & 1617.4 \\
98 & 260.53762 & 65.91680 & - & 110.7 \\
102 & 261.0004 & 65.70246 & - & 21.8 \\
103 & 260.29140 & 65.73343 & - & 52.3 \\
105 & 260.75950 & 65.69037 & - & 21.6 \\
111 & 260.86090 & 65.78809 & 8 & 810.4 \\
      \hline
    \end{tabular}
\end{table}

The Advanced CCD Imaging Spectrometer (ACIS) instrument on the Chandra X-ray observatory observed the NEP field 66 times between 2018 and 2023, totaling 1.8 Ms of observations (PI: Maksym, proposal numbers 19900666, 20900658, 21900294, and 22900038).
Unlike the NuSTAR and XMM-Newton observations, the Chandra observing program was not designed with distinct epochs.
In this work, we chose to group observations $\leq 5$ days apart into epochs.
When the median observation date of a Chandra epoch is within 5 days of a NuSTAR epoch, we link them together.
The observation IDs, epoch groups, and median observation dates are listed in Table \ref{tab:obs}.

The Chandra observations were reduced using \textsc{ciao} version 4.18.
From the \textsc{chandra-repro} routine, higher-level data products were created.
Next, the NuSTAR catalogs were cross-matched with the latest source catalog from Chandra observations (Maksym; private communication).
The catalog is not yet complete, and more NuSTAR sources will likely be matched with Chandra counterparts as the catalog is updated with more recent observations.
The catalog creation will be detailed in future work (Maksym et al. \textit{in prep}).
Table \ref{tab:ch_obs} gives the Chandra coordinates for the cross-matched sources.

For sources matched to Chandra counterparts, 5$\arcsec$ source regions and 6$\arcsec$--15$\arcsec$ background annuli were placed around the coordinates of the Chandra detection. 
From these regions, the \textsc{specextract} routine was used to extract source and background spectra.
The flags \textsc{weight = no} and \textsc{correctpsf = yes} were used, which together correct for photons scattered outside the aperture according to the Chandra PSF.

Once spectra were extracted, observations within epoch groups were then added using \textsc{addspec}\label{ft:addspec}.
Finally, the \textsc{grppha} routine from \textsc{heasoft} was used to assure a minimum of 20 counts per bin in the resulting spectra. 
For each source, we exclude Chandra epochs with $< 30$ net counts.
This is a stricter minimum than the NuSTAR and XMM-Newton thresholds.
For most Chandra epochs, only Chandra data are available, and a stricter count threshold maximizes the ability to constrain the model parameters when only the soft X-ray band is available.
The final net counts for Chandra matches are shown in Table \ref{tab:ch_obs} and Figure \ref{fig:cts_eps}.

Nine sources (IDs 1, 3, 8, 13, 15, 98, 102, 103, and 105) have $>30$ net counts when all Chandra observations are combined, but no individual epochs fall above this count threshold.
For these sources, we use the combined Chandra spectrum when obtaining initial, non-variable fits (Section \ref{sec:non_var}), but we do not use the Chandra spectra when assessing for variability (Section \ref{sec:var_fit}) due to low statistics in individual epochs.
Table \ref{tab:ch_obs} lists the Chandra net spectral counts of the combined epochs, as well as the number of epochs that pass the $>30$ net-count threshold.

\startlongtable
\tabletypesize{\tiny}
\begin{deluxetable*}{cccccccccccc}
\tablecaption{Properties of NuSTAR-selected AGN in the NEP-TDF\label{tab:agn_props}}
\tiny
\startdata
\\
\multicolumn{3}{c}{NuSTAR IDs$^a$} & XMM\tablenotemark{\tiny a} & Nu &  XMM & RA & DEC & $z$\tablenotemark{\tiny b} &  $F^{3-24}_{\mathrm{obs}}$  & $L^{3-24}_{\mathrm{obs}}$ &    \\
 ID & (Nu56) & (Nu89) & ID & net cts & net cts & [deg] & [deg]  &  & $10^{-14}$ [erg~s$^{-1}$cm$^{-2}$] & $10^{44}$ [erg~s$^{-1}$] &  Class\tablenotemark{\tiny c}    \\
 \\
 \hline
1 & 2 & 10 & 375 & 476.7 & 1373.5 & 261.004963 & 66.012388 & 0.8918 & 10.7$^{+\text{3.68}}_{-\text{3.37}}$ & 4.21$^{+\text{1.45}}_{-\text{1.33}}$ & Q (Z24) \\
2 & 14 & 23 & - & 278.1 & - & 260.359797 & 65.853886 & - & 2.89$^{+\text{1.2}}_{-\text{1.15}}$ & - & - \\
3 & 19 & 38 & 227 & 855.5 & 225.7 & 260.689412 & 65.740591 & 1.32 & 3.7$^{+\text{0.81}}_{-\text{0.78}}$ & 3.84$^{+\text{0.84}}_{-\text{0.82}}$ & Q (O26) \\
4 & 20 & 12 & 318 & 158.5 & 216.4 & 260.867239 & 65.982407 & 1.4252 & <7.15 & $<$8.97e-44 & Q (Z24) \\
5 & 21 & 14 & 305 & 408.9 & 645.8 & 260.823684 & 65.944847 & 0.4949 & 3.96$^{+\text{0.83}}_{-\text{0.8}}$ & 0.37$^{+\text{0.08}}_{-\text{0.08}}$ & G (Z24) \\
6 & 22 & 16 & 218 & 382.2 & 134.7 & 260.656285 & 65.931053 & 1.23 & 4.89$^{+\text{0.73}}_{-\text{0.71}}$ & 4.26$^{+\text{0.64}}_{-\text{0.62}}$ & Q (O26) \\
7 & 24 & 71 & 108 & 347.7 & 1466.7 & 260.42374 & 65.921001 & 0.5233 & 8.88$^{+\text{1.11}}_{-\text{1.06}}$ & 0.95$^{+\text{0.12}}_{-\text{0.11}}$ & Q (Z24) \\
8 & 26 & 31 & 378 & 184.3 & 201.8 & 261.025736 & 65.842349 & 1.8 & <2.15 & $<$4.8e-44 & Q (O26) \\
9 & 29 & 28 & 299 & 8447.9 & 7784.0 & 260.808941 & 65.796673 & 1.4411 & 27.9$^{+\text{0.8}}_{-\text{0.79}}$ & 35.99$^{+\text{1.04}}_{-\text{1.02}}$ & Q (Z24) \\
10 & 30 & 26 & 122 & 299.4 & 111.7 & 260.44349 & 65.839801 & 1.68 & 1.62$^{+\text{0.77}}_{-\text{0.75}}$ & 3.05$^{+\text{1.46}}_{-\text{1.41}}$ & Q (O26) \\
11 & 31 & 54 & 397 & 411.1 & 135.3 & 261.078906 & 65.815655 & 0.3759 & 4.89$^{+\text{1.17}}_{-\text{1.13}}$ & 0.24$^{+\text{0.06}}_{-\text{0.06}}$ & G (Z24) \\
12 & 35 & 34 & - & 146.2 & - & 260.53465 & 65.756721 & - & <2.68 & - & - \\
13 & 36 & 32 & 193 & 552.8 & 172.9 & 260.58976 & 65.734981 & 1.019 & 1.79$^{+\text{0.89}}_{-\text{0.86}}$ & 0.98$^{+\text{0.49}}_{-\text{0.47}}$ & Q (Z24) \\
14 & 37 & 42 & 379 & 299.5 & 307.3 & 261.035039 & 65.734455 & 0.16 & 3.49$^{+\text{1.58}}_{-\text{1.49}}$ & 0.02$^{+\text{0.01}}_{-\text{0.01}}$ & U (Z24) \\
15 & 41 & 19 & 266 & 346.0 & 486.9 & 260.751816 & 65.908622 & - & 2.3$^{+\text{0.6}}_{-\text{0.58}}$ & - & - \\
16 & 43 & 25 & 213 & 1259.9 & 1575.7 & 260.638622 & 65.84636 & 1.3369 & 2.35$^{+\text{0.52}}_{-\text{0.51}}$ & 2.52$^{+\text{0.56}}_{-\text{0.54}}$ & Q (Z24) \\
17 & 45 & 68 & 367 & 565.2 & 113.8 & 260.990613 & 65.824629 & 1.35 & 4.52$^{+\text{0.84}}_{-\text{0.82}}$ & 4.96$^{+\text{0.92}}_{-\text{0.90}}$ & Q (O26) \\
18 & 48 & 69 & - & 262.0 & - & 260.673401 & 65.659318 & - & 6.31$^{+\text{2.2}}_{-\text{2.08}}$ & - & - \\
19 & 53 & 72 & 279 & 983.6 & 258.3 & 260.778758 & 65.856746 & 0.8846 & 6.85$^{+\text{0.58}}_{-\text{0.56}}$ & 2.64$^{+\text{0.22}}_{-\text{0.22}}$ & Q (Z24) \\
20 & 54 & 55 & 327 & 547.4 & 367.5 & 260.893905 & 65.819897 & 0.7791 & 1.62$^{+\text{0.66}}_{-\text{0.64}}$ & 0.46$^{+\text{0.19}}_{-\text{0.18}}$ & G (Z24) \\
21 & 55 & 56 & 111 & 1778.1 & 2670.9 & 260.429729 & 65.766017 & 0.7813 & 15.9$^{+\text{1.26}}_{-\text{1.21}}$ & 4.52$^{+\text{0.36}}_{-\text{0.34}}$ & Q (Z24) \\
22 & 57 & 59 & 148 & 359.4 & 145.0 & 260.493365 & 65.736825 & 1.35 & 3.28$^{+\text{1.01}}_{-\text{0.97}}$ & 3.60$^{+\text{1.11}}_{-\text{1.06}}$ & U (Z24) \\
23 & 58 & 75 & 222 & 2586.7 & 663.2 & 260.671348 & 65.711891 & 0.1791 & 19.8$^{+\text{1.23}}_{-\text{1.19}}$ & 0.18$^{+\text{0.01}}_{-\text{0.01}}$ & G (Z24) \\
24 & 1 & - & - & 37.2 & - & 260.530704 & 66.059882 & - & 16.9$^{+\text{5.94}}_{-\text{5.38}}$ & - & - \\
25 & 3 & - & 207 & 183.4 & 83.8 & 260.615623 & 65.980155 & 0.0 & 2.8$^{+\text{1.04}}_{-\text{1.0}}$ & - & S (Z24) \\
26 & 4 & - & - & 424.4 & - & 260.74987 & 65.968707 & - & 2.35$^{+\text{0.96}}_{-\text{0.93}}$ & - & - \\
27 & 5 & - & - & 493.9 & - & 260.434776 & 65.961318 & - & 2.9$^{+\text{1.22}}_{-\text{1.16}}$ & - & - \\
28 & 6 & - & 154 & 728.1 & 233.8 & 260.522038 & 65.954027 & 0.62 & 7.39$^{+\text{0.99}}_{-\text{0.95}}$ & 1.19$^{+\text{0.16}}_{-\text{0.15}}$ & G (O26) \\
29 & 7 & - & 134 & 304.9 & 188.5 & 260.468713 & 65.93478 & 0.2699 & 2.28$^{+\text{0.96}}_{-\text{0.92}}$ & 0.05$^{+\text{0.02}}_{-\text{0.02}}$ & G (Z24) \\
30 & 8 & - & 208 & 160.3 & 79.0 & 260.622873 & 65.923493 & 0.81 & 1.29$^{+\text{0.68}}_{-\text{0.66}}$ & 0.40$^{+\text{0.21}}_{-\text{0.20}}$ & Q (O26) \\
31 & 9 & - & - & 114.6 & - & 261.188068 & 65.881698 & - & 7.26$^{+\text{2.7}}_{-\text{2.54}}$ & - & - \\
32 & 10 & - & 172 & 188.3 & 60.4 & 260.544655 & 65.887843 & - & 1.29$^{+\text{0.62}}_{-\text{0.61}}$ & - & - \\
33 & 11 & - & 254 & 357.7 & 290.9 & 260.73184 & 65.89017 & 0.6007 & 1.26$^{+\text{0.53}}_{-\text{0.52}}$ & 0.19$^{+\text{0.08}}_{-\text{0.08}}$ & Q (Z24) \\
34 & 12 & - & 129 & 322.9 & 104.7 & 260.466229 & 65.879465 & 5.35 & 2.26$^{+\text{0.75}}_{-\text{0.73}}$ & 68.96$^{+\text{22.98}}_{-\text{22.25}}$ & G (O26) \\
35 & 13 & - & 286 & 368.5 & 218.2 & 260.783715 & 65.872453 & - & 1.38$^{+\text{0.55}}_{-\text{0.54}}$ & - & - \\
36 & 15 & - & 163 & 141.3 & 114.8 & 260.539214 & 65.790189 & 2.2508 & 1.57$^{+\text{0.69}}_{-\text{0.67}}$ & 6.05$^{+\text{2.65}}_{-\text{2.58}}$ & Q (Z24) \\
37 & 16 & - & - & 352.2 & - & 261.103491 & 65.770554 & - & 5.21$^{+\text{1.87}}_{-\text{1.76}}$ & - & - \\
38 & 17 & - & - & 166.9 & - & 260.978502 & 65.723582 & - & 3.53$^{+\text{1.32}}_{-\text{1.26}}$ & - & - \\
39 & 18 & - & - & 180.2 & - & 260.403922 & 65.720876 & - & 4.58$^{+\text{1.76}}_{-\text{1.65}}$ & - & - \\
40 & 23 & - & 144 & 214.3 & 181.7 & 260.495899 & 65.907535 & 0.08 & <2.1 & $<$0.0 & U (Z24) \\
41 & 25 & - & - & 403.5 & - & 261.194464 & 65.859152 & - & 10.3$^{+\text{2.56}}_{-\text{2.43}}$ & - & - \\
42 & 27 & - & - & 414.6 & - & 260.523638 & 65.834538 & - & 1.37$^{+\text{0.62}}_{-\text{0.61}}$ & - & - \\
43 & 28 & - & 356 & 557.8 & 512.1 & 260.965651 & 65.836937 & 0.6726 & 3.15$^{+\text{0.77}}_{-\text{0.75}}$ & 0.62$^{+\text{0.15}}_{-\text{0.15}}$ & Q (Z24) \\
44 & 32 & - & 229 & 334.1 & 175.2 & 260.695612 & 65.808196 & 1.83 & <1.47 & $<$3.42e-44 & Q (O26) \\
45 & 33 & - & - & 126.1 & - & 261.11552 & 65.810138 & - & <6.96 & - & - \\
46 & 34 & - & 325 & 181.0 & 161.9 & 260.885542 & 65.781661 & 0.4031 & 1.12$^{+\text{0.69}}_{-\text{0.67}}$ & 0.06$^{+\text{0.04}}_{-\text{0.04}}$ & G (Z24) \\
47 & 38 & - & - & 247.2 & - & 261.006258 & 65.635478 & - & 25.6$^{+\text{6.8}}_{-\text{6.03}}$ & - & - \\
48 & 39 & - & 241 & 124.3 & 34.1 & 260.70263 & 66.014412 & 4.35 & <7.78 & $<$1.46e-42 & G (O26) \\
49 & 40 & - & - & 161.9 & - & 260.99109 & 65.922953 & - & <3.86 & - & - \\
50 & 42 & - & - & 122.6 & - & 261.054304 & 65.870312 & - & <3.03 & - & - \\
51 & 44 & - & - & 274.6 & - & 260.718727 & 65.845092 & - & <1.5 & - & - \\
52 & 46 & - & 281 & 254.6 & 106.8 & 260.769055 & 65.77412 & 0.6339 & <3.05 & $<$5.2e-45 & Q (Z24) \\
53 & 47 & - & - & 396.8 & - & 260.767207 & 65.821916 & - & <2.15 & - & - \\
54 & 49 & - & - & 199.7 & - & 260.982765 & 65.932518 & - & <3.86 & - & - \\
55 & 50 & - & - & 240.6 & - & 260.801946 & 65.929818 & - & <1.56 & - & - \\
56 & 51 & - & 183 & 227.9 & 20.5 & 260.557293 & 65.948639 & 0.47 & <3.46 & $<$2.9e-45 & U (Z24) \\
57 & 52 & - & - & 227.9 & - & 260.823766 & 65.895685 & - & 1.44$^{+\text{0.62}}_{-\text{0.6}}$ & - & - \\
58 & 56 & - & - & 242.3 & - & 260.566157 & 65.764297 & - & 1.56$^{+\text{0.77}}_{-\text{0.75}}$ & - & - \\
59 & 59 & - & - & 178.8 & - & 260.807485 & 65.705176 & - & <3.55 & - & - \\
60 & 60 & - & 248 & 319.6 & 0.0 & 260.726199 & 65.698978 & 0.44 & 3.08$^{+\text{1.12}}_{-\text{1.08}}$ & 0.22$^{+\text{0.08}}_{-\text{0.08}}$ & U (Z24) \\
61 & - & 1 & - & 80.9 & - & 260.430138 & 66.104772 & - & 5.09$^{+\text{1.8}}_{-\text{1.69}}$ & - & - \\
62 & - & 2 & - & 75.4 & - & 261.225616 & 66.092168 & - & 6.16$^{+\text{2.1}}_{-\text{1.92}}$ & - & - \\
63 & - & 3 & 131 & 239.3 & 482.3 & 260.465932 & 66.081661 & 0.913 & 11.1$^{+\text{1.72}}_{-\text{1.63}}$ & 4.63$^{+\text{0.72}}_{-\text{0.68}}$ & Q (S25) \\
64 & - & 4 & - & 105.9 & - & 261.208428 & 66.064208 & - & 5.0$^{+\text{1.75}}_{-\text{1.62}}$ & - & - \\
65 & - & 5 & 151 & 149.4 & 62.7 & 260.52171 & 66.068151 & - & 5.72$^{+\text{1.65}}_{-\text{1.56}}$ & - & - \\
66 & - & 6 & 214 & 59.4 & 121.6 & 260.654889 & 66.047252 & - & 4.66$^{+\text{1.69}}_{-\text{1.57}}$ & - & - \\
67 & - & 7 & - & 79.6 & - & 260.779362 & 66.043985 & - & 3.6$^{+\text{1.25}}_{-\text{1.17}}$ & - & - \\
68 & - & 8 & 386 & 106.0 & 102.2 & 261.051764 & 66.03986 & - & 4.43$^{+\text{1.3}}_{-\text{1.23}}$ & - & - \\
69 & - & 9 & - & 90.3 & - & 261.110572 & 66.035011 & - & 3.91$^{+\text{1.37}}_{-\text{1.29}}$ & - & - \\
70 & - & 11 & 127 & 359.2 & 59.3 & 260.450575 & 65.99763 & 0.63 & 4.32$^{+\text{1.27}}_{-\text{1.22}}$ & 0.71$^{+\text{0.21}}_{-\text{0.20}}$ & U (S25) \\
71 & - & 13 & 161 & 895.3 & 165.7 & 260.536841 & 65.951426 & - & 5.57$^{+\text{1.38}}_{-\text{1.32}}$ & - & - \\
72 & - & 15 & - & 53.9 & - & 260.179956 & 65.950854 & - & 5.5$^{+\text{1.51}}_{-\text{1.42}}$ & - & - \\
73 & - & 17 & - & 88.6 & - & 261.208412 & 65.924231 & - & 5.54$^{+\text{1.64}}_{-\text{1.54}}$ & - & - \\
74 & - & 18 & - & 341.3 & - & 260.932244 & 65.928634 & 0.3768 & 3.37$^{+\text{1.12}}_{-\text{1.08}}$ & 0.17$^{+\text{0.05}}_{-\text{0.05}}$ & G (W26) \\
75 & - & 20 & 134 & 229.4 & 204.9 & 260.47521 & 65.929418 & 0.2699 & 3.67$^{+\text{1.29}}_{-\text{1.24}}$ & 0.08$^{+\text{0.03}}_{-\text{0.03}}$ & G (S25) \\
76 & - & 21 & 237 & 229.6 & 81.3 & 260.716555 & 65.891657 & 0.0011 & 2.43$^{+\text{0.93}}_{-\text{0.89}}$ & 0.00$^{+\text{0.00}}_{-\text{0.00}}$ & G (S25) \\
77 & - & 22 & - & 115.5 & - & 261.130845 & 65.860866 & - & 3.01$^{+\text{1.07}}_{-\text{1.01}}$ & - & - \\
78 & - & 24 & - & 99.3 & - & 261.217563 & 65.852517 & - & 4.13$^{+\text{1.2}}_{-\text{1.14}}$ & - & - \\
79 & - & 27 & 238 & 239.6 & 19.9 & 260.696911 & 65.839056 & 1.46 & 2.33$^{+\text{0.74}}_{-\text{0.71}}$ & 3.11$^{+\text{0.99}}_{-\text{0.95}}$ & U (S25) \\
80 & - & 29 & 383 & 392.3 & 87.6 & 261.032178 & 65.830257 & - & 2.58$^{+\text{0.93}}_{-\text{0.89}}$ & - & - \\
81 & - & 30 & - & 213.9 & - & 260.543968 & 65.800541 & - & 2.1$^{+\text{0.79}}_{-\text{0.76}}$ & - & - \\
82 & - & 33 & - & 215.7 & - & 260.584422 & 65.76171 & - & 2.04$^{+\text{0.81}}_{-\text{0.78}}$ & - & - \\
83 & - & 35 & 51 & 143.0 & 291.2 & 260.247019 & 65.752449 & 2.2517 & 3.72$^{+\text{1.05}}_{-\text{1.0}}$ & 14.36$^{+\text{4.05}}_{-\text{3.86}}$ & Q (S25) \\
84 & - & 36 & 412 & 126.3 & 24.4 & 261.138436 & 65.779601 & - & 2.98$^{+\text{0.93}}_{-\text{0.88}}$ & - & - \\
85 & - & 37 & - & 221.7 & - & 260.7593 & 65.717589 & - & 4.14$^{+\text{1.06}}_{-\text{1.02}}$ & - & - \\
86 & - & 39 & 358 & 297.0 & 289.5 & 260.966839 & 65.71559 & 0.32 & 4.7$^{+\text{1.03}}_{-\text{1.0}}$ & 0.16$^{+\text{0.03}}_{-\text{0.03}}$ & U (S25) \\
87 & - & 40 & - & 172.9 & - & 260.606727 & 65.701668 & - & 3.02$^{+\text{1.14}}_{-\text{1.1}}$ & - & - \\
88 & - & 41 & - & 195.5 & - & 260.420482 & 65.723047 & 1.0811 & 3.34$^{+\text{1.01}}_{-\text{0.96}}$ & 2.12$^{+\text{0.64}}_{-\text{0.61}}$ & G (W26) \\
89 & - & 43 & - & 221.2 & - & 260.935306 & 65.723794 & - & 2.51$^{+\text{0.92}}_{-\text{0.89}}$ & - & - \\
90 & - & 44 & - & 84.4 & - & 260.002463 & 65.661076 & - & 26.8$^{+\text{8.01}}_{-\text{7.19}}$ & - & - \\
91 & - & 45 & 376 & 460.0 & 241.0 & 261.006377 & 65.641144 & - & 17.8$^{+\text{2.67}}_{-\text{2.52}}$ & - & - \\
92 & - & 46 & - & 77.5 & - & 260.556143 & 65.637381 & - & 9.43$^{+\text{3.21}}_{-\text{2.94}}$ & - & - \\
93 & - & 47 & - & 83.6 & - & 261.095875 & 66.091021 & - & 3.82$^{+\text{1.54}}_{-\text{1.45}}$ & - & - \\
94 & - & 48 & - & 72.7 & - & 260.908854 & 66.057688 & - & <3.26 & - & - \\
95 & - & 49 & - & 442.3 & - & 260.203822 & 65.997397 & - & 15.5$^{+\text{1.55}}_{-\text{1.47}}$ & - & - \\
96 & - & 50 & - & 53.8 & - & 261.139878 & 65.933912 & - & <5.53 & - & - \\
97 & - & 51 & - & 205.0 & - & 260.963974 & 65.897898 & - & <3.39 & - & - \\
98 & - & 52 & 164 & 512.6 & 377.6 & 260.533979 & 65.917278 & - & 3.91$^{+\text{1.24}}_{-\text{1.19}}$ & - & - \\
99 & - & 53 & 10 & 160.8 & 91.6 & 260.134804 & 65.891911 & 1.4216 & 7.28$^{+\text{1.59}}_{-\text{1.51}}$ & 9.08$^{+\text{1.98}}_{-\text{1.88}}$ & Q (S25) \\
100 & - & 57 & 6 & 251.4 & 23.0 & 260.098702 & 65.763353 & - & 11.8$^{+\text{1.79}}_{-\text{1.69}}$ & - & - \\
101 & - & 58 & - & 141.6 & - & 261.016829 & 65.755844 & - & 1.96$^{+\text{0.83}}_{-\text{0.8}}$ & - & - \\
102 & - & 60 & 372 & 203.0 & 600.5 & 261.00498 & 65.704644 & 1.6954 & 2.93$^{+\text{1.27}}_{-\text{1.2}}$ & 5.64$^{+\text{2.45}}_{-\text{2.31}}$ & Q (S25) \\
103 & - & 61 & 68 & 233.0 & 676.2 & 260.285696 & 65.734173 & 1.125 & 5.33$^{+\text{1.11}}_{-\text{1.05}}$ & 3.73$^{+\text{0.78}}_{-\text{0.73}}$ & Q (S25) \\
104 & - & 62 & 425 & 168.8 & 58.5 & 261.192466 & 65.700985 & - & 5.04$^{+\text{1.41}}_{-\text{1.33}}$ & - & - \\
105 & - & 63 & 270 & 174.0 & 226.0 & 260.754464 & 65.691638 & 2.1507 & 4.52$^{+\text{1.69}}_{-\text{1.59}}$ & 15.60$^{+\text{5.83}}_{-\text{5.49}}$ & Q (S25) \\
106 & - & 64 & 433 & 269.5 & 160.6 & 261.227579 & 65.722081 & 0.39 & 7.58$^{+\text{1.45}}_{-\text{1.38}}$ & 0.40$^{+\text{0.08}}_{-\text{0.07}}$ & U (S25) \\
107 & - & 65 & 55 & 144.5 & 185.2 & 260.265427 & 65.680148 & - & <8.73 & - & - \\
108 & - & 66 & 413 & 205.2 & 135.1 & 261.134087 & 65.66763 & - & 5.23$^{+\text{1.73}}_{-\text{1.63}}$ & - & - \\
109 & - & 67 & 31 & 133.0 & 39.7 & 260.195333 & 65.636141 & - & 14.4$^{+\text{4.81}}_{-\text{4.42}}$ & - & - \\
110 & - & 70 & 361 & 473.5 & 649.0 & 260.972775 & 66.052208 & 0.7831 & 14.6$^{+\text{1.48}}_{-\text{1.41}}$ & 4.17$^{+\text{0.42}}_{-\text{0.40}}$ & Q (S25) \\
111 & - & 73 & 317 & 475.0 & 717.2 & 260.863802 & 65.790025 & 1.7767 & <4.16 & $<$8.99e-44 & Q (W26) \\
112 & - & 74 & 233 & 299.1 & 103.2 & 260.709055 & 65.776254 & - & <2.37 & - & - \\
\hline
\enddata
\tablecomments{ \\
$^{a}$ NuSTAR IDs are listed from the total 112-source catalog (\citetalias{Silver26}), the Cycle 5+6 catalog (\citetalias{Zhao24}), and the Cycle 8+9 catalog (\citetalias{Silver25}). XMM-Newton IDs are given from the total Cycle 6+8+9 catalog (\citetalias{Silver26}).\\
$^{b}$ Photometric redshifts are rounded to two decimal places, while spectroscopic redshifts are rounded to four.\\
$^{c}$ S for Star, Q for Quasar, G for Galaxy, U for unidentified. Classifications are presented by \citetalias{Zhao24}, \citetalias{Silver25}, \cite{Ortiz26} (O26), and Willmer et al \textit{in prep} (W26).\\\
}
\end{deluxetable*}

\section{SPECTRAL ANALYSIS}\label{sec:analysis}
\begin{figure*}
    \centering
    \includegraphics[width=0.8\textwidth]{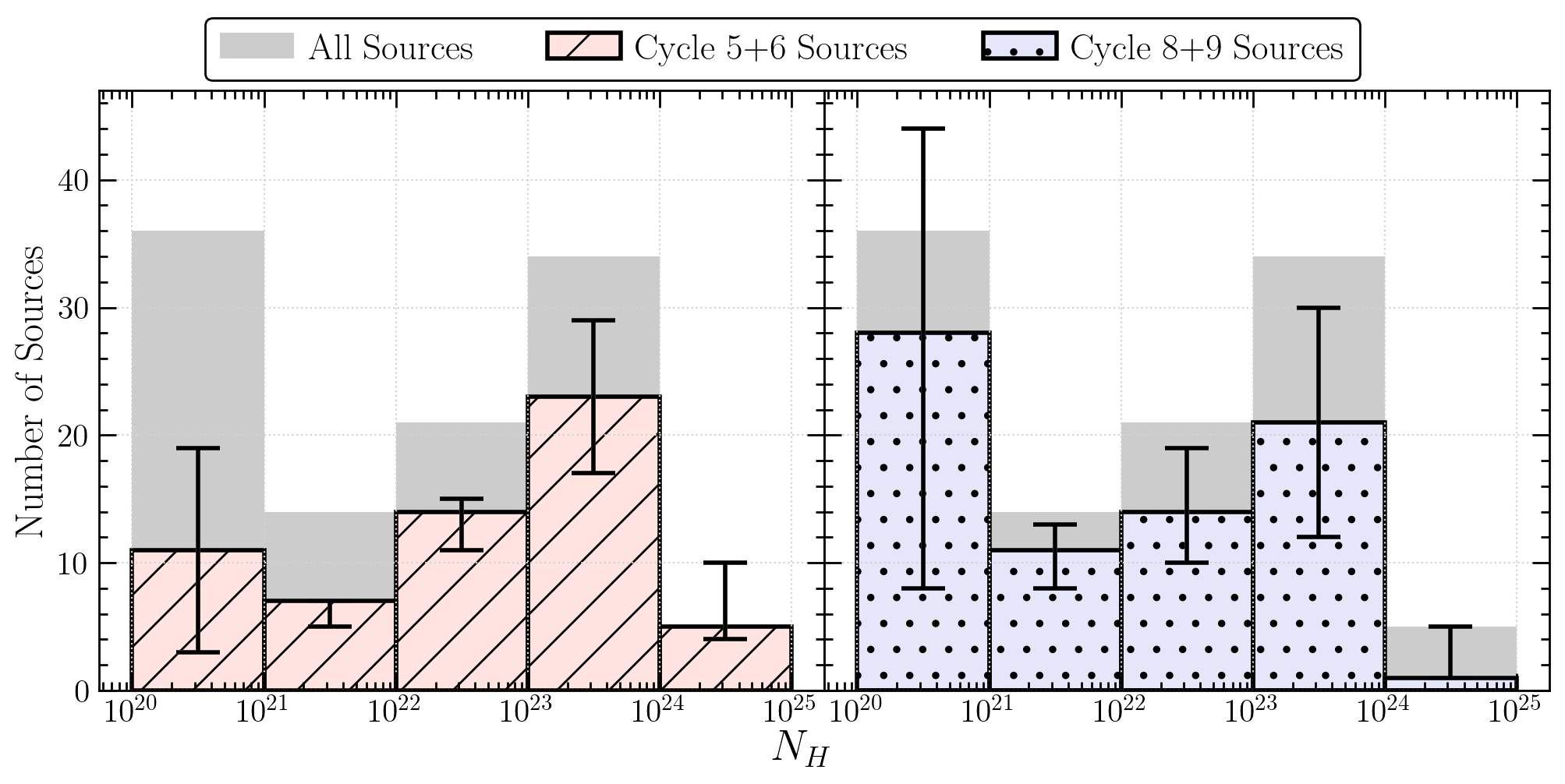}
    \includegraphics[width=0.8\textwidth]{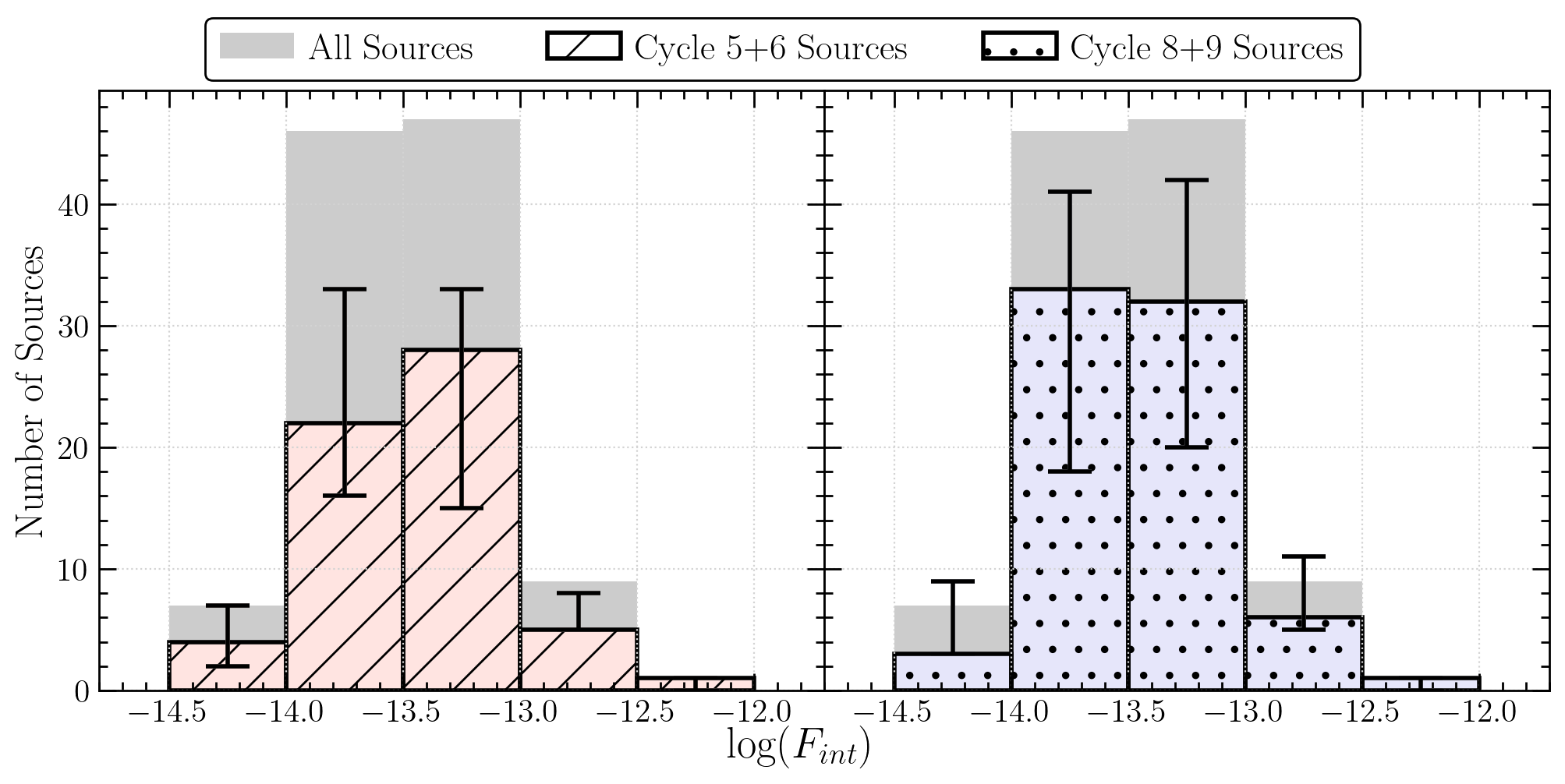}
    \caption{The observed $N_{\mathrm{H}}$ (top) and $F_{\mathrm{int}}$ (bottom) distribution for the entire NuSTAR NEP sample (grey histogram), the \citetalias{Zhao24} sample (left; pink, hatched), and the \citetalias{Silver25} sample (right; purple, dotted). The distributions shown here are calculated using standard \textsc{xspec} routines.} 
    \label{fig:non_var}
\end{figure*}

\subsection{X-ray Spectral Models}\label{sec:models}

The models used in this work have three additive components.
The primary component is a power-law that represents transmitted emission from the X-ray corona.
A secondary power-law models coronal emission that is Thompson scattered by photo-ionized material near the SMBH \citep[e.g.,][]{Matt19-hot_mirror}.

The third additive component accounts for thermal emission from collisionally ionized plasma within the host galaxy.
 \citetalias{Creech25} uses a \textsc{mekal} component for this thermal emission \citep{mekal1,mekal2,mekal3}.
However, \textsc{mekal} is not up to date with the latest atomic data.
While this has minor effects on the model fitting, we elected to use an \textsc{apec} component instead, as \textsc{apec} is linked to HEASOFT's atomic database.
The important parameters for this work are those of the transmitted power-law ($N_{\mathrm{H}}$, the photon index $\Gamma$, and intrinsic power-law flux $F_{\mathrm{int}}$), so this model change does not impact our ability to compare the analysis presented here with the results from \citetalias{Creech25}.

First, we employ a simplified ``baseline" model that represents the AGN as an obscured power-law. In \textsc{xspec}, the baseline model is expressed as follows:
\begin{equation}\label{eq:simp}
 \begin{split}
    \textsc{phabs}_{\textsc{Gal}} *( \textsc{zphabs}_{\textsc{tor}}*\textsc{zpowerlw}~ + \\
    f_{\textsc{scatt}} * \textsc{zpowerlw} + \textsc{apec).}
\end{split}
\end{equation}
Where the two absorption components represent photoelectric absorption from material within the Milky Way \citep[$\textsc{phabs}_{\textsc{Gal}}$;][which is $\sim 3 \times 10^{20}$~cm$^{-2}$ for all sources]{H_map90,H_map05,H_map16} and intrinsic absorption from the AGN torus or host galaxy ($\textsc{zTBabs}_{\textsc{tor}}$). 
The transmitted and scattered components are represented as redshifted power-laws (\textsc{zpowerlw}) with $\Gamma$ and the normalization ($n$) as the two free parameters.
However, instead of using $n$ to control the strength of the transmitted power-law, we chose to freeze $n$ to 1.0 and use a multiplicative \textsc{cflux} component (not shown in Equation 1), which calculates the flux of \textsc{zpowerlw} within a chosen energy band (we use the $2.0-10.0$~keV energy band for ease of comparison with other studies).
This allows easy calculation of intrinsic (unabsorbed) flux of the corona.
The parameters for the scattered component (both \textsc{cflux} and \textsc{zpolwerlw}) are linked to the transmitted component, with a constant $f_{\textsc{scatt}}$ controlling the scattering strength.

For unobscured sources, the baseline model described above is sufficient. 
However, for column densities log$(N_{\mathrm{H}}) \gtrsim 23$, effects of the obscuring torus itself (i.e., Compton scattering, cold reflection, florescent emission) become important, and a more complex torus model is needed to self-consistently account for those effects \citep[e.g.,][]{Yaqoob12_MYTorus, Balokovic18_borus, Buchner19_clumpy}.
To remedy this, we supplement the baseline model with the \textsc{uxclumpy} torus model by \cite{Buchner19_clumpy}.
\textsc{uxclumpy} is widely used for studies of obscured AGN \citep[e.g.,][]{Marchesi22-NGC1358, Akylas24-CT_local, Boorman25, Cox25, Pal25, Sengupta25, TA25-Mrk477}, and it follows the formalism of the \textsc{Clumpy} model by \citet{Nenkova08-clumpy1,Nenkova08-clumpy2}, which describes the IR emission from a gaussian distribution of dust clouds with various column densities.
In \textsc{xspec}, the \textsc{uxclumpy} model is expressed as follows:
 \begin{equation}\label{eq:clumpy}
 \begin{split}
 \textsc{phabs} *( \textsc{uxclumpy-cutoff}  + f_{\textsc{scatt}} * \\
 \textsc{uxclumpy-cutoff-omni} + \textsc{apec).}
\end{split}
\end{equation}
where a single absorption component represents Galactic absorption from the Milky Way.
\textsc{uxclumpy-cutoff} self-consistently includes the transmitted power-law alongside absorption and reflection from a clumpy torus that surrounds the central engine.
\textsc{uxclumpy-cutoff-omni} accounts for scattered emission, and its relative strength is controlled by $f_{\textsc{scatt}}$.
As with the baseline model, an \textsc{apec} component represents thermal emission from the host galaxy.  

For faint sources---such as those studied in this work---it is standard to fix the torus geometry parameters.
The covering fraction of the inner ring (CTKcover) is frozen at 0.4\footnote{As suggested by the \textsc{uxclumpy} github page: \url{https://github.com/JohannesBuchner/xars/blob/master/doc/ uxclumpy.rst}\label{ft:clump}}, while the inclination angle (Theta\textsc{\_}inc) and cloud dispersion (TORsigma) are set to $45 \degree$, which is the median value allowed by the model for both quantities.
CTKcover, Theta\textsc{\_}inc, and TORsigma are subtle parameters that cannot be constrained without high SNR data, and they are generally not degenerate with the other parameters in the modeling.
Therefore, freezing them does not change the results.
In the end, the only free parameters from \textsc{uxclumpy} are $N_{\mathrm{H}}$, $\Gamma$, $f_{\mathrm{scatt}}$, and the normalization of the transmitted emission from the corona ($n$).


\subsection{Fitting Procedure}\label{sec:non_var}

In \citetalias{Creech25}, the non-variable scenario (i.e., model parameters were not allowed to change over time) was assumed for the entire analysis of the \citetalias{Zhao24} catalog.
As a result, spectra from all epochs were combined using \textsc{addspec}\footref{ft:add} to improve model fitting (FPMA, FPMB, PN, M1, and M2 spectra are all kept separate. In this work, ACIS is separate as well).
We apply this methodology to the \citetalias{Silver25} catalog to find initial parameter estimates with no variability applied to the modeling.

\citetalias{Creech25} utilizes both standard \textsc{xspec} routines and the Bayesian X-ray Analysis package \citep[BXA;][]{Buchner14_BXA} to perform model fitting.
BXA is computationally expensive, especially when many parameters are left free, and the standard \textsc{xspec} routines are sufficient for the goals of this work, which is to identify and characterize variable candidates.

For all model fitting, we utilize the Cash statistic \citep[C;][]{Cash79}, which is the standard statistic for fitting poisson data.
When using C, it is ideal to fit source and background models simultaneously. 
However, given the complexity of the parameter space---in particular, the complexity of the NuSTAR background parameterization---simultaneous background and source fitting is not practical for this work.
We instead fit background-subtracted source spectra (using background spectra from the extraction procedures described in Section \ref{sec:DR}).
In \textsc{xspec}, using background-subtracted spectroscopy automatically results in a modified version of C---sometimes called the W statistic---being used\footnote{See the \textsc{xspec} user manual: \url{https://heasarc.gsfc.nasa.gov/docs/software/xspec/manual/XSappendixStatistics.html}}.
This does not change the interpretation of the statistic, and we continue to refer to the fitting statistic as C.
Section 3.3 of \citetalias{Creech25} details the fitting procedure, which we summarize here. 
The following steps are repeated for each individual source.
 
First, the main parameters of the transmitted power-law ($N_{\mathrm{H}}$, $\Gamma$, and $F_{\mathrm{int}}$ (baseline model) or $n$ (\textsc{uxclumpy} model)), are constrained one at a time, freezing other parameters.
All three parameters are then freed and constrained together.
In cases where $\Gamma$ is unconstrained or exceeds the expected physical bounds ($1.5 < \Gamma  < 2.2$), it is fixed to 1.8, which is the average value for AGN \citep[e.g.,][]{Nandra94,Ricci17-BAT}.\footnote{When fitting the \textsc{uxclumpy} model, this procedure is first carried out with hard-band-only data, then soft-band data is incrementally included in order to avoid falling into local minima in the parameter space, as recommended by the \textsc{uxclumpy} github page.}

The next step of the fitting procedure is to test for the presence of a scattered or thermal component.
These two components affect the soft part of the spectrum and have high degeneracy with one another.
As a result, they can only be constrained if the $0.1$--$3.0$~keV band is accessible; therefore, both components are excluded for sources without XMM-Newton or Chandra spectra.
When soft-band spectra exist, $f_\mathrm{scatt}$ and the normalization of the thermal component $n_\mathrm{soft}$ are freed one at a time.
Introducing a new component improves the fit if C decreases more than the decrease in degrees of freedom ($\Delta \mu - \Delta C > 0$).
Using this criteria---and in response to the high degeneracy between $f_\mathrm{scatt}$ and $n_\mathrm{soft}$---we chose to include only the component that most minimized C, with visual inspection to assure that the included component did not result in an unreasonable model fit. 
In cases where neither thermal nor scattered components significantly improved the fit, both were excluded.

The above procedure is repeated for both the baseline and the \textsc{uxclumpy} model. 
For all sources newly discovered in the \citetalias{Silver25} NuSTAR catalog, the non-variable parameter estimates are reported in Table \ref{tab:pars} in Appendix \ref{sec:a_fit}. 
Table 7 by \citetalias{Creech25} gives the best-fit parameters for the \citetalias{Zhao24} sources.
The observed $N_{\mathrm{H}}$ and $F_\mathrm{int}$ distributions for both catalogs is shown in Figure \ref{fig:non_var} (baseline model).
To obtain the uncertainties on the number of AGN in each bin ($n_\mathrm{AGN}$), the parameter distributions are calculated for three scenarios.
First, we assume the best-fit parameter values for the entire sample; this becomes the best-fit distribution.
Next, we assume the upper bound of the parameter uncertainties.
Finally, the lower bounds are assumed for the entire sample.
For each bin, the lowest/highest values of $n_\mathrm{AGN}$ (among all three scenarios) become the lower/upper bounds on the number of AGN in that bin.


\subsection{Variable Model Fitting}\label{sec:var_fit}

\begin{figure}
    \centering
    \includegraphics[width=0.49\textwidth]{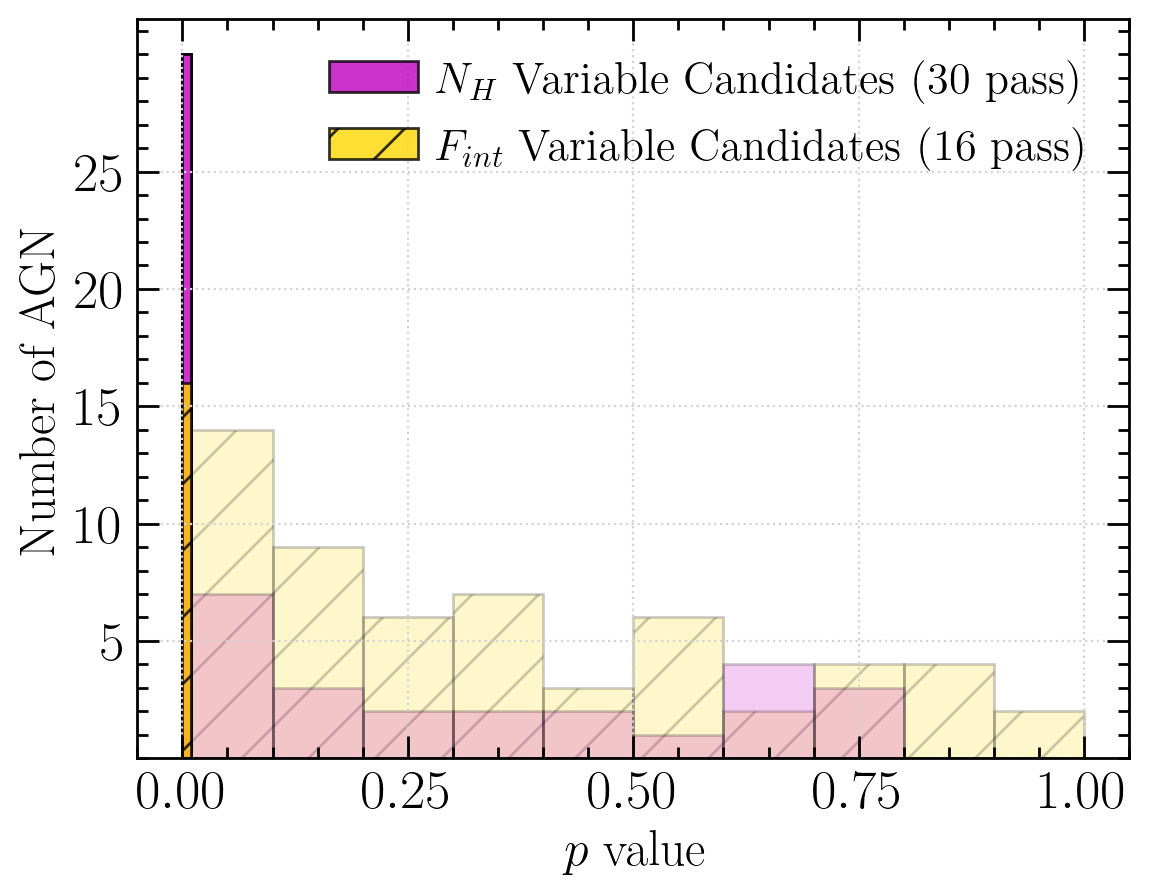}
    \caption{The p values for $N_{\mathrm{H}}$ (purple, solid) and $F_{\mathrm{int}}$ (gold, hatched) variable candidates. Sources that pass the test are placed in a single bin below the $p=0.01$ threshold.}
    \label{fig:par_p}
\end{figure}

\begin{figure}
    \centering
    \includegraphics[width=0.49\textwidth]{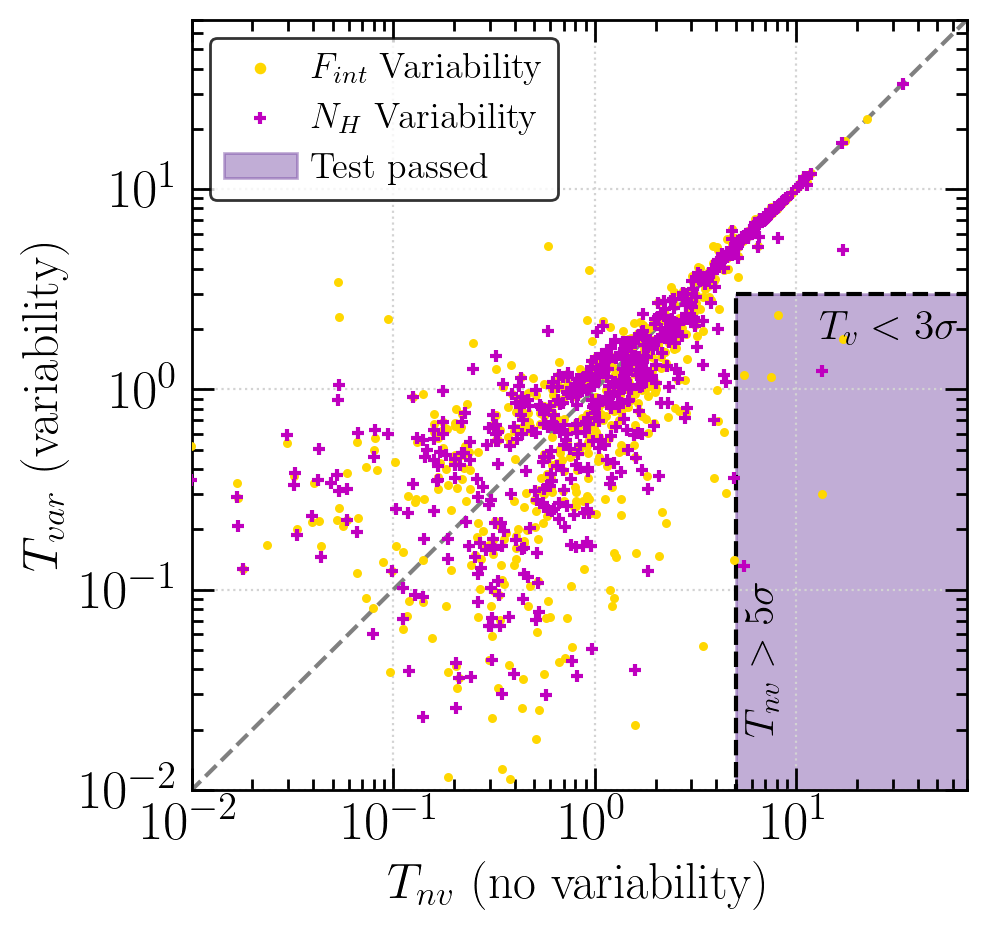}
    \caption{Comparing the tension between the data and model ($T$; Equation \ref{eq:T}) for variable (y axis) and non-variable (x axis) models. The purple region denotes the parameter space where variability is considered a significant improvement to the fit. Each point represents an individual epoch for a single source.}
    \label{fig:T}
\end{figure}

Three versions of the baseline model are employed to test for variability:

\begin{itemize}

\item \textbf{Scenario 1: No variability:} All free parameters are linked across epochs.

\item \textbf{Scenario 2: Obscuration variability:} $N_{\mathrm{H}}$ is allowed to vary between epochs.

\item \textbf{Scenario 3: Intrinsic flux variability:} $F_{\mathrm{int}}$ is allowed to vary between epochs.

\end{itemize} 

Below, we describe how the best-fit model parameters and statistics are acquired for each of the three scenarios.

\textbf{No variability:} 
The non-variable model fit statistics acquired in Section \ref{sec:non_var} cannot be directly compared to variable models.
This is because, in Section \ref{sec:non_var}, spectra from all epochs are added.
When modeling variability, the epochs must instead be kept separate. 
The spectra from each epoch are then loaded into separate data groups, allowing parameters to either be linked or left free to vary, as appropriate.
This methodology changes the number of spectral bins---and, as a result, the number of degrees of freedom---which makes comparison to the results of Section \ref{sec:non_var} non-trivial.
To alleviate this issue, we first apply the best-fit models from Table \ref{tab:pars} and \citetalias{Creech25} to a non-variable model with epochs retained as separate spectra.
This model is identical to the one used in Section \ref{sec:non_var}, except that parameters are linked across all epochs instead of applied to a single spectrum per instrument.
Next, the free parameters are re-fit in order to find the true minimum value for C when uncombined spectra are used. The parameter values resulting from this re-fitting are consistent with the original model fits obtained using co-added spectra.
The resulting statistics are pulled aside for comparison with the variable models.

\textbf{Obscuration variability:} 
In order to test for changes in the obscuring column density, we begin with the non-variable model described above, then free $N_{\mathrm{H}}$ across all epochs. 
All other parameters remain linked.
Once constraints on all free parameters are found, the statistics are pulled aside for comparison with the other scenarios.
We hereafter refer to this scenario as $N_{\mathrm{H}}$ variability.

\textbf{Intrinsic flux variability:} 
Once again, we begin with the non-variable model described above.
For the baseline model, we free $F_{\mathrm{int}}$ (from \textsc{cflux}) across all epochs.
For \textsc{uxclumpy}, $n$ is allowed to vary instead. 
Constraints on the transmitted power-law strength are obtained, and statistics are pulled aside.
We hereafter refer to this scenario as $F_{\mathrm{int}}$ variability.

Due to the faint nature of the NEP X-ray sample, many sources have low SNR and, as a result, are overfit ($C < \mu$) in one or more epochs.
This issue is exacerbated by more complex models.
For this reason, we do not initially test for simultaneous $N_{\mathrm{H}}$ and $F_{\mathrm{int}}$ variability.
Any sources that that pass the variability tests (described in the next section) for both scenarios are later fit to the more complex model that allows both parameters to change.

Similarly, we chose to carry out the variability analysis for the baseline model only, as the \textsc{uxclumpy} model is needlessly complex for most sources.
Then, for sources that pass the variability tests, \textsc{uxclumpy} is used to confirm that the detected variability is not dependent on the model used.
This methodology is discussed further in Section \ref{sec:var}.


\section{IDENTIFYING VARIABLE CANDIDATES}\label{sec:var}

As discussed in section \ref{sec:non_var}, changing a model improves the fit if C decreases more than the decrease degrees of freedom ($\Delta \mu - \Delta C > 0$).
Under this metric, 73 sources are identified as variability candidates (8 $N_{\mathrm{H}}$-only variable candidates,  17 $F_{\mathrm{int}}$-only variable candidates, 48 candidates for both scenarios).
However, this is not a robust statistical test.
In order to further test these variability candidates, we utilize two statistical estimators presented by \citet{TA23}.
One estimator tests the significance of parameter deviation from the average.
The other utilizes the Cash statistic to estimate the tension between the model and spectral data.

\subsection{Test 1: Parameter Variability}\label{sec:p_test}

The first estimator discussed here tests the significance of the parameter deviations from the average.
This was done by assuming a null hypothesis of no variability (i.e., the value of the parameter is constant over time) and calculating the significance of the parameter's deviation in each epoch (summed over the number of epochs $n_\mathrm{epoch}$) via a $\chi^2$ computation:
\begin{center}
\begin{equation}\label{eq:p}
    \chi^2 = \sum_{n=1}^{n_\mathrm{epochs}} \frac{( x_i - \langle x \rangle )^2 }{\delta(x_i)^2},
\end{equation}
\end{center}
where $x_i$ is the best-fit parameter value ($N_{\mathrm{H}}$, $F_\mathrm{int}$, or $n$, as appropriate) at that epoch, $\langle x \rangle$ is the average parameter value (taken from epochs where both upper and lower bounds are constrained), and $\delta(x_i)$ is the uncertainty of $x_i$.
However, the uncertainties on $x_i$ are asymmetric ($\sigma_+ \neq \sigma_-$), requiring caution while calculating $\delta(x_i)$. 
Without access to the full probability distribution, we instead follow \cite{Barlow03,TA23} and utilize a simple formalism that describes the probability distribution function as two straight lines meeting at the best-fit value.
Under this model, $\delta(x_i)$ is equal to either $\sigma_+$ or $\sigma_-$, depending on which side of the error-bar ``points" to the average value of the model. 
In the case of an upper limit, $\sigma_- = x_i -  x_\mathrm{min}$, where $x_\mathrm{min}$ is the minimum value allowed by the model.
Similarly, in the case of a lower limit,  $\sigma_+ = x_\mathrm{max} - x_i$, where $x_\mathrm{max}$ is the maximum value allowed by the model. 
For the obscuration-variable model, $x_\mathrm{min}$ and $x_\mathrm{max}$ occur at $N_{\mathrm{H}}=10^{19},10^{26}$~cm$^{-2}$.
For the flux-variable model, $x_\mathrm{min}$ and $x_\mathrm{max}$  are either log$(F_{2-10})=-100,100$ (baseline model) or $n= 0, 10^{24}$ (\textsc{uxclumpy} model).
In all cases, epochs where $x_i$ is unconstrained are not included in this analysis.

Once the cumulative $\chi^2$ value is found, a p-value ($p$) was calculated using the function \textsc{scipy.stats.chi2.cdf}, where the degrees of freedom is $\mu = n_\mathrm{epochs} - 1$.
Sources with $p < 0.01$---which corresponds to a significance of $\sim 2.3\sigma$---are classified as variable candidates (Figure \ref{fig:par_p}).
Using this test, we find 30 candidates for $N_{\mathrm{H}}$ variability and 16 candidates for $F_{\text{int}}$ variability.

\subsection{Test 2: Tension Between Data and Model}\label{sec:t_test}

The next estimator proposed by \cite{TA23} uses the fit statistic to calculate a tension $T$, which measures the number of standard deviations that the model strays from a perfect fit.
In \cite{TA23}, most of the sources are fit using a $\chi^2$ statistic, and the reduced $\chi^2$ value is used in the tension calculation.
However, bins with zero variance are common for the sources in this work, rendering $\chi^2$ invalid as a fitting statistic.
We instead use a reduced Cash statistic $C/\mu$ as a proxy for $\chi^2_\mathrm{red}$.
From this, $T$ can be calculated:
\begin{center}
\begin{equation}\label{eq:T}
    T = \frac{|1-C_\mathrm{red}|}{\sigma},
\end{equation}
\end{center}
where $1-C_\mathrm{red}=0$ would indicate a perfect fit to the model, and $\sigma$ is the standard deviation for the $C_\mathrm{red}$ distribution. 
When $T < 3 \sigma$ for two models, they cannot be reliably differentiated from one another \cite[e.g.,][]{Andrae10-Chisq}.
In response, we consider variability to be a possible improvement when $T_{\text{var}} < 3 \sigma$ and $T_{\text{nv}} > 3 \sigma$, where higher thresholds of $T_{\text{nv}}$ are considered as stricter criteria.
Under the strictest threshold used in this work, variability is considered a significant improvement to the fit when $T_{\text{var}} < 3 \sigma$ and $T_{\text{nv}} > 5 \sigma$.
This calculation is performed for each individual epoch, and the results are shown in Figure \ref{fig:T}.

There are two important caveats to the calculation of $T$.
First, the exact value of $\sigma$ is non-trivial to calculate; this is because the probability density function for the Cash statistic is not known.
As a result, a generalized analytical value of the standard deviation does not exist.
For many systems, $C_\mathrm{red}$ approaches the $\chi^2_\mathrm{red}$ distribution at $\gtrsim 10$ counts \cite[e.g.,][]{Kaastra17,Bonamente20-Cash}, which has a variance of $\sigma^2 = 2/\mu$.
We chose to utilize $\sigma = \sqrt{2/\mu}$ for the Tension calculation; therefore, the results of this Tension test should be taken as an approximation and may not be valid for the faintest sources in this work.

Second, it is common in this sample for sources to have $C < \mu$ for one or more epochs.
This indicates that the model is overfitting the data for that epoch.
In these cases, further complexity is unnecessary and increases the calculated tension.
This scenario is demonstrated by points falling above the $T_{\text{var}} = T_{\text{nv}}$ line in Figure \ref{fig:T}; these epochs do not pass the Tension test.

Under the strictest threshold for $T_{nv}$, four sources (AGN 5, 7, 9, and 21) pass the Tension test for one or more epochs. 
Two (AGN 9, 21) pass the test for $F_\mathrm{int}$ variability, and the others (AGN 5, 7) pass for both types of variability.
Under the least strict threshold, 29 sources pass the Tension test (9 $N_{\mathrm{H}}$ variable candidates, 15 $F_\mathrm{int}$ variable candidates, and 5 candidates for both).

\tabletypesize{\footnotesize}
\begin{deluxetable*}{cccccccccc}
\tablecaption{Results of Variability Tests for Final Candidates\label{tab:var_info}}
\startdata
\\
ID & $N_\mathrm{eps}$ & C/DOF$_\mathrm{nv}$\tablenotemark{\small a} & Var. par. & Model & $\Delta$C/ $\Delta$DOF\tablenotemark{\small c} &  $N_\mathrm{var-eps}$        & $N_\mathrm{var-eps}$        & $N_\mathrm{var-eps}$ &  $p$-value\tablenotemark{\small d}  \\
 &     &        &   ( $N_{\mathrm{H}}$/$F_\mathrm{int}$)   &   (b/c)\tablenotemark{\small b}      &                      & $(T_\mathrm{nv} >3\sigma)$ & $(T_\mathrm{nv} >4\sigma)$  & $(T_\mathrm{nv} >5\sigma)$ & \\
\hline
5 & 19 & 1281.1, 1293.0/2426 & $N_{\mathrm{H}}$ & b & -(59.1/18) & 2 & 2 & 2 & $\sim0$ \\
& & & & c & -(64.2/18) & 2 & 2 & 2 & $\sim0$ \\
 &  &  & $F_\mathrm{int}$ & b & -(69.3/18) & 2 & 2 & 2 & $3.2 \times 10^{-04}$ \\
& & & & c & -(71.7/18) & 2 & 2 & 2 & $9.2 \times 10^{-04}$ \\
\hline
7 & 19 & 1637.0, 1632.8/3931 & $F_\mathrm{int}$ & b & -(47.9/18) & 1 & 1 & 1 & $3.1 \times 10^{-07}$ \\
& & & & c & -(47.9/18) & 1 & 1 & 1 & $\sim0$ \\
 &  &  & $N_{\mathrm{H}}$ & b & -(52.4/18) & 1 & 1 & 1 & $7.9 \times 10^{-10}$ \\
& & & & c & -(49.9/18) & 1 & 1 & 1 & $\sim0$ \\
\hline
9 & 50 & 9956.1, 9973.9/11190 & $F_\mathrm{int}$ & b & -(926.0/49) & 3 & 3 & 1 & $\sim0$ \\
& & & & c & -(926.6/49) & 3 & 2 & 1 & $\sim0$ \\
\hline
21 & 42 & 5960.2, 5955.7/8590 & $F_\mathrm{int}$ & b & -(744.3/41) & 1 & 1 & 1 & $\sim0$ \\
& & & & c & -(753.9/41) & 1 & 1 & 1 & $\sim0$ \\
\hline
23 & 29 & 3221.3, 3201.3/4297 & $N_{\mathrm{H}}$ & b & -(141.8/28) & 2 & 0 & 0 & $1.7 \times 10^{-12}$ \\
& & & & c & -(144.9/28) & 2 & 0 & 0 & $5.0 \times 10^{-06}$ \\
\hline
43 & 14 & 1298.1, 1284.4/1085 & $F_\mathrm{int}$ & b & -(112.5/13) & 3 & 1 & 0 & $2.6 \times 10^{-11}$ \\
& & & & c & -(104.9/13) & 3 & 1 & 0 & $\sim0$ \\
\hline
71 & 8 & 573.3, 585.2/414 & $F_\mathrm{int}$ & b & -(57.8/7) & 1 & 0 & 0 & $1.9 \times 10^{-06}$ \\
& & & & c & -(55.1/7) & 1 & 0 & 0 & $1.5 \times 10^{-04}$ \\
 &  &  & $N_{\mathrm{H}}$ & b & -(30.7/7) & 1 & 0 & 0 & $\sim0$ \\
& & & & c & -(61.1/7) & 1 & 0 & 0 & $\sim0$ \\
\hline
\enddata
\tablecomments{ \\
$^{a}$ Non-variable fit statistics listed for the baseline and \textsc{uxclumpy} models: C$_b$,C$_c$/DOF.\\\
$^{b}$ b=baseline model, c=\textsc{uxclumpy} model\\\
$^{c}$ (C$_\mathrm{var}-$C$_\mathrm{nv})/($DOF$_\mathrm{var}-$DOF$_\mathrm{nv})$\\\
$^{d}$ At high values of $\chi^2$, the $p$ value becomes smaller than the resolution of the \textsc{scipy.stats.chi2.cdf} function, which is indicated by $p \sim 0$. \\\
}
\end{deluxetable*}

\begin{figure*}
    \centering
    \includegraphics[width=0.9\textwidth]{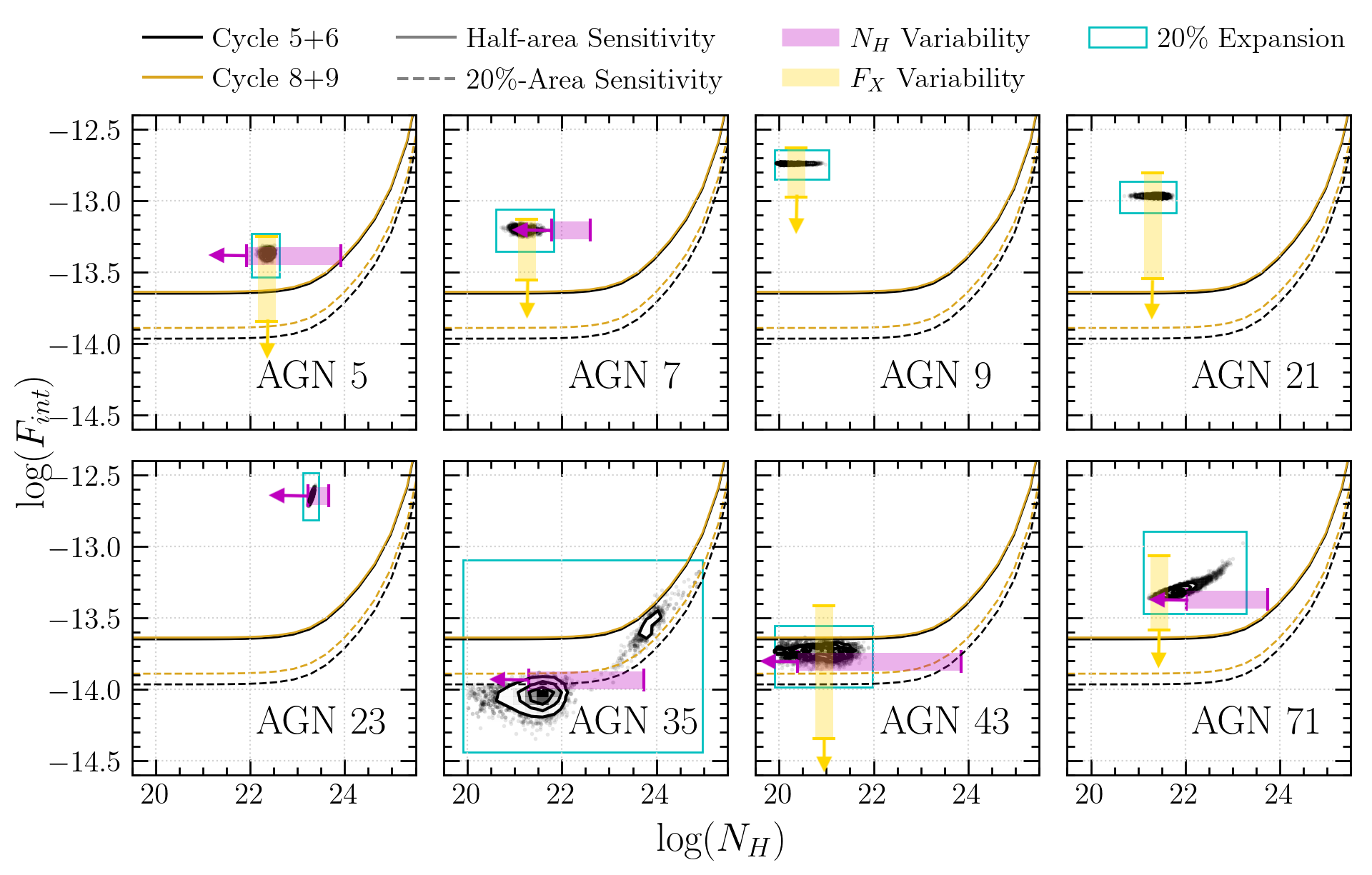}
    \caption{The non-variable $N_{\mathrm{H}}$ and $F_{\mathrm{int}}$ posteriors obtained from BXA (see \citetalias{Creech25}) compared with the extent of the observed $N_{\mathrm{H}}$ (pink bars) and $F_\mathrm{int}$ (yellow bars) variability.  In order to estimate the severity of cross-calibration uncertainties, the cyan box gives a rough sketch of the parameter space if the posteriors are expanded by $20 \%$. These constraints are compared to the half area (solid lines) and 20$\%$ area (dashed lines) flux limits of the NuSTAR Cycle 5+6 (black lines) and Cycle 8+9 (gold lines) surveys} 
    \label{fig:contours}
\end{figure*}

\subsection{Addressing Systematics}\label{sec:sys}

In order to bolster the results of the statistical tests described in Sections \ref{sec:p_test} and \ref{sec:t_test}---which utilized the baseline model---each source that passes both the $p$ and Tension tests were fit for variability using the \textsc{uxclumpy} model as well (Equation \ref{eq:clumpy}).
Three sources (AGNs 38, 87, and 111) that appear as variable candidates under the baseline model do not pass the tests when \textsc{uxclumpy} is used, so these sources are not classified as variable. 
In all three cases, this discrepancy is due to increased uncertainties when utilizing the more complex \textsc{uxclumpy} model. 
One source (AGN 23) passes under the $N_{\mathrm{H}}$ variable scenario when using \textsc{uxclumpy}, but not the $F_\mathrm{int}$ variable scenario, so only the former case is considered.

Next, the variability of the remaining candidates is compared against the constraints on $N_{\mathrm{H}}$ and $F_\mathrm{int}$ obtained from the Bayesian X-ray Analysis package \citep[BXA;][see \citetalias{Creech25} for the methodology]{Buchner14_BXA}.
The variable regions are denoted by colorful bars in Figure \ref{fig:contours}, which are created by determining the minimum and maximum values of $N_{\mathrm{H}}$ and/or $F_\mathrm{int}$ across epochs.
For all sources, at least one epoch has an upper limit on $N_{\mathrm{H}}$ and/or $F_\mathrm{int}$; as a result, the bars in Figure \ref{fig:contours} should be considered upper limits on the variability range.

From Figure \ref{fig:contours}, it is possible to estimate the strength of the variability results by comparing the variability range to the non-variable posteriors.
For example, the variability displayed by AGN 35 is entirely engulfed by the uncertainties on its posterior; as a result, we do not consider this source to be a variable candidate moving forward.
In two other sources (AGN 7 and 71), the scale of variability is comparable to the extent of their non-variable constraints.
We chose to consider these candidates in our analysis, with the caveat that this variability is not definitive. 

In some cases, the $N_{\mathrm{H}}$ variability does not overlap with the non-variable best-fit (albeit upper limits encompass the best-fit value; e.g. AGN 7).
In fact, the time-averaged value of $N_{\mathrm{H}}$ (when assuming variability) is in all cases higher than the non-variable best-fit value for $N_{\mathrm{H}}$ (see Appendix \ref{sec:a_var} for plots that demonstrate this effect), although this effect is less drastic when using the \textsc{uxclumpy} model, which accounts for the reflection components that become important at high obscuration (note, for example, the difference between the non-variable baseline and \textsc{uxclumpy} fits for AGN 71; Appendix \ref{sec:a_var}).
The discrepancy between the average and non-variable $N_{\mathrm{H}}$ values is likely caused by the relatively unobscured epochs contributing more photons and thus dominating the co-added spectra used in the non-variable model fitting.

Figure \ref{fig:contours} points out an additional selection effect built into the spectral analysis procedure that could bias this work against detecting variability, especially $N_{\mathrm{H}}$ variability. 
An estimate of the half-area and $20\%$-area flux sensitivity curves (3--24~keV energy band) are plotted within the $N_{\mathrm{H}}$-$F_\mathrm{int}$ parameter space (see Section \ref{sec:s26} for a description of the calculation).
With the exception of AGN 43, the variability constraints shown in Figure \ref{fig:contours} do not extend below the 20$\%$-area sensitivity curves.
If a source goes through a particularly strong obscuration event and/or becomes very faint---especially during a CT-like epoch---the spectral quality decreases, which may cause that epoch to dip below the minimum count threshold (i.e., fall below the survey sensitivity) and get excluded from the analysis. 

Next, the inclusion of Chandra-only epochs opens the possibility of cross-instrument calibration errors, which could mimic spectral variability. 
Comparing NuSTAR-only epochs with XMM+NuSTAR epochs introduces a similar issue. 
The cross-calibration error of NuSTAR compared with XMM-Newton is $\sim 20 \%$, and the estimated NuSTAR-Chandra uncertainty is $\sim 15 \%$ \citep[][T\"{u}mer et al. \textit{in prep}, Inoue et al. \textit{in prep}]{Madsen17}. 
In most of the variable candidates, the magnitude of variability exceeds known cross-calibration uncertainties.
We demonstrate the possible effects of this uncertainty by plotting a rough $20\%$ expansion of the parameter space in Figure \ref{fig:contours}.
Furthermore, all of the final candidates pass the parameter variability test when Chandra-only and NuSTAR-only epochs are excluded (i.e., only XMM-Newton + NuSTAR epochs are considered, which use the full 0.5--24~keV band), demonstrating that observed variability in the modeling is likely not due to cross-calibration errors alone.
However, in AGN 9, all epochs that pass the Tension test are Chandra-only; while this caveat is important to note, this source also demonstrates $F_\mathrm{int}$ variability that is significantly more extensive than the $20 \%$ error regions shown in Figure \ref{fig:contours}.
We determine that cross-calibration is most likely not a strong contributor to the detected variability in the candidates.

\subsection{Test Results}\label{sec:both_tests}

After AGN 35 is excluded, seven sources pass the $p$ and the Tension tests for both the baseline and \textsc{uxclumpy} models. 
These variable candidates are shown in Table \ref{tab:var_info}.
Four of the candidates (AGN 5, 7, 9, and 21) pass the strictest criteria for the Tension test ($T_\mathrm{var} < 3 \sigma$, $T_\mathrm{var}  > 5 \sigma$).
Three additional sources display improved model tension ($T_\mathrm{var} < 3 \sigma$, $T_\mathrm{nv} > 3 \sigma$; AGN 23, 43, and 71) for at least one epoch.
However, $T_\mathrm{nv}$ is below the strict $5 \sigma$ threshold for these sources.
As a result, the variable status of these three candidates is not definitive.
Figures showing the spectral and parameter variability of all seven candidates are presented in Appendix \ref{sec:a_var}.

\begin{figure}
    \centering
    \includegraphics[width=0.4\textwidth]{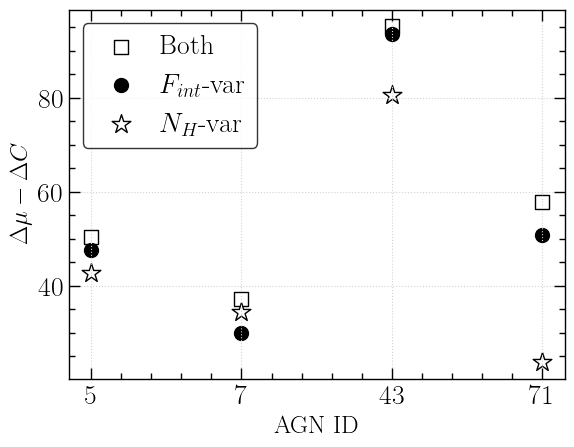}
    \caption{Comparison of $\Delta \mu - \Delta C$ (where higher positive values indicate greater improvement to the fit) for the sources that display both $N_{\mathrm{H}}$ and $F_{\mathrm{int}}$ variability. Allowing both parameters to vary improves the fit compared to the single-parameter variable models.}
    \label{fig:del_both}
\end{figure}

Four of the sources in Table \ref{tab:var_info} (AGN 5, 7, 43, and 71) are candidates for both $N_{\mathrm{H}}$ and $F_{\mathrm{int}}$ variability.
In all cases, fitting these sources to a model that allows both parameters to vary causes slight improvement to the the fit (Figure \ref{fig:del_both}).
It is thus possible that AGN 5, 7, 35, and 71 are varying in both X-ray flux and line-of-sight obscuration.
However, especially when there are few counts available, there is strong degeneracy between $N_{\mathrm{H}}$ and $F_{\mathrm{int}}$ at $N_{\mathrm{H}} > 10^{23}$~cm$^{-2}$ (i.e., see AGN 35 and 71 in Figure \ref{fig:contours}). 
For the sources that pass the statistical tests for both scenarios, it is therefore not feasible to distinguish which is the dominant cause of variability.

\begin{figure*}
    \centering
    \includegraphics[width=0.4\textwidth]{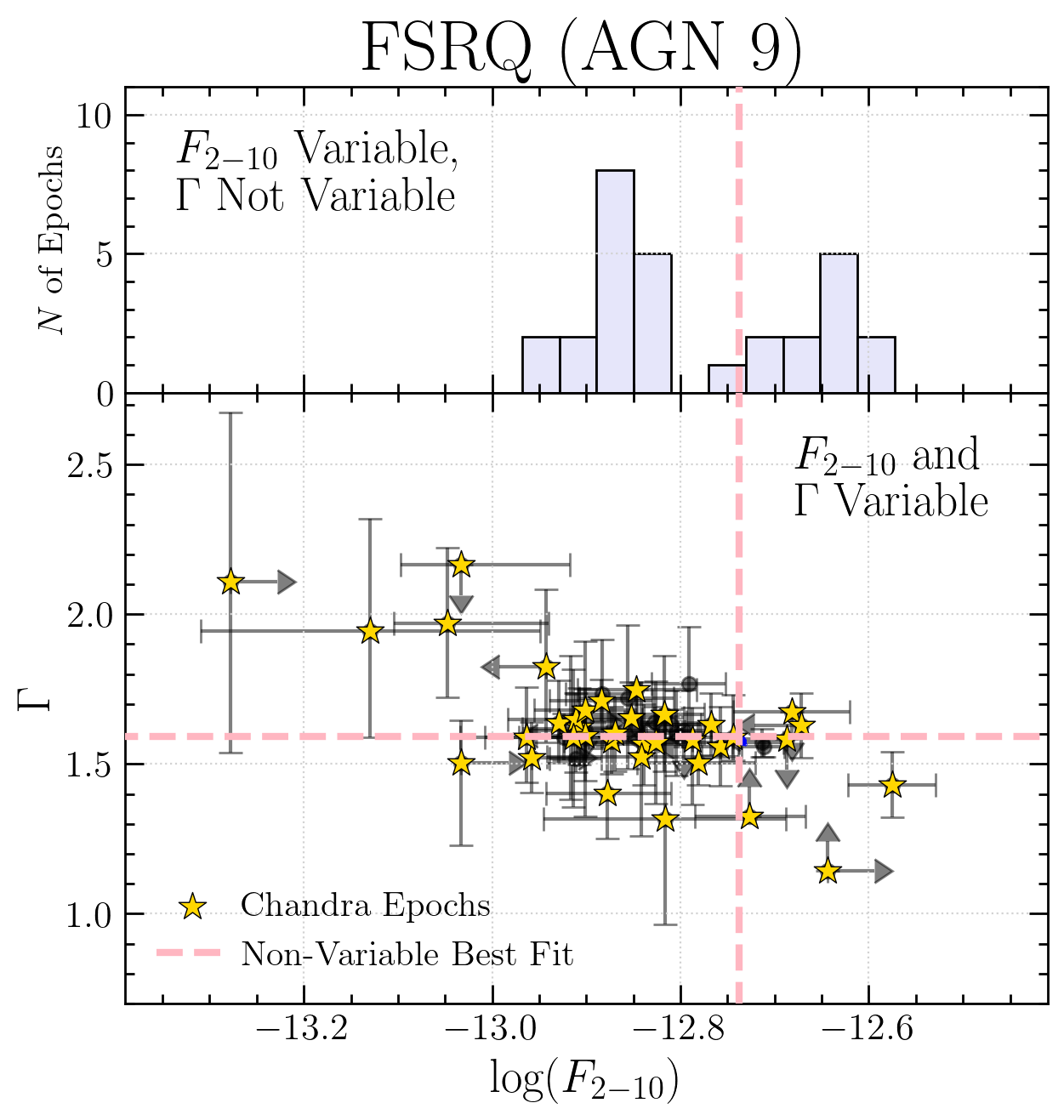}
    \includegraphics[width=0.4\textwidth]{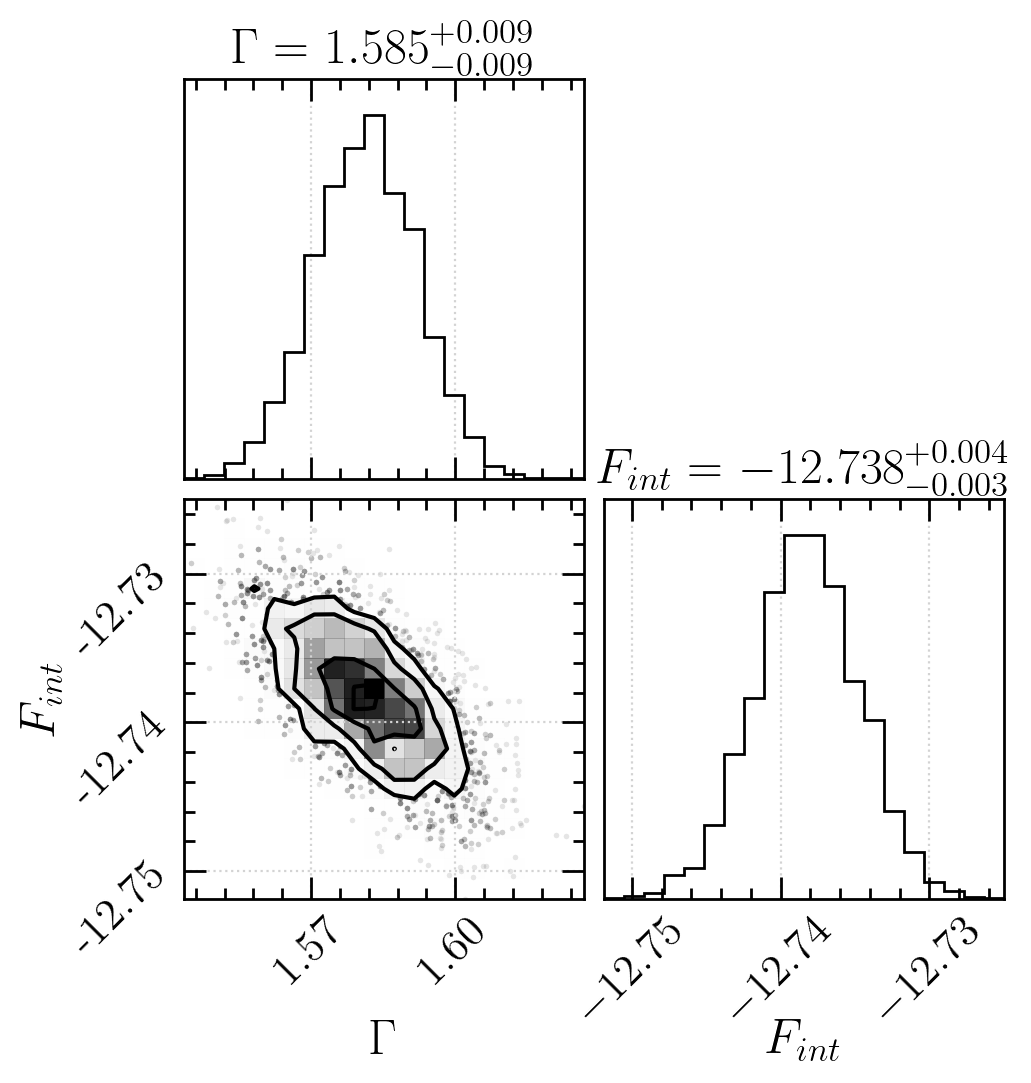}
    \caption{\textbf{Left, Bottom:} The photon index ($\Gamma$) as a function of 2-10 keV flux for observations of AGN 9, which is an FSRQ and the brightest X-ray source in the NEP-TDF. Epochs with Chandra data only (no NuSTAR spectra) are denoted by yellow stars. Dashed lines indicate the best-fit values of $\Gamma$ and $F_{\mathrm{int}}$ when the non-variable scenario is assumed. \textbf{Left, Top:} A histogram of the flux for  AGN 9 when $\Gamma$ is not allowed to vary. The dashed line shows the best-fit $F_{\mathrm{int}}$. \textbf{Right:} A corner plot showing the degeneracy between $F_{\mathrm{int}}$ and $\Gamma$ when the non-variable scenario is assumed (Fit taken from \citetalias{Creech25}).} 
    \label{fig:fsrq}
\end{figure*}

One of the most notable variable sources is AGN 9, which passes the above tests for both $F_\mathrm{int}$ and $N_{\mathrm{H}}$ variability; however, this source has been previously identified as a flat spectrum radio quasar \citep[e.g.,][]{SDSS02_QSOs}, which are not expected to have dust obscuration.
Thus, we should not see $N_{\mathrm{H}}$ variability for AGN 9. 
It is possible that the $N_{\mathrm{H}}$ variability identified in the modeling is due to changes in $\Gamma$, which indicates variability in the jet.
In order to test this, both $\Gamma$ and $F_{\mathrm{int}}$ were allowed to vary across epochs.
Figure \ref{fig:fsrq} shows the results, and this FSRQ may demonstrate ``harder-when-brighter" trend observed in some blazars \citep[e.g.,][]{Zhang05-blazar,Zhang06-blazar,Bhatta18-blazars}.
However, the correlation is more likely due to degeneracy between $\Gamma$ and $F_{\mathrm{int}}$, which is shown by the corner plot shown in Figure \ref{fig:fsrq}.

\section{DISCUSSION}\label{sec:discussion}
\begin{figure*}
    \centering
    \includegraphics[width=0.31\textwidth]{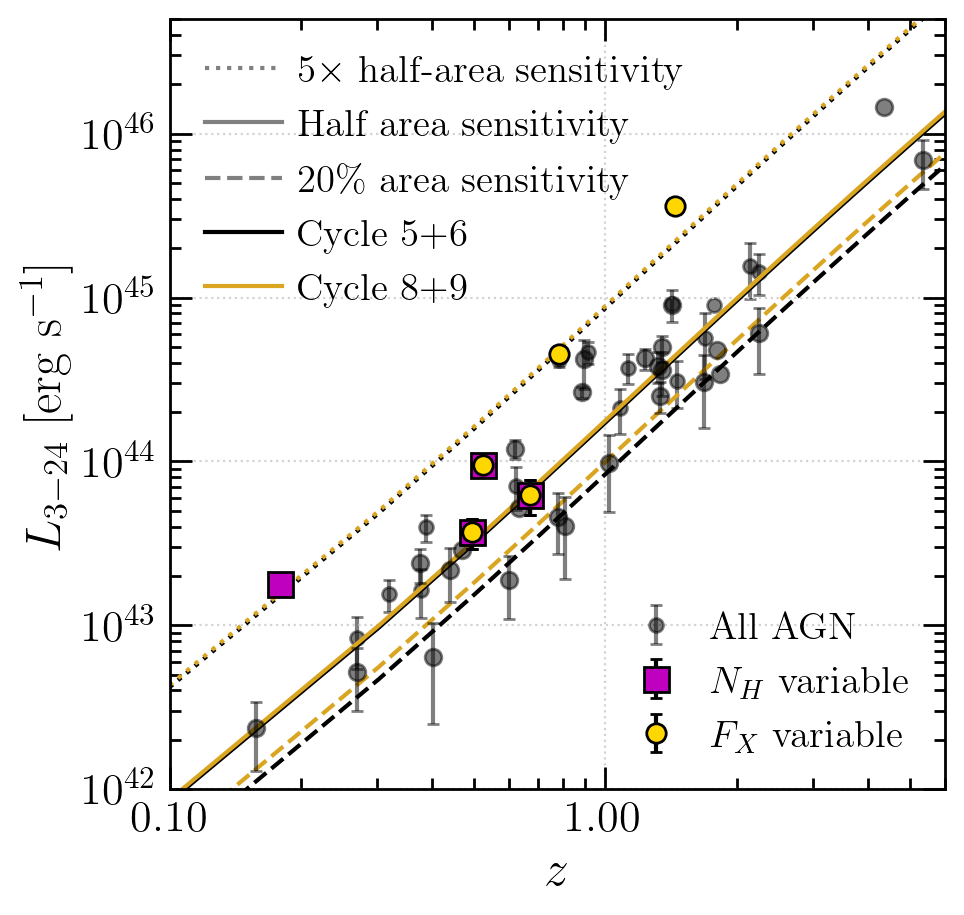}
    \includegraphics[width=0.30\textwidth]{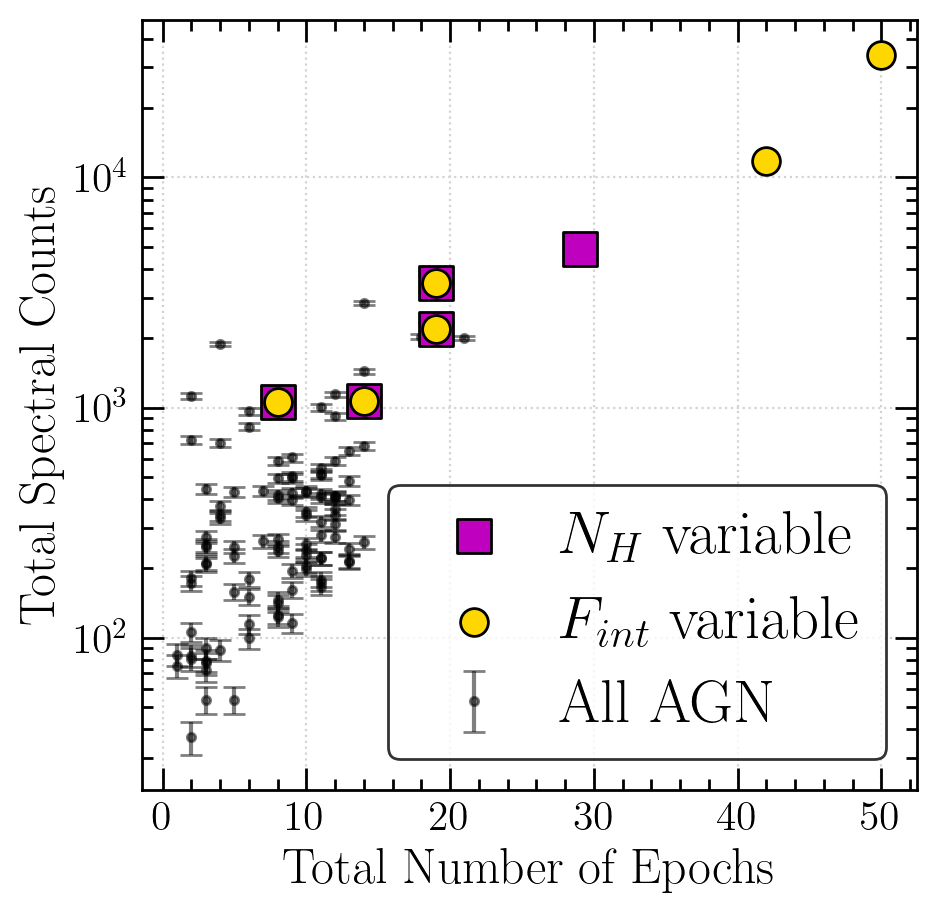}
    \includegraphics[width=0.367\textwidth]{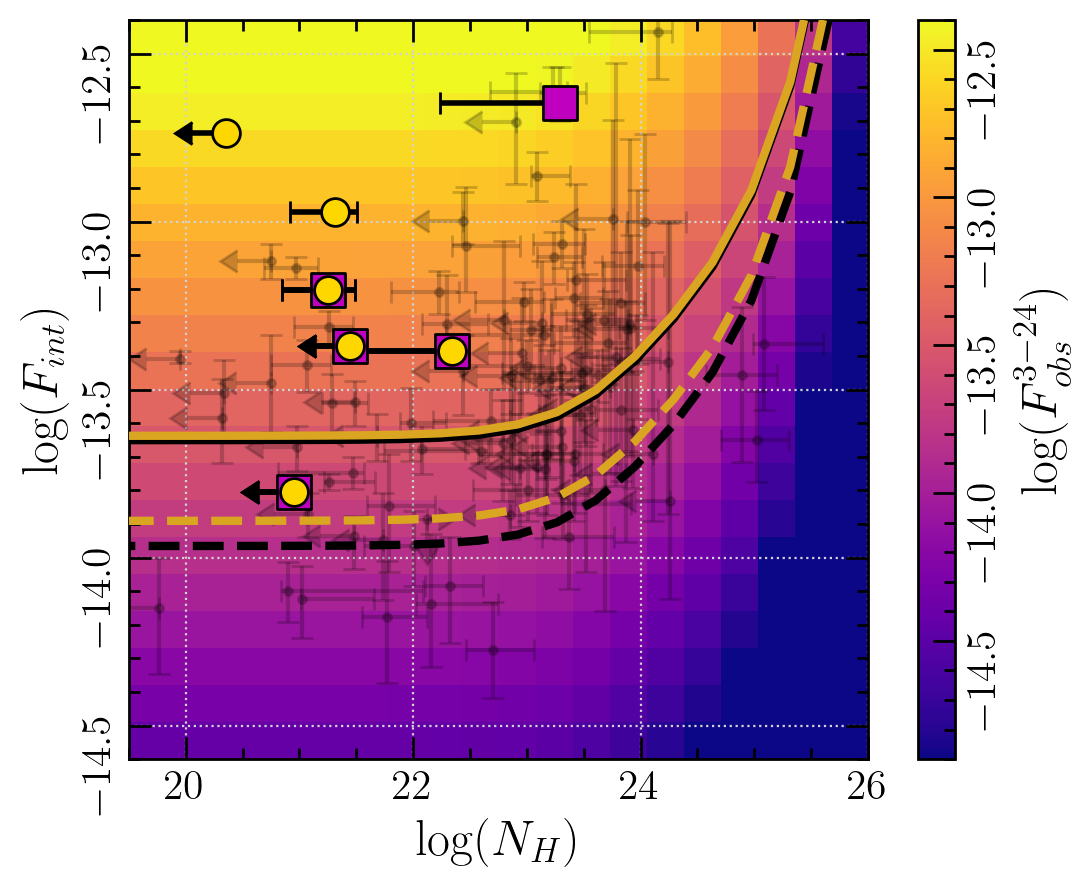}
    \caption{X-ray properties of the $F_{\mathrm{int}}$ (yellow circles) and $N_{\mathrm{H}}$ (pink squares) variable candidates compared to the total NuSTAR-selected sample (black points). \textbf{Left:} The redshifts and 3--24~keV luminosities of sources with multi-wavelength counterparts. Sources are compared with the sensitivity limits of the NuSTAR Cycle 5+6 (black lines) and Cycle 8+9 (gold lines) surveys. \textbf{Middle:} The total spectral counts across all epochs as a function of the number of available epochs per source. \textbf{Right:} The best-fit parameter values of $N_{\mathrm{H}}$ and $F_{\mathrm{int}}$ when no variability is assumed. The background color denotes the observed 3--24~keV flux that is expected given the parameter values at that location, and this is compared to the half area (solid lines) and 20$\%$ area (dashed lines) flux limits of the NuSTAR Cycle 5+6 (black lines) and Cycle 8+9 (gold lines) surveys.}
    \label{fig:var_props}
\end{figure*}


\begin{figure*}
    \centering
    \includegraphics[width=0.99\textwidth]{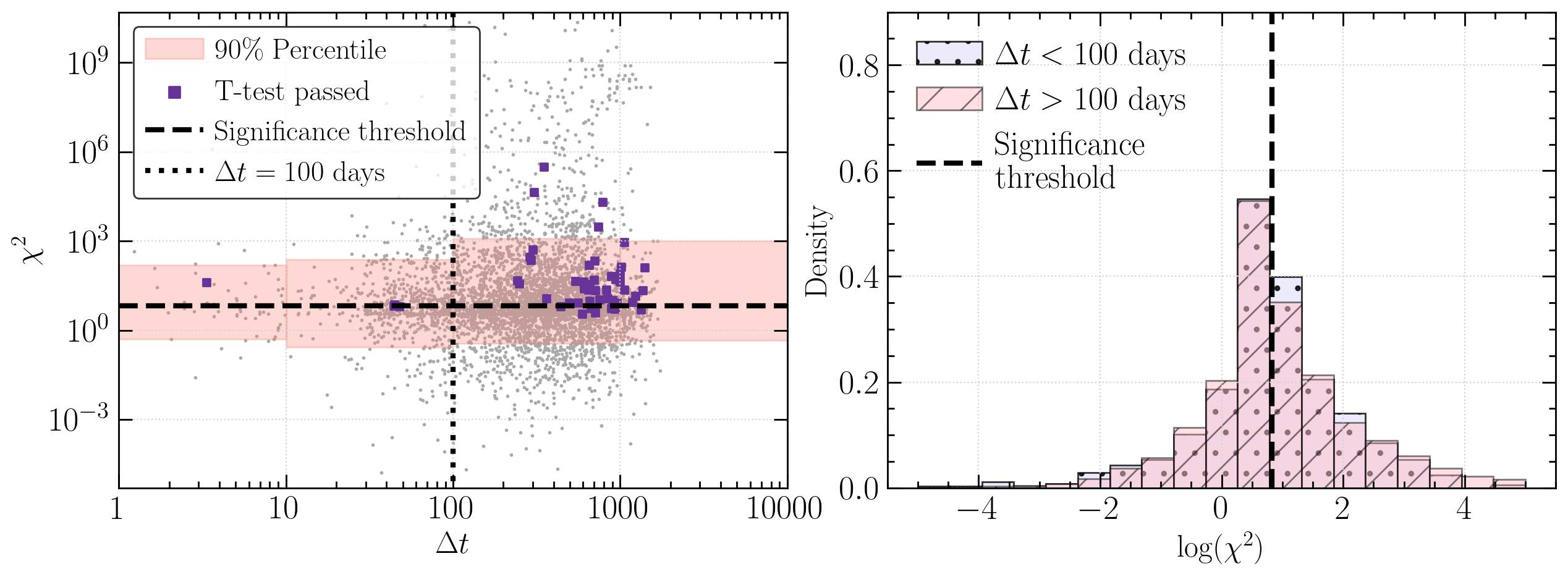}
    \caption{\textbf{Left:} Comparison between the $\chi^2$ value of $N_{\mathrm{H}}$ variability and the difference in time ($\Delta t$) between observation pairs. Observation pairs that pass the Tension test in Section \ref{sec:t_test} are denoted by purple squares, while the entire sample is shown by the grey background points. The pink shaded region shows the 10-90$^\text{th}$ percentile of the $\chi^2$ distribution in 1~dex $\Delta t$ bins. \textbf{Right:} A normalized histogram of $\chi^2$ split into $\Delta t$ below (purple; dotted) and above (pink; hashed) 100 days, where variability from a dusty torus is expected to occur on timescales $\Delta t > 100$ days. In both plots, a black dashed line denotes the $\chi^2 = 6.65$ \reviewchange{($p=0.01$)} significance threshold.}
    \label{fig:delT}
\end{figure*}

\tabletypesize{\footnotesize}
\begin{deluxetable*}{cccccccccc}
\tablecaption{Spectroscopic Variability Results of Photometric Variability Candidates from \citetalias{Silver26} \label{tab:S26_comp}}
\startdata
\\
ID & $\sigma_\mathrm{var}$\tablenotemark{\footnotesize a} & C/DOF$_\mathrm{nv}$ & $N_\mathrm{eps}$ & Var. par. & $\Delta$C/ $\Delta$DOF\tablenotemark{\footnotesize b} & $N_\mathrm{var-eps}$   & $N_\mathrm{var-eps}$        & $N_\mathrm{var-eps}$ &  $p$-value\tablenotemark{\footnotesize c} \\
      &  3-24~keV    &    &           &      &        &     $(T_\mathrm{nv} >3\sigma)$ &$(T_\mathrm{nv} >4\sigma)$  &$(T_\mathrm{nv} >5\sigma)$ & \\
\hline
9 & 8.5 & 9956.1/11190 & 50 & $N_{\mathrm{H}}$ & -(294.5/49) & 2 & 2 & 0 & $\sim0$ \\
& & & & $F_{\mathrm{int}}$ & -(926.0/49) & 3 & 3 & 1 & $\sim0$ \\
23 & 3.67 & 3221.3/4297 & 29 & $N_{\mathrm{H}}$ & -(141.8/28) & 2 & 0 & 0 & $1.4 \times 10^{-10}$ \\
& & & & $F_{\mathrm{int}}$ & -(162.8/28) & 2 & 0 & 0 & $8.1 \times 10^{-06}$ \\
63 & 3.48 & 177.5/177 & 2 & $N_{\mathrm{H}}$ & -(1.2/1) & 0 & 0 & 0 & - \\
& & & & $F_{\mathrm{int}}$ & -(10.0/1) & 0 & 0 & 0 & 0.03 \\
73 & 3.47 & 78.4/73 & 5 & $N_{\mathrm{H}}$ & -(-1.0/3) & 1 & 0 & 0 & $2.1 \times 10^{-10}$ \\
& & & & $F_{\mathrm{int}}$ & -(-4.6/3) & 1 & 0 & 0 & 0.1 \\
\hline
\enddata
\tablecomments{ \\
$^{a}$ The significance of 3-24~keV photometric variability calculated by \citetalias{Silver26} .\\\
$^{b}$ (C$_\mathrm{var}-$C$_\mathrm{nv})/($DOF$_\mathrm{var}-$DOF$_\mathrm{nv})$\\\
$^{c}$ A dash indicates that no epochs were fully constrained, disallowing a \reviewchange{$p$} calculation.\\\
}
\end{deluxetable*}

\subsection{Spectral Properties of the NEP-TDF NuSTAR Sample}

Assuming the non-variable scenario (Section \ref{sec:non_var}), Figure \ref{fig:non_var} shows the observed $N_{\mathrm{H}}$ and $F_{\mathrm{int}}$ distributions for the NEP-TDF NuSTAR survey, as obtained from the standard \textsc{xspec} routines using the baseline model.
The observed CT fractions ($f_\mathrm{CT} = N_\mathrm{CT}/N_\mathrm{Total}$) are consistent across subsamples ($f_\mathrm{CT} = 0.05_{-0.01}^{+0.07}$ for the entire NEP NuSTAR-selected sample, $0.08_{-0.02}^{+0.08}$ for the \citetalias{Zhao24} sample, and $0.01_{-0.00}^{+0.05}$ for the \citetalias{Silver25} sample).
However, flux-limited surveys---including the X-ray surveys of the NEP-TDF---are biased against detecting absorbed sources near the flux limit, and this must be corrected in order to extrapolate the underlying $N_{\mathrm{H}}$ distribution and the true value of $f_\mathrm{CT}$ \citep[e.g.,][]{Burlon11-BAT,Zappacosta18-NUSTAR_extragal_hard-band}. 
This process is non-trivial and beyond the scope of this work.
Additionally, the small area of the NEP-NuSTAR surveys (0.19 deg$^2$ for the Cycle 5+6 survey, 0.30 deg$^2$ for the Cycle 8+9 survey) renders the samples susceptible to cosmic variance \citep{Driver10-cosmic_var}.
\citetalias{Creech25} estimates the intrinsic $N_{\mathrm{H}}$ distribution using the \citetalias{Zhao24} sample; future studies may expand on their results to include the newly-discovered \citetalias{Silver25} sources as well.

\subsection{Comparison to \citetalias{Silver26} }\label{sec:s26}

Table \ref{tab:S26_comp} lists the sources that display $>3 \sigma$ photometric variability in their NuSTAR observations (\citetalias{Silver26}) alongside the results of the spectroscopic variability tests done in this work. 
Of the four photometric variables, only AGN 9 is definitively variable via spectroscopic means; however, AGN 23 also shows evidence of spectroscopic variability.
These two sources both have notable multi-wavelength properties (Section \ref{sec:var_mw}).

The other two variable sources identified by \citetalias{Silver26}  do not have detectable spectroscopic variability.
However, the photometric variability of these sources does not exceed $2.5 \sigma$ in any band, and each has fewer than 5 epochs for model fitting.
In contrast, almost all of the sources with measurable spectroscopic variability have $>10$ epochs (Table \ref{tab:var_info}).
As \cite{Cox25} finds (and as Figure \ref{fig:var_props} (middle panel) suggests), the likelihood of detecting spectral variability is systematically dependent on the number of available epochs; therefore, we do not rule out that these sources are variable, but more observations and higher quality spectroscopic data are needed.

Additionally, this work identifies five variability candidates (AGN 5, 7, 21, 43, and 71) without significant photometric variability in their NuSTAR data (\citetalias{Silver26}).
Each source has an explanation for why spectroscopic variability may be detected.
First, \citetalias{Silver26}  analyzed NuSTAR and XMM-Newton data separately, while this work carries out a joint analysis.
As a result, AGN 5, 21, and 43---which are matched to variable XMM-Newton sources (IDs 305, 111, and 356, respectively)---are also flagged as variable here.
Meanwhile, AGN 7 has $13$ Chandra epochs, which were not included in the \citetalias{Silver26}  analysis.
Lastly, AGN 71 shows possible count rate variability (see Appendix \ref{sec:a_var}) that was not strong enough to pass the significance thresholds used by \citetalias{Silver26} .

\subsection{X-ray Properties of Variable Candidates}\label{sec:s26}

In addition to the above results, \citetalias{Silver26}  finds that NuSTAR sources with photometric variability are preferentially brighter than $5 \times$ the sensitivity limit of the NEP X-ray surveys.
As Figure \ref{fig:var_props} (left panel) shows, spectroscopic analysis yields a similar result.
The three candidates that fall below the $5 \times$ sensitivity curve (AGN 5, 7, 43) are either matched with bright, variable XMM-Newton sources (AGN 5 and 43) or have supplemental multi-epoch Chandra data (AGN 7). 
Additionally, the variable sources tend to have the greatest number of epochs and total spectral counts when compared to the full sample (middle panel of Figure \ref{fig:var_props}).
All of the variable candidates have $>1000$ total spectral counts split across $\gtrsim 10$ epochs.
Together, these results support the need for both high-SNR spectra and multi-year monitoring in order to continue the search for X-ray variability in faint AGN.

In order for variability analyses to be informative for population synthesis models \citep[e.g.,][]{Comastri95-CXB,Gilli07-CXB,Ueda14-CXB,Annana19-Accretion_HistoryI,Gerolymatou25}, the sample and variable candidates must be compared against the sensitivity of the survey.
Calculating the full, redshift and luminosity-dependent selection function of a survey is non-trivial \citep[e.g.,][]{Peca23} and beyond the scope of this paper.
We instead perform a simple calculation here.
After binning the $N_{\mathrm{H}}$--$F_{\mathrm{int}}$ parameter space covered by the sample, the baseline model (Equation \ref{eq:simp}) is used in order to estimate the observed 3--24 keV flux that is expected within that $N_{\mathrm{H}}$--$F_{\mathrm{int}}$ bin.
For this calculation, a modest scattered component ($f_{scatt} = 0.001$) and the median redshift of the full NEP sample ($z = 0.88$) are assumed.
The results are given by the colormap in the leftmost panel of Figure \ref{fig:var_props}.
Contours are drawn at the half-area and 20$\%$-area flux limits of both the Cycle 5+6 and Cycle 8+9 NuSTAR surveys (on both Figures \ref{fig:contours} and \ref{fig:var_props}), allowing for a comparison of the model parameters, estimated survey limits, and the extent of detectable variability.
As shown in Figure \ref{fig:var_props}, the unabsorbed fluxes spanned by the variable candidates appear to be representative of the sample.
However---with the notable exception of AGN 23---the variable candidates mostly occupy the unobscured parameter space.
This result likely reflects the intrinsic challenge of measuring faint, obscured sources due to the attenuation of X-ray photons.
Assessing variability lowers the statistics of the modeling by splitting epochs into separate spectra, which exacerbates the challenge presented by obscured sources.
Ultimately, all but one candidate falls well above the half-area flux limit, and the upper limit of the variability itself almost never extends below the $20\%$-area sensitivity (Figure \ref{fig:contours}).

The observed variability presented in this work does not provide compelling evidence that any sources underwent changing-look events during the NEP observation campaigns.
Within uncertainties, the observed $N_{\mathrm{H}}$ variability does not suggest a change in obscuration regimes (i.e., CT to Compton thin), as required for a changing-obscuration classification.
Similarly, no sources show clear $F_{\mathrm{int}}$ variability beyond an order of magnitude; thus, we do not see evidence that the X-ray coronae were destroyed or significantly changed by changing-state events.
Given that changing-look events are rare and the NEP X-ray surveys cover a small area of the sky, it is unsurprising that no changing-look AGN appear in the sample.




\subsection{Multi-Wavelength Properties of Variable Candidates}\label{sec:var_mw}

Excluding AGN 71, all of the variable candidates in Table \ref{tab:var_info} match with multi-wavelength counterparts, and two are coincident with JWST NIRcam observations that allow their galaxy properties to be studied in detail. 
Cross-matching and classifications are presented by \citetalias{Zhao24}---who classify objects by cross-correlating optical counterparts against SDSS template spectra\footnote{https://classic.sdss.org/dr5/algorithms/spectemplates/}---and \cite{Ortiz26} (\citetalias{Ortiz26} hereafter).
\citetalias{Ortiz26} infers galaxy types via SED inferences made with X-CIGALE \citep{Boquien19-CIGALE,Yang20-CIGALE} using all available broadband photometry from space (JWST \citep[][Jansen et al \textit{in prep}]{Windhorst23-NEP}); HST \citep{OBrien24}; WISE \citep{unWISE19}) and the ground (HSC \citep{HSC18}, SDSS \citep{SDSS_DR17}, MMT \citep{Willmer23-MMT}, JPAS \citep{Hernan-Caballero23-JWST_NEP_z}). 
Of particular interest is the $f_{\mathrm{AGN}}$ parameter, which is the the fractional contribution of the AGN SED to the 0.1-30 $\mu$m bolometric luminosity.
Generally, sources with $f_{\mathrm{AGN}} > 0.2$ are inferred to be QSOs, but due to large uncertainties \citep[$\sim 5$-10$\%$;][]{Ciesla15}, we require $f_{\mathrm{AGN}} > 0.3$ for QSO classification; otherwise, the source is classified as a composite galaxy.

Of the variable sources, the most notable are AGN 9 and AGN 23, which both display spectroscopic and photometric variability.
As discussed in previous sections, AGN 9 is a a well-studied FSRQ \citep[SDSS J172314.14 + 654746.1, ][]{SDSS02_QSOs}, and it is the brightest X-ray source in the NEP-TDF.
In addition to displaying the most significant X-ray variability, AGN 9 is also a known radio variable \cite[e.g.,][]{Hyun23-NEP}.
In order to better understand the physical drivers behind this variability (i.e., the properties of the relativistic jet), a multi-epoch SED fitting campaign is needed.
AGN 9 falls outside the footprint of the existing JWST NIRCam campaign within the NEP-TDF; therefore, future imaging and spectroscopic JWST coverage could fill in valuable details on the SED that would allow the physical properties of this FSRQ to be unraveled. 
Such a campaign would be especially informative if paired with simultaneous X-ray observations with NuSTAR and/or XMM-Newton.

Meanwhile, AGN 23 is hosted by a Seyfert galaxy with a strong AGN contribution to its SED \citep[$f_{\mathrm{AGN}}=0.70$; ][\citetalias{Ortiz26}]{Ortiz24-NEP_IR} and a bright point-source core identified in JWST NIRCam observations, which is interpreted as thermal emission from the dusty torus \citep{Windhorst23-NEP,Ortiz24-NEP_IR}.
As a result, this source was identified as a CT candidate.
While its X-ray hardness ratio supports the CT scenario (see \citetalias{Zhao24}), AGN 23 was spectroscopically identified to have $\langle N_{\mathrm{H}} \rangle = 1.62_{-0.23}^{+0.26} \times 10^{23}$~cm$^{-2}$ (\citetalias{Creech25}), falling $\sim 1$ order of magnitude below the typical CT threshold ($10^{24}$~cm$^{-2}$).
Now, \citetalias{Silver26}  and this work further refine the picture by showing that AGN 23 likely displays $N_{\mathrm{H}}$ variability, although $\Delta N_{\mathrm{H}}$ never exceeds an order of magnitude.
$N_{\mathrm{H}}$ variability is not definitively confirmed with the statistical tests used here; however, it is heavily supported by the significant photometric variability found by \citetalias{Silver26} .

Similar to AGN 23, the NIR counterpart to AGN 43---which shows evidence of both $N_{\mathrm{H}}$ and $F_{\mathrm{int}}$ variability---has bright AGN emission in NIRCam observations \citep[$f_{\mathrm{AGN}}=0.60$;][\citetalias{Ortiz26}]{Ortiz24-NEP_IR}.
This source is unobscured in the X-ray band, which is consistent with its strong UV and optical point-source emission in HST F435W and F275W observations \citep{Bowling26}.
The host galaxy is an elliptical at $z=0.673$, and it lacks the 3 GHz emission that would be indicative of dust-obscured star formation at such redshifts \citep{Hodge20-mm_SFR,Hyun23-NEP}.
Therefore, AGN 23 is likely hosted by a quiescent galaxy. 

AGN 5 was identified as a composite galaxy by \citetalias{Zhao24}, and this classification is supported by SED fitting ($f_{\mathrm{AGN}}=0.26$; \citetalias{Ortiz26}).
AGN 7 was identified as a quasar by both \citetalias{Zhao24} and \citetalias{Ortiz26} ($f_{\mathrm{AGN}}=0.75$).
Finally, AGN 21---which was identified as a quasar by \citetalias{Zhao24}---has a low AGN contribution ($f_{\mathrm{AGN}}=0.21$), so the classification of this source is unclear.


\subsection{Variability Timescale}

The timescale of $N_{\mathrm{H}}$ variability probes the structure of the obscuring material as it moves along the corona line-of-sight.
In particular, shorter timescales ($\Delta t \lesssim 100$ days) would point to obscuring material closer to the SMBH---such as clouds within the BLR---while the more distant material in the dusty torus lead to preferentially longer-timescale variability.
Previous works have found that $N_{\mathrm{H}}$ variability is more likely to occur on timescales $> 100$ days, which is consistent with variability caused primarily by clumps in the dusty torus \citep[e.g.,][]{TA23,Pizzetti25,TA25-Mrk477,Cox25}.
In order to test this phenomenon on the NEP NuSTAR sample, we follow the methodology presented by \citet{Cox25}, who find that pairs with $\Delta t > 100$ days are more likely to display significant $N_{\mathrm{H}}$ variability.
Within the NEP NuSTAR sample, the sources identified as initial $N_{\mathrm{H}}$ variability candidates (from the $\Delta \mu - \Delta C$ calculation) yield 4411 total observation pairs.
The time separation $\Delta t$ and the $\chi^2$ of $N_{\mathrm{H}}$ variability (Equation \ref{eq:p}) are calculated for each observation pair.
For sources with redshift measurements, the timescale is adjusted according to cosmological time dilation $\Delta t = \Delta t_{obs}/(1+z)$.
Sources without redshift measurements are assigned $z=0$, which gives an upper limit on $\Delta t$.

Figure \ref{fig:delT} compares the results to the threshold for variability significance ($\chi^2 > 6.65$, which corresponds to $p=0.01$ for a sample size of two).
No significant correlation is seen between $\chi^2$ and $\Delta t$;  a Spearman's rank correlation gives a statistic of 0.03. 
To further test for a relationship between $N_{\mathrm{H}}$ variability and $\Delta t$, we follow \citet{Cox25} and split the NEP observation pairs into short ($\Delta t < 100$ days) and long ($\Delta t \geq 100$ days) timescale groups.
A two-sample KS test on the $\chi^2$ distributions for short- and long-timescale $N_{\mathrm{H}}$ pairs suggests that they are significantly different ($p=9.6 \times 10^{-4}$), and a slightly higher percentage of long-timescale pairs fall above the $\chi^2$ significance threshold ($48.8 \%$, compared to $46.7 \%$ of short-timescale pairs).
$92 \%$ of epoch-pairs that pass the Tension test (Section \ref{sec:t_test}) have timescales $\Delta t \geq 100$ days.
Together, these findings suggest that the dusty torus may hold only slight dominance, if any, over the BLR in driving $N_{\mathrm{H}}$ variability in the faint AGN population.
However, due to the necessity of grouping observations into epochs for this work, there are significantly fewer short-timescale pairs (724 total) than long-timescale pairs (3871 pairs). 
Thus, the full parameter space is not well explored, and these results are not conclusive.


Looking at bright sources in the local universe, \citet{TA23} finds that the change in obscuring column density $\Delta N_{\mathrm{H}}$ is positively correlated with $\Delta t$ on timescales $< 10$~days.
The $N_{\mathrm{H}}$ values of the NEP observation pairs are generally too poorly constrained to conclusively investigate this trend, and all but seven of the epoch pairs in this work are $> 10$~days apart.
Nonetheless, we test for trends between $\Delta N_{\mathrm{H}}$ and $\Delta t$ here.
$\Delta N_{\mathrm{H}}$ and its uncertainties are calculated using the \textsc{add}\textsc{\_}\textsc{asym}\footnote{\url{https://github.com/anisotropela/add_asym}} python routine \citep{Laurensen19}, which follows the formalism of \citet{Barlow03} to add asymmetric errors.
Next, a Spearman's rank correlation test is used to search for a monotonic relation between $\Delta N_{\mathrm{H}}$ and $\Delta t$.
The resulting statistics do not yield evidence that high $\Delta N_{\mathrm{H}}$ occurs on longer timescales, regardless of the chosen subsample.
The subsamples tested here include:
\begin{itemize}
\item All observation pairs (Spearman rank correlation $=0.12$)
 \item Observation pairs with fully constrained $\Delta N_{\mathrm{H}}$ (0.16)
 \item Observation pairs belonging to the final $N_{\mathrm{H}}$-variability candidates (0.24)
 \item Observation pairs with $\chi^2 > 6.65$ (0.12).
\end{itemize}

Given the large uncertainties on $\Delta N_{\mathrm{H}}$---which span multiple orders of magnitude for most observation pairs---the lack of evidence for a correlation between $\Delta N_{\mathrm{H}}$ and $\Delta t$ does not reflect physical reality; instead, it is most likely a result of the poor constraints on these models.
The $\Delta N_{\mathrm{H}}$-$\Delta t$ relationship should be further investigated with higher-quality spectroscopic data.

\section{SUMMARY AND CONCLUSIONS}\label{sec:conclusion}
This work searches for spectroscopic variability in NuSTAR identified AGN in the JWST-North Ecliptic Pole Time Domain Field \citep[][\citetalias{Zhao24}, \citetalias{Silver25}]{Windhorst23-NEP}.
This unique campaign is the first deep time-domain survey to be conducted simultaneously in the hard and soft X-ray bands.
As a result, the NEP-TDF provides an excellent opportunity seek out AGN variability in a faint, largely unexplored parameter space.

Following the methodology proposed by \citet{TA23}, we utilize NuSTAR, XMM-Newton, and Chandra spectra to identify the physical drivers behind variability in the NEP sample, searching specifically for changes in the brightness of the corona ($F_{\mathrm{int}}$ variability) and changes in the line-of-sight column density due to obscuring material in the circumnuclear region ($N_{\mathrm{H}}$ variability).
Our findings are summarized below.

\begin{itemize}
    \item We perform the first spectral analysis of the 52 newly-discovered, NuSTAR-selected sources in the \citetalias{Silver25} catalog, resulting in complete spectral information for the entire NuSTAR survey of the NEP-TDF. 
    
    \item We present the observed $N_{\mathrm{H}}$ and $F_{\mathrm{int}}$ distributions of the NEP-TDF NuSTAR sample (Figure \ref{fig:non_var}). The observed Compton thick fraction of the total sample is $f_\text{CT} = 0.05^{-0.01}_{+0.07}$. Future work may use the spectral analysis presented here in order to calculate the intrinsic (absorption-corrected) $f_\text{CT}$ and $N_{\mathrm{H}}$ distribution, building off of previous work in the NEP field (\citetalias{Creech25}). 

    \item Of the 112 NuSTAR-identified sources in the NEP field, we identify seven variable candidates. Two display possible $N_{\mathrm{H}}$ variability, two show possible $F_{\mathrm{int}}$ variability, and three show both. Of these sources, three (AGN 7, 9, and 21) pass at the strictest thresholds of the statistical tests. For caveats to the variability measurements, see Section \ref{sec:sys}.
    
      \item Two sources with photometric variability (from \citetalias{Silver26}) are spectroscopic variability candidates as well (AGN 9; $F_{\mathrm{int}}$ variability, and AGN 23; $N_{\mathrm{H}}$ variability). Both of these sources have notable multi-wavelength properties. AGN 9 is an established FSRQ and is variable in multiple wavelength ranges. AGN 23 is an obscured AGN in a seyfert galaxy with a bright NIR core \citep[as observed by JWST NIRCam,][O26]{Windhorst23-NEP,Ortiz24-NEP_IR}. AGN 9 has not yet been observed with JWST, and multi-epoch MIRI observations of AGN 23 would allow a close study of the reprocessed IR emission from the dusty torus. Both cases demonstrate the benefit of a continued JWST time-domain campaign in the NEP-TDF.
      
      \item Compared to the entire NuSTAR-selected sample, the seven spectroscopic variability candidates preferentially have the greatest number of epochs and spectral counts, and all the candidates have both NuSTAR and XMM-Newton data. This demonstrates the continued need for deep, broad-band X-ray campaigns with $>1000$ total spectral counts per source, split across multiple ($\gtrsim10$) visits. By following this strategy, future campaigns will continue detecting and unraveling $N_{\mathrm{H}}$ variability in faint AGN.
      
      \item We estimate the NuSTAR-NEP survey sensitivity in $N_{\mathrm{H}}$--$F_{\mathrm{int}}$ parameter space, and this sensitivity is compared to the detected variable candidates (Figures \ref{fig:contours} and \ref{fig:var_props}), with the goal of informing population synthesis models on how variability can affect surveys' abilities to detect the faint AGN population, which dominates the cosmic SMBH accretion history \citep[e.g][]{Annana19-Accretion_HistoryI,Gerolymatou25}. 
      
      \item  $N_{\mathrm{H}}$ variability in the NEP-TDF is significant slightly more often on timescales $\geq 100$ days (Figure \ref{fig:delT}). This suggests that the torus may hold only a slight dominance, if any, over the BLR in driving $N_{\mathrm{H}}$ variability in the faint AGN population. In comparison, $N_{\mathrm{H}}$ variability in the local universe shows a much stronger tendency to occur on $>100$ day timescales \citep[e.g.,][]{Cox25,TA25-Mrk477}. Investigating wether this difference is driven by observational effects or physical properties of the faint AGN population may be an interesting avenue for future work.
\end{itemize}
    
The NEP-TDF X-ray survey is a one-of-a-kind campaign designed to detect variability using simultaneous soft and hard X-ray observations, and it is the deepest NuSTAR survey to date.
However, the observational constraints on faint AGN in the NEP-TDF lead to large uncertainties in spectral modeling, which has limited the ability of this work to confirm spectroscopic variability. 
This demonstrates that a new, broad-band X-ray telescope with high angular resolution \citep[such as the previously proposed X-ray probe-class mission HEX-P;][]{Boorman24-HEXP,Civano24-HEXP,Garcia24-HEXP,Kammoun24-HEXP,Pfeifle24-HEXP,Piotrowska24-HEXP} would revolutionize our ability to detect and monitor variable AGN, therefore improving our understanding of the environments of growing SMBHs.


\section*{Acknowledgments}
The material is based upon work supported by NASA under award numbers 80GSFC21M0002 and 80NSSC24K1482.
This work made use of data from the NuSTAR mission, which is led by the California Institute of Technology, managed by the Jet Propulsion Laboratory, and funded by the National Aeronautics and Space Administration.
Additionally, this work made use of data from XMM-Newton, an ESA science mission with instruments and contributions directly funded by ESA Member States and NASA.
This research also employs a list of Chandra datasets, obtained by the Chandra X-ray Observatory, contained in~\dataset[doi:10.25574/cdc.651]{https://doi.org/10.25574/cdc.651}.

This research used data obtained with the Dark Energy Spectroscopic Instrument (DESI). DESI construction and operations is managed by the Lawrence Berkeley National Laboratory. This material is based upon work supported by the U.S. Department of Energy, Office of Science, Office of High-Energy Physics, under Contract No. DE–AC02–05CH11231, and by the National Energy Research Scientific Computing Center, a DOE Office of Science User Facility under the same contract. Additional support for DESI was provided by the U.S. National Science Foundation (NSF), Division of Astronomical Sciences under Contract No. AST-0950945 to the NSF’s National Optical-Infrared Astronomy Research Laboratory; the Science and Technology Facilities Council of the United Kingdom; the Gordon and Betty Moore Foundation; the Heising-Simons Foundation; the French Alternative Energies and Atomic Energy Commission (CEA); the National Council of Humanities, Science and Technology of Mexico (CONAHCYT); the Ministry of Science and Innovation of Spain (MICINN), and by the DESI Member Institutions: www.desi.lbl.gov/collaborating-institutions. The DESI collaboration is honored to be permitted to conduct scientific research on I’oligam Du’ag (Kitt Peak), a mountain with particular significance to the Tohono O’odham Nation. Any opinions, findings, and conclusions or recommendations expressed in this material are those of the author(s) and do not necessarily reflect the views of the U.S. National Science Foundation, the U.S. Department of Energy, or any of the listed funding agencies.

This work also made extensive use of Astropy \citep{astropy13,astropy18,astropy22}, numpy \citep{numpy}, and SciPy \citep{SciPy}.
We would also like to thank the High Energy Astrophysics Science Archive Research Center (HEASARC) team; this work would not have been possible without their data and software.

Finally, we would like to thank the anonymous reviewer for their helpful comments, which improved the quality and clarity of this paper.

\bibliography{NEP_bib}{}
\bibliographystyle{aasjournal}

\appendix
\section{NuSTAR source extraction regions}\label{sec:a_r}

\tabletypesize{\footnotesize}
\begin{deluxetable*}{ccccccccc||ccccccccc}\label{tab:snr_size}
\tablecaption{NuSTAR Spectral Extraction Region Sizes for Sources Discovered in Cycle 8+9}
\startdata
 \\
  \multicolumn{2}{c}{}  &
 \multicolumn{2}{c}{FPMA} &
 \multicolumn{2}{c}{} & 
 \multicolumn{2}{c}{FPMB}  & &
  \multicolumn{2}{c}{} & 
 \multicolumn{2}{c}{FPMA} &
 \multicolumn{2}{c}{} & 
 \multicolumn{2}{c}{FPMB}  & \\
  \hline
  & &
  \multicolumn{2}{c}{Cycle} & & &
 \multicolumn{2}{c}{Cycle} & &
 & & 
  \multicolumn{2}{c}{Cycle} & & &
 \multicolumn{2}{c}{Cycle} &
 \\
  ID & 5 & 6 & 8 & 9 & 5 & 6 & 8 & 9 &
  ID & 5 & 6 & 8 & 9 & 5 & 6 & 8 & 9
  \\
  \hline
 61 & - & - & 25" & 20" & - & - & 20" & 20" & 103 & 20" & 25" & 20" & 20" & 35" & 15" & 25" & 35" \\
 62 & - & - & - & 20" & - & - & - & 40" & 105 & 15" & 25" & 25" & 20" & 15" & 25" & 20" & 20" \\
 63 & - & - & 30" & 15" & - & - & 25" & 25" & 107 & 20" & 35" & 35" & 25" & 35" & 35" & 50" & 35" \\
 64 & - & - & 30" & 20" & - & - & 40" & 30" & 110 & - & - & 45" & 30" & - & - & 40" & 30" \\
 65 & 35" & - & 35" & 15" & 20" & - & 20" & 25" & 108 & 40" & 30" & 25" & 20" & 45" & 20" & 65" & 20" \\
 66 & - & - & 25" & 20" & - & - & 25" & 20" & 111 & 20" & 20" & 30" & 30" & 20" & 20" & 20" & 20" \\
 67 & - & - & 25" & 30" & - & - & 30" & 20" & 72 & - & - & 35" & 15" & - & - & 20" & 20" \\
 68 & - & - & 20" & 25" & - & - & 20" & 25" & 74 & 20" & 30" & 20" & 25" & 25" & 20" & 35" & 30" \\
 69 & - & - & 30" & 25" & - & - & 20" & 20" & 76 & 20" & 25" & 25" & 20" & 20" & 20" & 20" & 25" \\
 70 & 35" & 40" & 45" & 30" & 35" & 30" & 40" & 30" & 78 & - & 20" & 20" & 20" & - & 10" & 20" & 20" \\
 71 & 20" & 20" & 20" & 20" & 20" & 20" & 20" & 20" & 80 & 25" & 40" & 30" & 20" & 30" & 25" & 20" & 30" \\
 73 & 10" & - & 20" & 20" & 20" & - & 15" & 25" & 82 & 10" & 20" & 20" & 20" & 15" & 20" & 20" & 25" \\
 75 & 25" & 20" & 20" & 20" & 25" & 25" & 25" & 25" & 84 & 25" & 25" & 20" & 25" & 10" & 10" & 25" & 25" \\
 77 & 25" & 10" & 20" & 15" & 35" & 15" & 25" & 30" & 86 & 20" & 20" & 20" & 20" & 20" & 20" & 25" & 25" \\
 79 & 15" & 25" & 20" & 20" & 15" & 25" & 20" & 20" & 88 & 20" & 25" & 25" & 20" & 20" & 25" & 35" & 20" \\
 81 & 15" & 20" & 20" & 20" & 20" & 20" & 20" & 20" & 90 & - & - & - & 35" & - & - & - & 45" \\
 83 & 45" & 20" & 20" & 20" & 40" & 15" & 25" & 25" & 92 & 10" & 20" & 45" & - & 30" & 20" & 65" & - \\
 85 & 25" & 25" & 20" & 20" & 20" & 25" & 30" & 25" & 94 & 15" & - & 20" & 20" & 25" & - & 15" & 25" \\
 87 & 25" & 40" & 20" & 30" & 40" & 35" & 40" & 25" & 96 & 10" & - & 15" & 10" & 10" & - & 20" & 20" \\
 89 & 35" & 20" & 20" & 20" & 20" & 25" & 20" & 30" & 98 & 30" & 40" & 20" & 30" & 35" & 40" & 20" & 15" \\
 91 & 25" & 20" & 15" & 45" & 20" & 25" & 25" & 25" & 100 & - & - & 25" & 25" & - & - & 35" & 20" \\
 93 & - & - & 25" & 25" & - & - & 20" & 30" & 102 & 20" & 25" & 30" & 25" & 20" & 30" & 20" & 35" \\
 95 & - & - & 40" & 40" & - & - & 30" & 20" & 104 & 45" & 25" & 15" & 20" & 35" & 25" & 30" & 20" \\
 97 & 15" & 35" & 20" & 15" & 25" & 10" & 45" & 30" & 106 & 30" & 25" & 25" & 25" & 35" & 40" & 35" & 25" \\
 99 & - & - & 20" & 20" & - & - & 20" & 25" & 109 & - & - & 70" & 45" & - & - & 40" & 20" \\
 101 & 25" & 20" & 25" & 15" & 10" & 25" & 25" & 10" & 112 & 20" & 30" & 45" & 25" & 15" & 35" & 30" & 25" \\
\enddata
\end{deluxetable*}

When choosing a region size for extracting NuSTAR spectra, we first find the radius ($r$ in arc-seconds) that maximizes the signal-to-noise ratio using the equation presented by \citet{Zappacosta18-NUSTAR_extragal_hard-band}:

\begin{equation}
   \text{SNR}(<r) = \frac{N(<r)}{\sqrt{N(<r) + 2B(<r)}}.
\end{equation}

In this equation, $N(<r)$ is the total number of counts within a circular region of radius $r$ centered around the source, and $B(<r)$ is the number of background counts within that same region \citep[extracted from the background maps produced by \textsc{nuskybgd};][]{Wik14-nuskybgd}. 
However, sources with low angular separation can lead to confusion in the SNR calculation, which leads to overlapping extraction regions and contamination between sources.
To mitigate this issue, region sizes were manually adjusted when necessary.
Table \ref{tab:snr_size} shows the resulting region sizes for all newly-discovered sources from the \citetalias{Silver25} catalog.
Dashes are used to indicate sources outside the FOV for a given observation cycle.

\begin{figure*}
    \centering
    \includegraphics[width=0.95\textwidth]{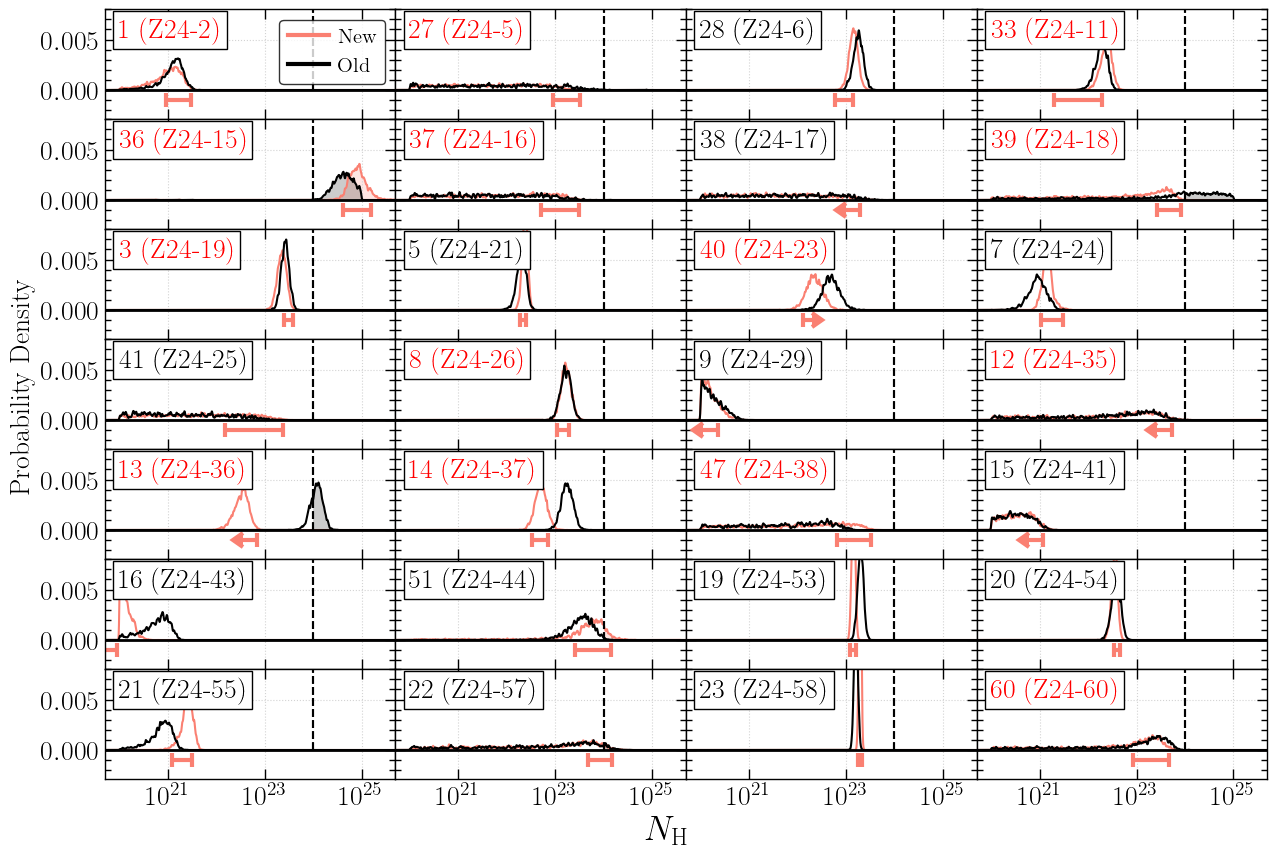}
    \caption{A comparison of new (pink) and old (black) $N_{\mathrm{H}}$ probability distribution curves for sources whose fits have been updated from \citetalias{Creech25}. Error-bars under each curve indicate the new $N_{\mathrm{H}}$ constraints found using standard \textsc{xspec} routines. The NuSTAR IDs are listed in the upper left-hand corners of the subplots (with the corresponding IDs from the \citetalias{Zhao24} catalog in parentheses), where red labels indicate sources that did not pass the data quality cuts required by \citetalias{Creech25}. Vertical dashed lines indicate the typical threshold used for Compton Thick column densities ($N_{\mathrm{H}} > 10^{24}$~cm$^{-2}$).}
    \label{fig:c56_BXA}
\end{figure*}

\startlongtable
\tabletypesize{\tiny}
\begin{deluxetable*}{lccccccccccccc}
\tablecaption{Best-Fit Model Parameters for Newly-Discovered \citetalias{Silver25} Sources \label{tab:pars}}
\tiny
\startdata
\\
 & &
 \multicolumn{6}{c}{\textbf{baseline}} &
  \multicolumn{6}{c}{\textbf{clumpy}} \\
  \hline
  & pass & & nH & & & & & & nH & & & &\\
  ID\tablenotemark{\tiny a} & (y/n)\tablenotemark{\tiny b}  & $\Gamma$ & ($10^{22}$~cm$^{-2}$) & $-$log($F$)\tablenotemark{\tiny c} & $n$\tablenotemark{\tiny d} & $f$\tablenotemark{\tiny e} & C/DOF & $\Gamma$ & ($10^{22}$~cm$^-2$) & $\kappa$\tablenotemark{\tiny f} & $n$\tablenotemark{\tiny d} & $f$\tablenotemark{\tiny e} & C/DOF \\
  \hline
61 & n & $1.80$ & $\leq 9.0$ & $13.4_{-0.2}^{+0.2}$ &  -  &  -  & 89.9/97 & $\geq 1.24$ & $\leq 8.0$ & $\leq 12.5$ &  -  &  -  & 89.6/96 \\
62 & n & $1.80$ & $27.2_{-23}^{+48}$ & $13.2_{-0.3}^{+0.3}$ &  -  &  -  & 131.9/97 & $1.80$ & $24.5_{-22}^{+46}$ & $2.6_{-1.8}^{+3.6}$ &  -  &  -  & 132.2/97 \\
63 & y & $2_{-0.08}^{+0.11}$ & $\leq 0.1$ & $13.1_{-0.1}^{+0.1}$ &  -  &  -  & 309.3/326 & $2.05_{-0.08}^{+0.09}$ & $\leq 0.0$ & $21.1_{-2.1}^{+5.8}$ &  -  &  -  & 308.2/326 \\
64 & n & $2.13_{-0.8}^{+1.1}$ & $\leq 6.3$ & $13.3_{-0.2}^{+0.2}$ &  -  &  -  & 88.8/101 & $\geq 1.29$ & $\leq 4.9$ & $\leq 35.8$ &  -  &  -  & 88.4/101 \\
65 & n & $1.66_{-0.46}^{+0.63}$ & $2_{-1.2}^{+2.5}$ & $13.3_{-0.1}^{+0.1}$ &  -  &  -  & 189.5/206 & $1.79_{-0.51}^{+0.56}$ & $2.1_{-1.3}^{+2.2}$ & $1.8_{-1.2}^{+3.5}$ &  -  &  -  & 189.5/206 \\
66 & n & $1.80$ & $\leq 0.0$ & $14.1_{-0.2}^{+0.1}$ &  -  &  -  & 410.4/112 & $1.80$ & $\leq 0.0$ & $0.2_{-0.1}^{+0.2}$ &  -  &  -  & 408.8/112 \\
67 & y & $1.80$ & $\leq 36.6$ & $13.6_{-0.2}^{+0.3}$ &  -  &  -  & 112.3/100 & $1.80$ & $\leq 44.1$ & $1.2_{-0.7}^{+1.4}$ &  -  &  -  & 144.6/135 \\
68 & n & $1.7_{-0.3}^{+0.35}$ & $0.2_{-0.2}^{+0.2}$ & $13.5_{-0.1}^{+0.1}$ &  -  &  -  & 185.5/169 & $1.72_{-0.26}^{+0.35}$ & $0.2_{-0.1}^{+0.2}$ & $\leq 1.2$ &  -  &  -  & 185.2/169 \\
69 & n & $1.80$ & $\leq 16.5$ & $13.5_{-0.2}^{+0.2}$ &  -  &  -  & 106.7/136 & $1.80$ & $\leq 8.3$ & $0.8_{-0.3}^{+1.5}$ &  -  &  -  & 108.3/136 \\
70 & y & $1.80$ & $33.9_{-16}^{+26}$ & $13.3_{-0.1}^{+0.1}$ &  -  &  -  & 583.7/573 & $1.80$ & $29.1_{-14}^{+24}$ & $4.9_{-1.4}^{+2.4}$ &  -  &  -  & 583.0/573 \\
71 & y & $\leq 0.5$ & $\leq 0.3$ & $13.4_{-0}^{+0}$ &  -  &  -  & 552.1/425 & $1.80$ & $17.3_{-2.9}^{+3.3}$ & $4.6_{-0.5}^{+0.5}$ &  -  & $\geq 9.2$ & 593.8/425 \\
72 & y & $1.80$ & $\leq 75.8$ & $13.3_{-0.4}^{+0.3}$ &  -  &  -  & 77.1/66 & $1.80$ & $\leq 76.9$ & $2_{-1.2}^{+3.3}$ &  -  &  -  & 76.9/66 \\
73 & n & $1.80$ & $\leq 13.6$ & $13.3_{-0.2}^{+0.2}$ &  -  &  -  & 77.0/82 & $1.80$ & $\leq 10.6$ & $1.7_{-1.1}^{+2.5}$ &  -  &  -  & 76.4/82 \\
74 & n & $1.95_{-0.73}^{+1.4}$ & $\leq 32.1$ & $13.5_{-0.2}^{+0.5}$ &  -  &  -  & 484.5/499 & $1.97_{-0.65}^{+0.86}$ & $\leq 22.1$ & $\leq 27.7$ &  -  &  -  & 483.5/499 \\
75 & n & $1.80$ & $0.1_{-0.1}^{+1.1}$ & $14.1_{-0.1}^{+0.1}$ & $\leq 0.2$ &  -  & 412.7/425 & $1.80$ & $0.1_{-0.1}^{+1}$ & $0.2_{-0}^{+0.2}$ & $0.1_{-0}^{+0.1}$ &  -  & 412.3/425 \\
76 & n & $1.80$ & $0.6_{-0.4}^{+0.7}$ & $14.2_{-0.2}^{+0.2}$ &  -  &  -  & 463.6/450 & $1.80$ & $0.6_{-0.4}^{+0.6}$ & $0.3_{-0.2}^{+0.1}$ &  -  &  -  & 462.7/450 \\
77 & y & $1.80$ & $\leq 127.9$ & $13.7_{-0.2}^{+0.6}$ &  -  &  -  & 168.5/176 & $1.80$ & $\leq 50.0$ & $\leq 2.6$ &  -  &  -  & 167.6/176 \\
78 & n & $1.80$ & $\leq 29.3$ & $13.6_{-0.2}^{+0.3}$ &  -  &  -  & 121.6/123 & $1.80$ & $\leq 26.5$ & $0.9_{-0.6}^{+0.7}$ &  -  &  -  & 121.4/123 \\
79 & y & $1.80$ & $23.4_{-14}^{+35}$ & $13.9_{-0.2}^{+0.1}$ &  -  &  -  & 436.0/508 & $1.80$ & $19.8_{-12}^{+34}$ & $2.1_{-0.8}^{+1.1}$ &  -  &  -  & 435.4/508 \\
80 & y & $1.80$ & $48.1_{-22}^{+29}$ & $13.3_{-0.2}^{+0.2}$ &  -  & $\geq 4.5$ & 540.1/597 & $2.12_{-0.23}^{+0.31}$ & $58.1_{-20}^{+34}$ & $6.2_{-1.6}^{+10}$ &  -  & $\geq 3.5$ & 540.2/596 \\
81 & n & $1.80$ & $\leq 12.2$ & $13.7_{-0.1}^{+0.2}$ &  -  &  -  & 279.2/299 & $1.80$ & $\leq 9.4$ & $1_{-0.3}^{+1}$ &  -  &  -  & 278.3/299 \\
82 & y & $1.80$ & $15.6_{-13}^{+25}$ & $13.6_{-0.2}^{+0.2}$ &  -  &  -  & 350.0/309 & $1.80$ & $14.1_{-12}^{+22}$ & $1_{-0.4}^{+0.7}$ &  -  &  -  & 350.3/309 \\
83 & n & $1.50$ & $1.2_{-1.2}^{+1.7}$ & $13.7_{-0.1}^{+0.1}$ & $8.6_{-2.7}^{+2.6}$ &  -  & 303.8/333 & $1.6_{-0.17}^{+0.28}$ & $1.6_{-1.5}^{+2.2}$ & $3.1_{-0.9}^{+2.9}$ & $13.1_{-3.4}^{+3.6}$ &  -  & 303.6/332 \\
84 & n & $1.80$ & $0.6_{-0.4}^{+1}$ & $13.8_{-0.2}^{+0.2}$ &  -  &  -  & 175.6/191 & $1.80$ & $0.6_{-0.4}^{+0.7}$ & $0.6_{-0.3}^{+0.3}$ &  -  &  -  & 175.0/191 \\
85 & n & $1.80$ & $\leq 7.2$ & $13.7_{-0.1}^{+0.1}$ &  -  &  -  & 266.3/324 & $1.74_{-0.68}^{+0.16}$ & $\leq 7.1$ & $0.6_{-0.2}^{+3.5}$ &  -  &  -  & 266.2/323 \\
86 & n & $1.18_{-0.15}^{+0.16}$ & $\leq 0.1$ & $13.7_{-0.1}^{+0.1}$ & $0.1_{-0.1}^{+0.1}$ &  -  & 491.2/456 & $1.80$ & $0.6_{-0.4}^{+0.2}$ & $1.2_{-0.6}^{+0.2}$ & $0.3_{-0.1}^{+0.1}$ &  -  & 513.0/457 \\
87 & n & $1.80$ & $\leq 181.0$ & $13.8_{-0.3}^{+0.8}$ &  -  &  -  & 410.7/434 & $\geq 1.02$ & $\geq 0.0$ & $\leq 306.3$ &  -  &  -  & 407.8/433 \\
88 & y & $1.80$ & $\leq 85.2$ & $13.6_{-0.2}^{+0.2}$ &  -  &  -  & 310.1/269 & $1.80$ & $\leq 64.8$ & $4_{-1.1}^{+4.3}$ &  -  &  -  & 309.3/269 \\
89 & n & $1.80$ & $\leq 174.0$ & $13.4_{-0.5}^{+0.4}$ & $84.4_{-56}^{+45}$ &  -  & 349.9/353 & $1.80$ & $\geq 191.9$ & $\leq 28.0$ &  -  &  -  & 359.7/354 \\
90 & n & $1.80$ & $\leq 8.0$ & $12.7_{-0.2}^{+0.1}$ &  -  &  -  & 123.0/77 & $1.80$ & $\leq 6.8$ & $6.4_{-2.5}^{+9.4}$ &  -  &  -  & 122.5/77 \\
91 & y & $1.80$ & $16.9_{-4.7}^{+5.8}$ & $12.6_{-0.1}^{+0.1}$ &  -  & $3.8_{-1}^{+1.2}$ & 480.8/286 & $1.80$ & $17.6_{-5.4}^{+6.6}$ & $10.6_{-2.1}^{+2.4}$ &  -  & $\geq 9.1$ & 484.5/286 \\
92 & n & $1.80$ & $\leq 56.6$ & $13_{-0.3}^{+0.3}$ &  -  &  -  & 121.6/118 & $1.80$ & $\leq 54.8$ & $3.6_{-2.7}^{+5.7}$ &  -  &  -  & 121.5/118 \\
93 & n & $\geq 0.95$ & $\leq 61.1$ & $13.4_{-0.4}^{+0.9}$ &  -  &  -  & 85.4/120 & $1.73_{-0.71}^{+0.71}$ & $\leq 42.7$ & $0.5_{-0.1}^{+2.6}$ &  -  &  -  & 85.3/120 \\
94 & y & $1.80$ & $\leq 48.3$ & $13.8_{-0.4}^{+0.4}$ & $96.2_{-69}^{+99}$ &  -  & 93.8/108 & $1.80$ & $\leq 4.0$ & $1.8_{-1.6}^{+0.9}$ &  -  &  -  & 99.5/109 \\
95 & y & $1.95_{-0.29}^{+0.32}$ & $\leq 2.7$ & $13_{-0.1}^{+0.1}$ &  -  &  -  & 247.1/275 & $2.04_{-0.3}^{+0.29}$ & $\leq 1.9$ & $8.8_{-5.5}^{+6.2}$ &  -  &  -  & 245.8/275 \\
96 & n & $1.80$ & $\leq 7.6$ & $13.6_{-0.2}^{+0.2}$ &  -  &  -  & 74.8/77 & $1.80$ & $\leq 6.0$ & $\leq 2.1$ &  -  &  -  & 74.6/77 \\
97 & n & $1.80$ & $\leq 6.2$ & $13.7_{-0.2}^{+0.1}$ &  -  &  -  & 359.0/402 & $\geq 1.48$ & $\leq 8.7$ & $\leq 15.1$ &  -  &  -  & 357.7/401 \\
98 & n & $1.80$ & $0.3_{-0.1}^{+0.1}$ & $13.7_{-0}^{+0}$ &  -  &  -  & 933.5/1110 & $1.80$ & $0.3_{-0.1}^{+0.1}$ & $0.5_{-0.1}^{+0.1}$ &  -  &  -  & 932.5/1110 \\
99 & n & $1.80$ & $11_{-7.2}^{+17}$ & $13.3_{-0.1}^{+0.1}$ & $2.4_{-1.6}^{+2}$ & $\geq 3.7$ & 177.4/190 & $1.80$ & $4_{-2.4}^{+3.1}$ & $7.2_{-1.2}^{+2.1}$ & $4.5_{-2.8}^{+3}$ &  -  & 179.0/191 \\
100 & y & $1.8_{-0.47}^{+0.56}$ & $2.9_{-2.2}^{+5.7}$ & $13.1_{-0.1}^{+0.2}$ &  -  &  -  & 221.8/208 & $1.84_{-0.31}^{+0.55}$ & $2.5_{-1.8}^{+5.1}$ & $3.1_{-1.7}^{+8}$ &  -  &  -  & 221.3/208 \\
101 & n & $1.80$ & $\leq 14.3$ & $13.7_{-0.2}^{+0.2}$ &  -  &  -  & 241.1/248 & $\geq 0.0$ & $\leq 8.7$ & $\leq 11.3$ &  -  &  -  & 240.5/247 \\
102 & n & $1.63_{-0.13}^{+0.16}$ & $\leq 0.5$ & $13.6_{-0.1}^{+0.1}$ & $1.4_{-0.8}^{+1.3}$ &  -  & 511.3/716 & $1.78_{-0.09}^{+0.13}$ & $\leq 0.0$ & $3.9_{-0.3}^{+3.9}$ &  -  &  -  & 517.7/717 \\
103 & y & $1.82_{-0.13}^{+0.15}$ & $\leq 0.3$ & $13.5_{-0.1}^{+0.1}$ & $\leq 0.9$ &  -  & 704.5/841 & $1.84_{-0.11}^{+0.12}$ & $0.1_{-0.1}^{+0.1}$ & $4_{-1.9}^{+3.8}$ & $0.6_{-0.5}^{+0.6}$ &  -  & 701.5/841 \\
104 & n & $1.73_{-0.2}^{+0.23}$ & $\leq 0.0$ & $13.5_{-0.1}^{+0.1}$ &  -  &  -  & 262.6/248 & $1.73_{-0.19}^{+0.23}$ & $\leq 0.0$ & $0.9_{-0.2}^{+0.4}$ &  -  &  -  & 262.1/248 \\
105 & n & $1.80$ & $10.3_{-3.4}^{+4.7}$ & $13.8_{-0.1}^{+0.1}$ & $2.5_{-0.9}^{+1}$ &  -  & 435.2/514 & $1.80$ & $9.2_{-3.1}^{+4.3}$ & $4.5_{-0.7}^{+1}$ & $3.6_{-1.4}^{+1.5}$ &  -  & 436.0/514 \\
106 & y & $1.95_{-0.25}^{+0.28}$ & $1.7_{-0.6}^{+0.8}$ & $13.2_{-0.1}^{+0.1}$ & $\leq 0.1$ &  -  & 306.6/326 & $2_{-0.26}^{+0.3}$ & $1.6_{-0.7}^{+0.7}$ & $5.5_{-2.7}^{+4.6}$ &  -  &  -  & 306.5/327 \\
107 & n & $1.72_{-0.18}^{+0.2}$ & $\leq 0.0$ & $13.6_{-0.1}^{+0.1}$ & $\leq 0.1$ &  -  & 342.8/441 & $\geq 0.0$ & $0.1_{-0}^{+0}$ & $1.2_{-0.2}^{+0.2}$ &  -  &  -  & 373.2/442 \\
108 & n & $1.80$ & $0.1_{-0.1}^{+0.1}$ & $13.4_{-0.1}^{+0.1}$ &  -  &  -  & 330.9/353 & $1.62_{-0.15}^{+0.28}$ & $0.1_{-0}^{+0.1}$ & $\leq 2.1$ &  -  &  -  & 329.4/352 \\
109 & n & $1.80$ & $\leq 0.1$ & $13.5_{-0.1}^{+0.1}$ &  -  &  -  & 271.1/230 & $1.80$ & $\leq 0.1$ & $1.1_{-0.3}^{+1.7}$ &  -  &  -  & 271.1/230 \\
110 & y & $1.80$ & $0.1_{-0}^{+0.1}$ & $13.1_{-0}^{+0}$ &  -  &  -  & 549.2/517 & $1.62_{-0.06}^{+0.07}$ & $\leq 0.1$ & $5.3_{-0.6}^{+2.5}$ & $\leq 0.5$ &  -  & 532.7/515 \\
111 & n & $1.80$ & $0.2_{-0.2}^{+0.3}$ & $13.8_{-0}^{+0}$ & $\leq 1.0$ &  -  & 1174.9/1135 & $1.80$ & $0_{-0}^{+0}$ & $5.9_{-0.8}^{+0.4}$ & $\leq 2.5$ &  -  & 1163.2/1135 \\
112 & y & $1.80$ & $1_{-0.4}^{+0.5}$ & $14_{-0.1}^{+0.1}$ &  -  &  -  & 668.0/703 & $1.80$ & $0.9_{-0.3}^{+0.5}$ & $0.3_{-0.1}^{+0.2}$ &  -  &  -  & 667.9/703 \\
\enddata
\tablecomments{ \\
$^{a}$ The listed NuSTAR IDs are from the total 112-source catalog (\citetalias{Silver26}). For the corresponding IDs in the \citetalias{Silver25} catalog, see Table \ref{tab:agn_props}. \\
$^{b}$ Whether this object pass the 8-24 keV maximum likelihood threshold. \\
$^{c}$ $-\log(F_{2-10}/\mathrm{erg\,cm^{-2}\,s^{-1}})$: Flux from 2–10 keV. \\
$^{d}$ Normalization of the \textsc{mekal} model: $\frac{10^{-14}}{4 \pi [D_A (1 + z)]^{2}} \int n_e n_H dV$ \\
$^{e}$ Factor of scattered power law emission (omni component of \textsc{uxclumpy}) \\
$^{f}$ Normalization of the \textsc{powerlaw} model: $10^{-3}$~photons~keV$^{-1}$~cm$^{-2}$s$^{-1}$ at 1 keV
}
\end{deluxetable*}

\begin{figure}
    \centering
      \includegraphics[width=0.45\textwidth]{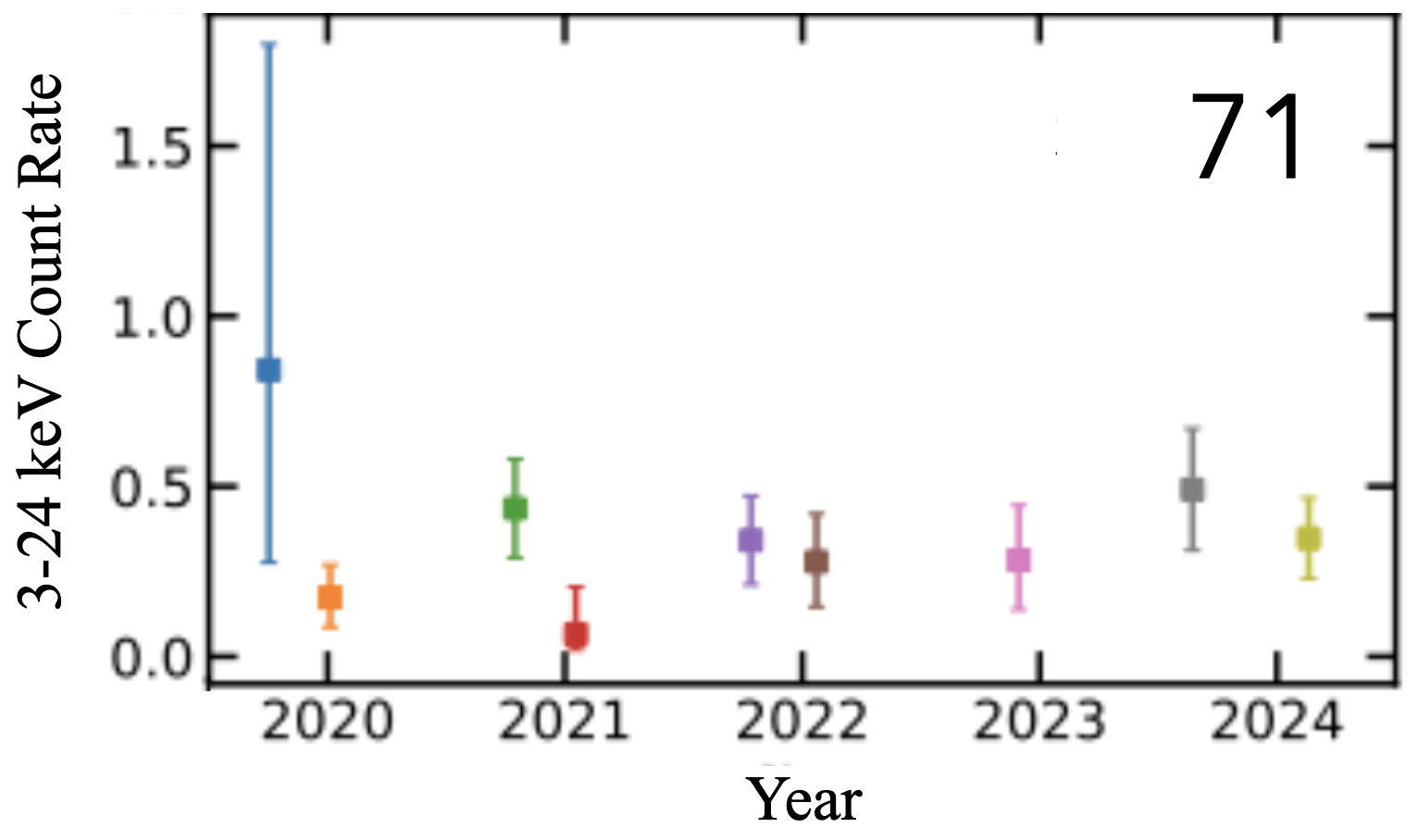}                   
\caption{The 3-24~keV count rate as a function of time for AGN 71, which was identified as a spectroscopic variability candidate (Table \ref{tab:var_info}) but lacks significant photometric variability in the \citetalias{Silver26} analysis.}
    \label{fig:phot}
\end{figure}

\section{Fit Details}\label{sec:a_fit}


Assuming the non-variable scenario, Table \ref{tab:pars} shows the best-fit parameters, $90\%$ uncertainties, and fit statistics for each newly discovered source in the \citetalias{Silver25} catalog.
Frozen parameters are given without uncertainties, and dashes indicate that a given model component was not included for that source.
For best-fit model information for the \citetalias{Zhao24} NuSTAR catalog, see \citetalias{Creech25}.

\subsection{Updated Fits from \citetalias{Creech25}}

In this work, the best-fit models for 28 of the sources analyzed by \citetalias{Creech25} have been updated for various reasons.
We describe these reasons below and show that the Compton thick fraction calculated by \citetalias{Creech25} remains unchanged.

As discussed in Appendix \ref{sec:a_r}, sources close together can lead to artificially large region sizes that contain source counts from multiple AGN.
In \citetalias{Creech25}, only the \citetalias{Zhao24} catalog was considered.
As a result, some of the original region sizes used for spectral extraction overlap with sources discovered by \citetalias{Silver25}.
For this work, we have manually adjusted the region sizes of these sources.
This effects 13 of the sources analyzed by \citetalias{Creech25}, 5 of which passed the data quality cuts and were used for results and conclusions.
The new region sizes for these sources are listed in Table \ref{tab:snr_size_c56}.

Three additional sources from the \citetalias{Zhao24} catalog matched with new optical counterparts from Willmer et al \textit{in prep.}, yielding photometric redshifts.
($z= 0.08, 0.16, 1.35$ for AGN 40, 14, 22, respectively).
Only one of these (AGN 22) passes the data quality cuts used by \citetalias{Creech25}, and the new best-fit $N_{\mathrm{H}}$ for that source is effectively unchanged (Figure \ref{fig:c56_BXA}), and therefor does not affect the results of that work.

Twelve \citetalias{Zhao24} sources were recently matched to Chandra counterparts (Table \ref{tab:ch_obs}).
Of these twelve sources, only one (AGN 13) has a significantly changed $N_{\mathrm{H}}$ distribution.
This source does not pass the quality cuts required by \citetalias{Creech25}; thus, the inclusion of Chandra data does not affect the results of that work.

Lastly, AGN 12 was matched with an XMM-Newton source discovered in the newer \citetalias{Silver25} catalog.
The new fit and constraints for this source are virtually identical to those found by \citetalias{Creech25}.

For all sources with new spectral fits, Figure \ref{fig:c56_BXA} compares the new and old probability distributions derived for $N_{\mathrm{H}}$ using the Bayesian X-ray Analysis package \citep[][see \citetalias{Creech25} for a detailed description of the analysis steps]{Buchner14_BXA}.
Of the six sources that pass the data quality cuts, only one (AGN 16) has a significantly changed $N_{\mathrm{H}}$ distribution, but this object is unobscured in both cases.
The primary goal of \citetalias{Creech25} was to derive the fraction of Compton Thick ($N_{\mathrm{H}} > 10^{24}$cm$^{-2}$) sources.
Thus, the changes to AGN 16 does not affect their results.

\tabletypesize{\footnotesize}
\begin{deluxetable*}{cc|cccc|cccc||cc|cccc|cccc}\label{tab:snr_size_c56}
\tablecaption{Updated NuSTAR Spectral Extraction Region Sizes (Cycle 5+6 Catalog)}
\startdata
 \\
  \multicolumn{3}{c}{}  &
 \multicolumn{2}{c}{FPMA} &
 \multicolumn{2}{c}{} & 
 \multicolumn{2}{c}{FPMB}  & &
  \multicolumn{3}{c}{} & 
 \multicolumn{2}{c}{FPMA} &
 \multicolumn{2}{c}{} & 
 \multicolumn{2}{c}{FPMB}  & \\
  \hline
  & pass & &
  \multicolumn{2}{c}{Cycle} & & &
 \multicolumn{2}{c}{Cycle} & &
 & pass & &
  \multicolumn{2}{c}{Cycle} & & &
 \multicolumn{2}{c}{Cycle} &
 \\
  ID & (y/n)\tablenotemark{\tiny a}  & 5 & 6 & 8 & 9 & 5 & 6 & 8 & 9 &
  ID & (y/n)\tablenotemark{\tiny a}  & 5 & 6 & 8 & 9 & 5 & 6 & 8 & 9
  \\
  \hline
 8 & n & 15" & 25" & 15" & 25" & 15" & 25" & 20" & 25" & 38 & y & 29" & 15" & 15" & 20" & 20" & 20" & 10" & 20" \\
 16 & y & 25" & 20" & 25" & 10" & 15" & 20" & 20" & 20" & 39 & n & 10" & 20" & 25" & 20" & 15" & 15" & 20" & 15" \\
 27 &  n & 20" & 20" & 20" & 20" & 20" & 20" & 20" & 20" & 41 & n & - & 20" & 10" & 25" & - & 25" & 10" & 15" \\
 28 & n & 20" & 20" & 20" & 20" & 20" & 20" & 20" & 20" & 47 & y & 20" & 20" & 20" & - & 20" & 20" & 20" & - \\
 33 & n & 20" & 20" & 25" & 25" & 20" & 20" & 25" & 20" & 51 & y & 20" & 15" & 39" & 45" & 10" & 30" & 15" & 15" \\
 36 & y &20" & 30" & 20" & 10" & 20" & 20" & 15" & 10" & 60 & n & 20" & 20" & 20" & 20" & 20" & 20" & 15" & 15" \\
 37 & n & 15" & 25" & 30" & 35" & 20" & 10" & 30" & 25" & & & & &  \\
\enddata
\tablecomments{ \\
$^{a}$ Whether this object passes the maximum likelihood and signal to noise cuts described by \citetalias{Creech25}. \\
}
\end{deluxetable*}

\section{Variability Figures}\label{sec:a_var}

\begin{figure*}[ht]
    \centering
     \includegraphics[width=0.7\textwidth]{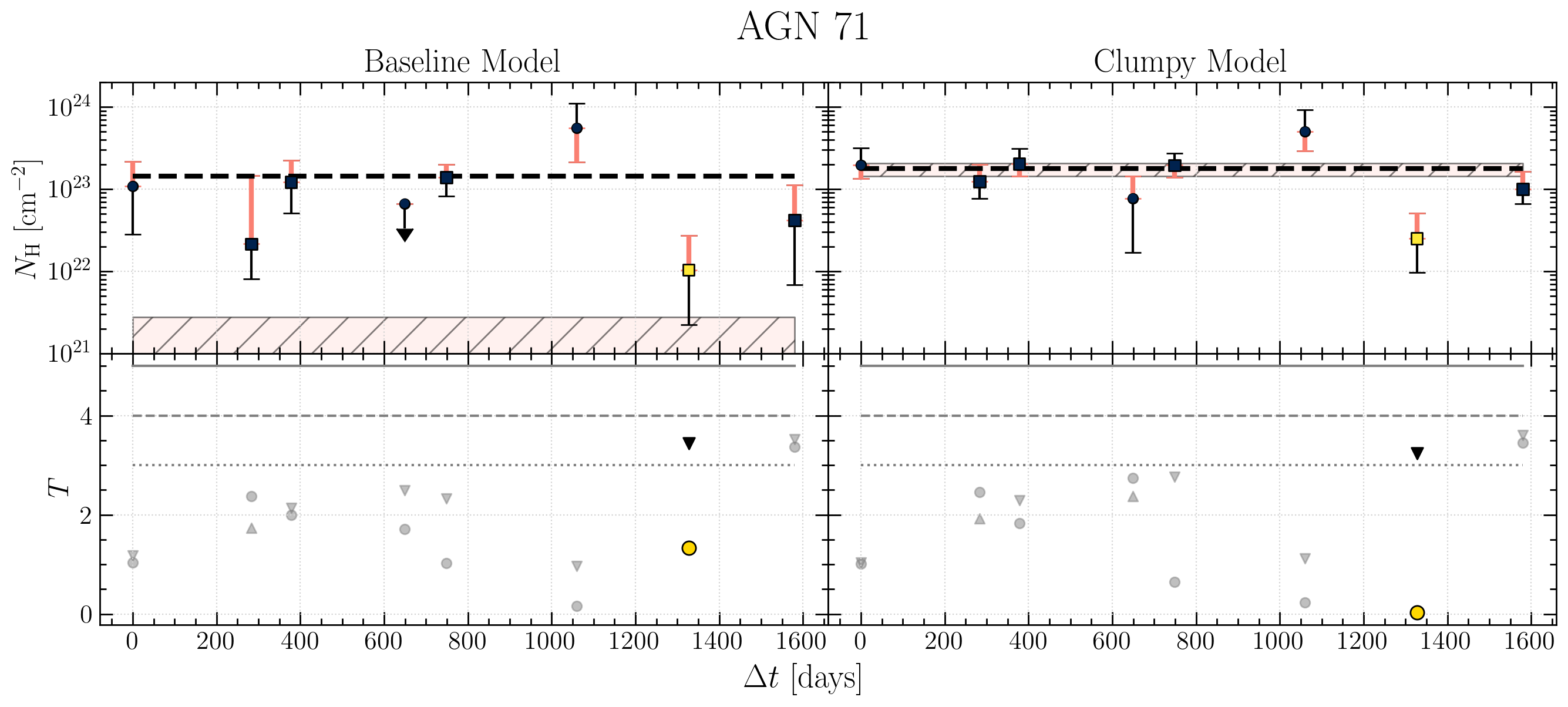}        
     \caption{Parameter variability (top panels) and model tension (bottom panels) for each of the $N_{\mathrm{H}}$ variable candidates identified in Section \ref{sec:var}. The colorful side of the error-bars for the parameter values become $\delta(p_i)$ in Equation \ref{eq:p}. The parameter values are compared with the time-averaged value (dashed line) and the $90\%$ uncertainties on the non-variable best fit (shaded region). Yellow points indicate $T_\text{nv} > 3 \sigma$ and $T_\text{var} < 3 \sigma$ for that epoch (see Section \ref{sec:t_test} for a description of the Tension test).}
    \label{fig:var_pars_vo}
\end{figure*}

\begin{figure*}[ht]
    \centering
     \includegraphics[width=0.7\textwidth]{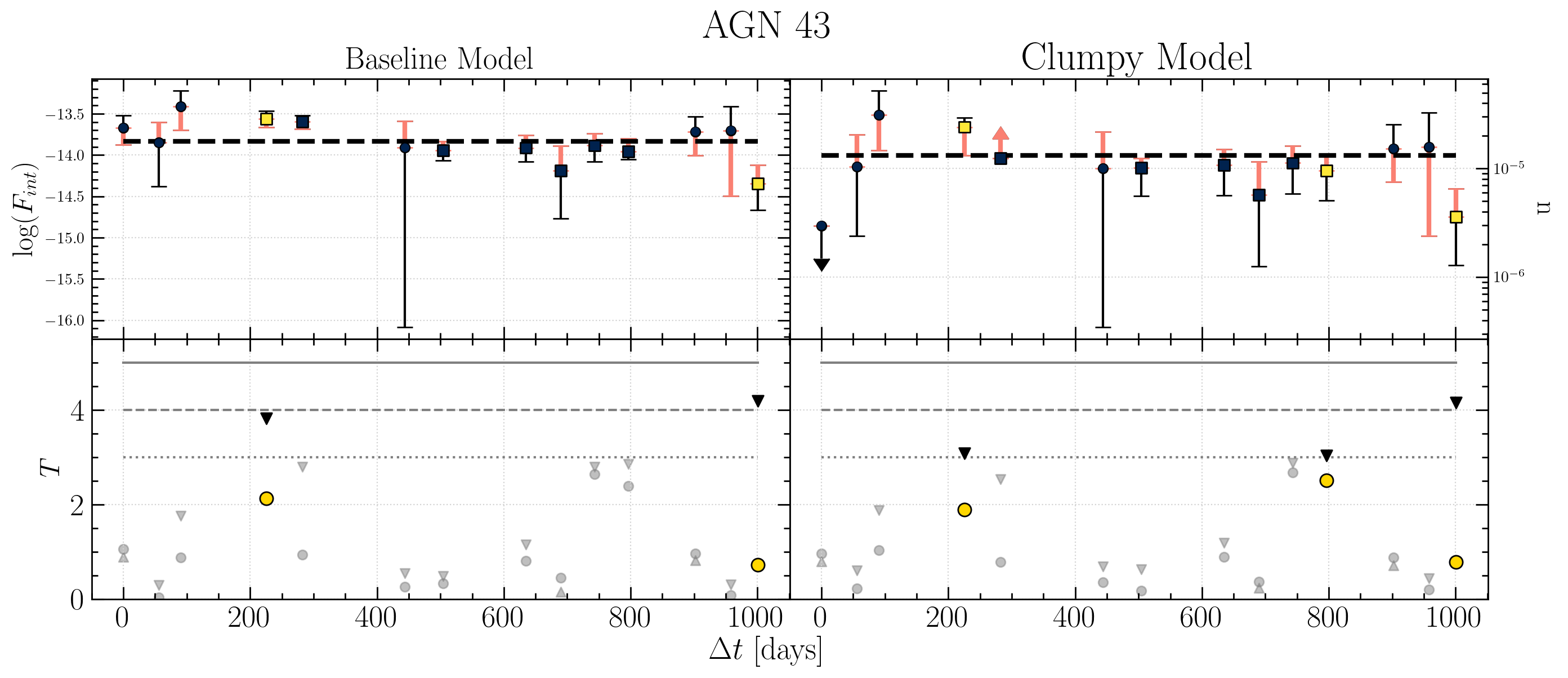}   
     \includegraphics[width=0.7\textwidth]{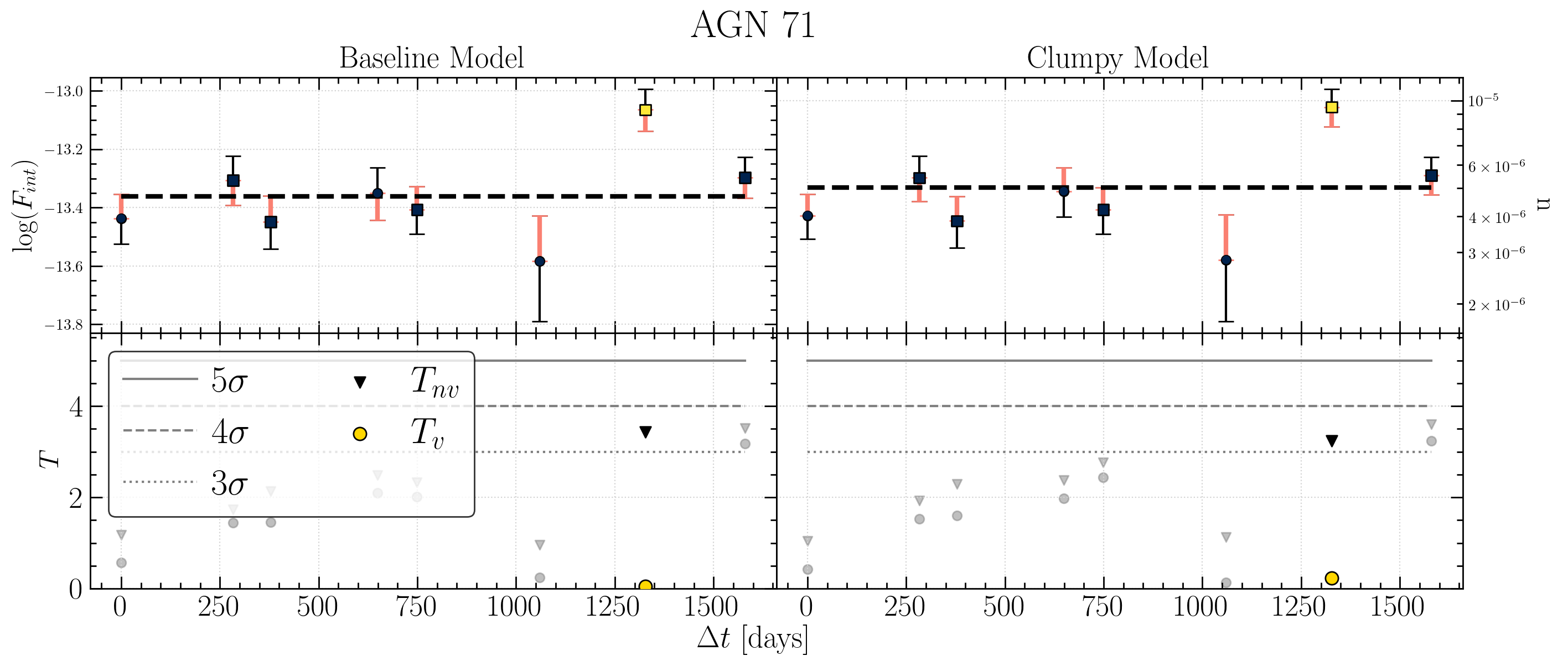}        
     \caption{Parameter variability (top panels) and model tension (bottom panels) for each of the $F_\text{int}$ variable sources identified in Section \ref{sec:var}. The colorful side of the error-bars for the parameter values become $\delta(p_i)$ in Equation \ref{eq:p}. Yellow points indicate that $T_\text{nv} > 3 \sigma$  and $T_\text{var} < 3 \sigma$ for that epoch (see Section \ref{sec:t_test} for a description of the Tension test).}
    \label{fig:var_pars_va}
\end{figure*}

\begin{figure*}
    \centering
    \includegraphics[width=0.4\textwidth]{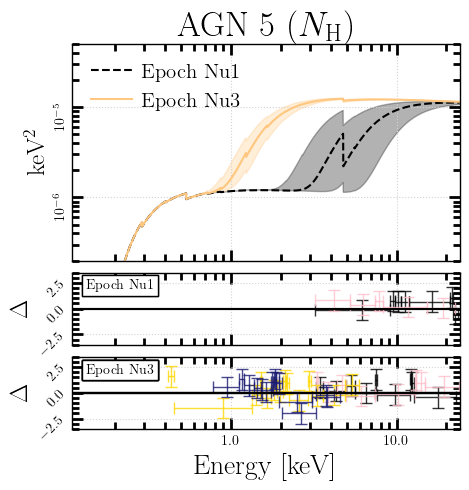}
      \includegraphics[width=0.4\textwidth]{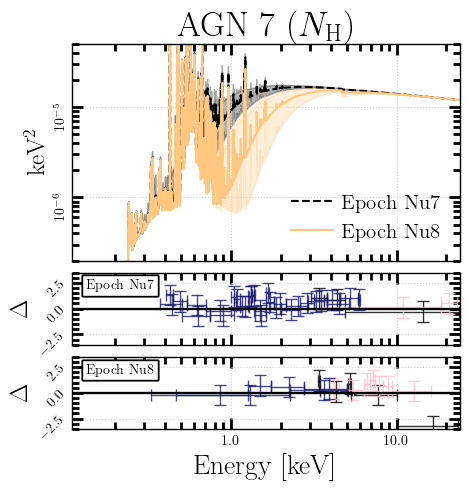}   \includegraphics[width=0.4\textwidth]{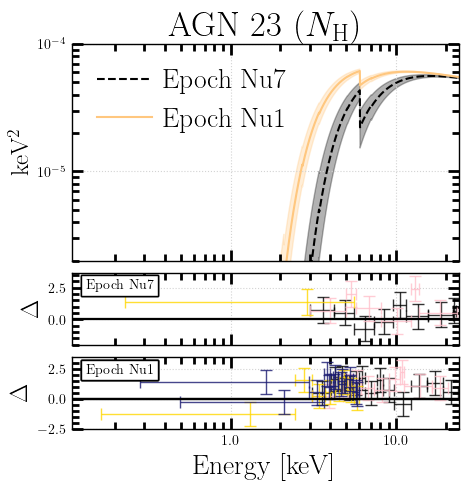}  
       \includegraphics[width=0.4\textwidth]{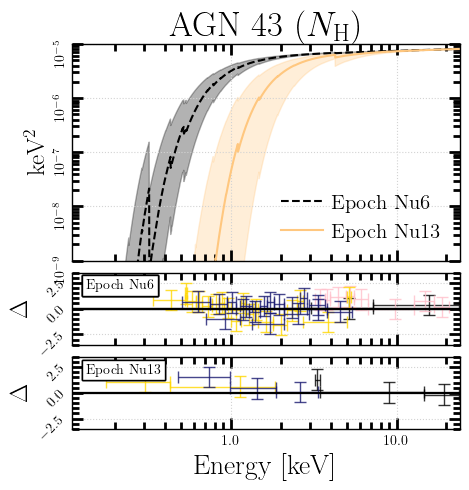}            
      \includegraphics[width=0.4\textwidth]{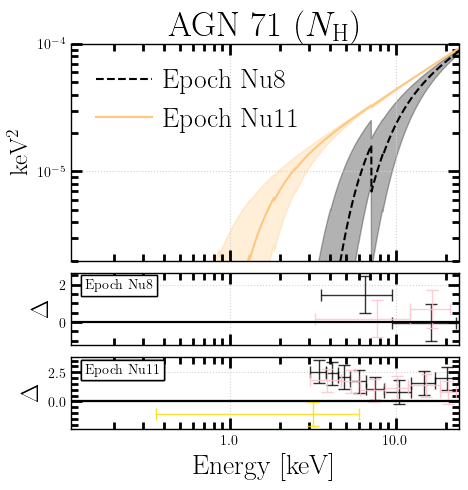}   
   \caption{The folded theoretical models compared with delta-chi error ($\Delta$) for selected epochs of each $N_{\mathrm{H}}$ variable candidate identified in Section \ref{sec:var}. The source catalog and ID number are listed in the title of each panel.}
    \label{fig:var_spec_vo}
\end{figure*}

\begin{figure*}
    \centering
    \includegraphics[width=0.4\textwidth]{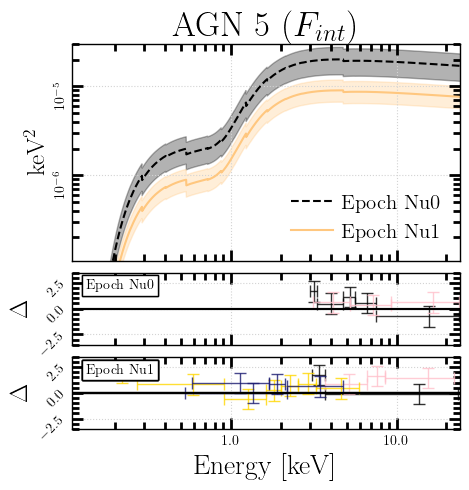}  
     \includegraphics[width=0.4\textwidth]{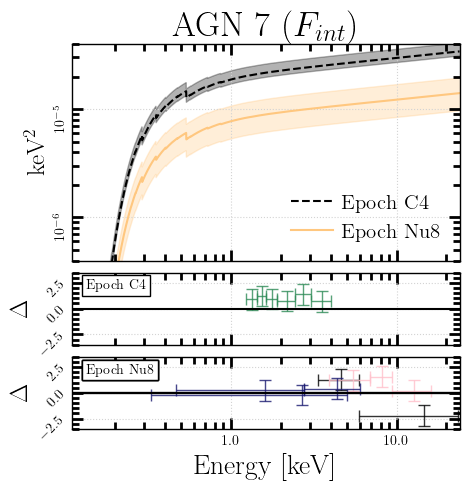}   \includegraphics[width=0.4\textwidth]{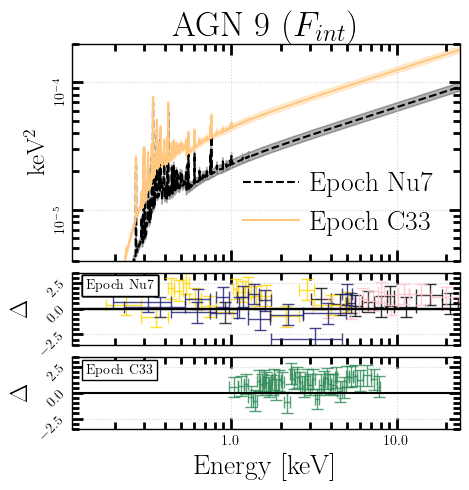}             
      \includegraphics[width=0.4\textwidth]{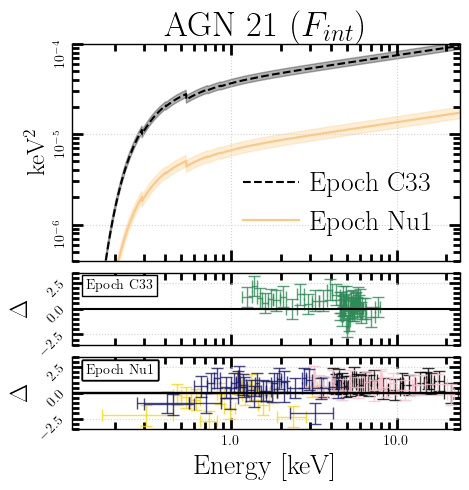}    \includegraphics[width=0.4\textwidth]{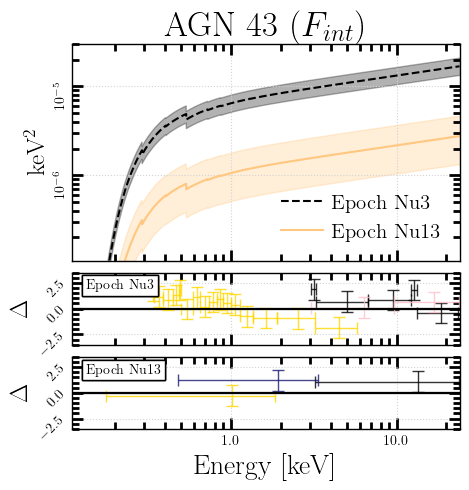} 
       \includegraphics[width=0.4\textwidth]{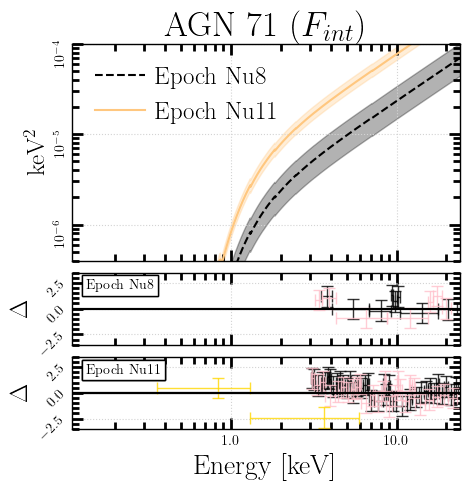}    
   \caption{The folded theoretical models compared with delta-chi error ($\Delta$) for selected epochs of each $F_\text{int}$ variable candidate identified in Section \ref{sec:var}. The source catalog and ID number are listed in the title of each panel.}
    \label{fig:var_spec_va}
\end{figure*}

Figure \ref{fig:phot} shows the 3-14~keV count rates (from \citetalias{Silver26}) as a function of time for two spectroscopic variability candidates that did not have significant photometric variability, nor an obvious explanation for this discrepancy.
The variable parameter values for the final candidates are plotted in Figures \ref{fig:var_pars_vo} ($N_{\mathrm{H}}$ variable candidates) and \ref{fig:var_pars_va} ($F_\text{int}$ variable candidates).
Finally, Figures \ref{fig:var_spec_vo} ($N_{\mathrm{H}}$ variable candidates) and \ref{fig:var_spec_va} ($F_\text{int}$ variable candidates) show the spectral models and residuals for selected epochs with notable variability.

\begin{figure*}
    \centering
     \includegraphics[width=0.7\textwidth]{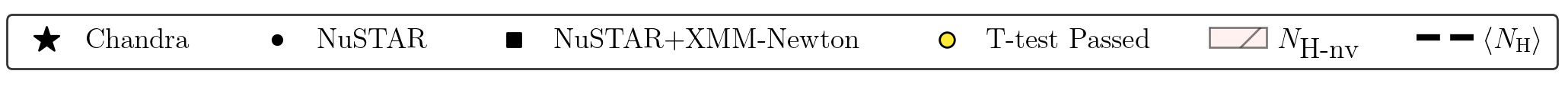}
     \includegraphics[width=0.7\textwidth]{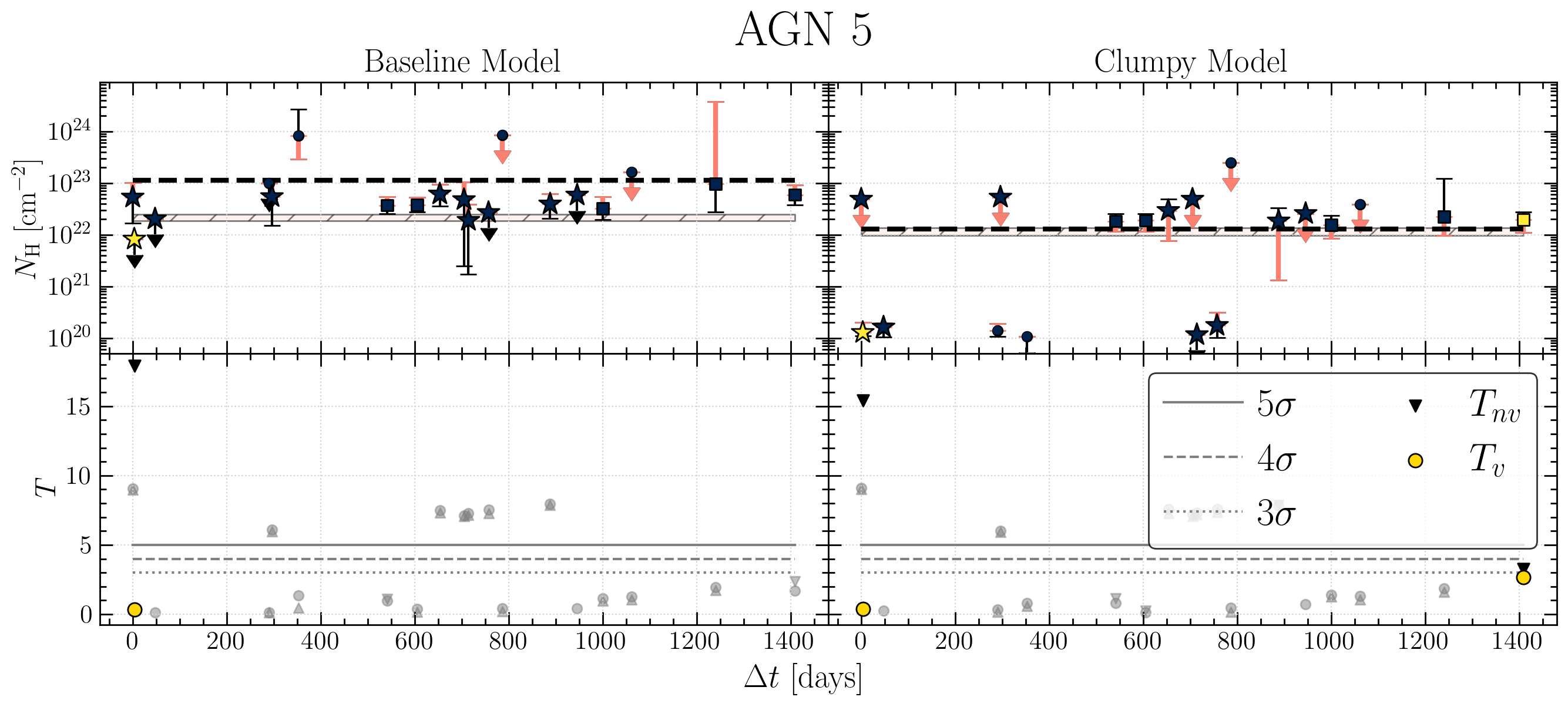}    
     \includegraphics[width=0.7\textwidth]{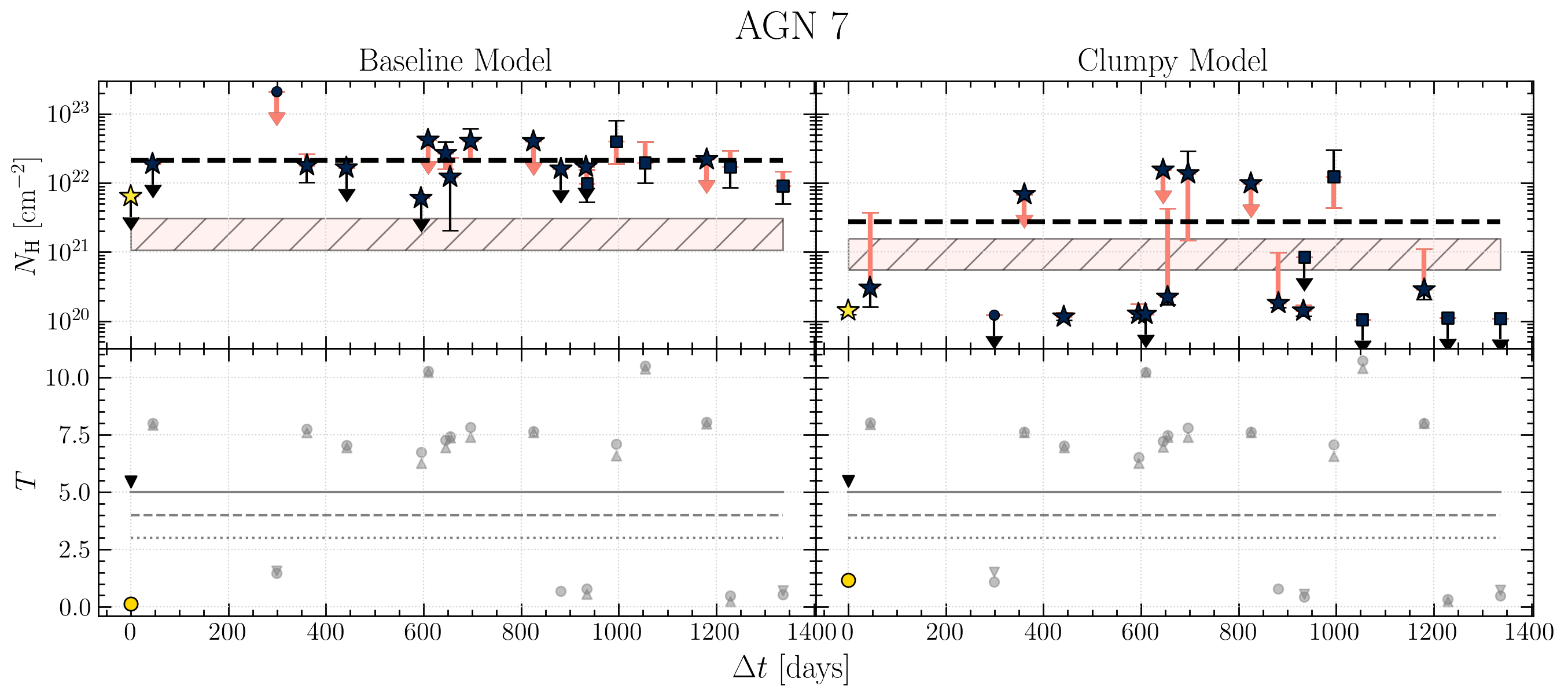}  
     \includegraphics[width=0.7\textwidth]{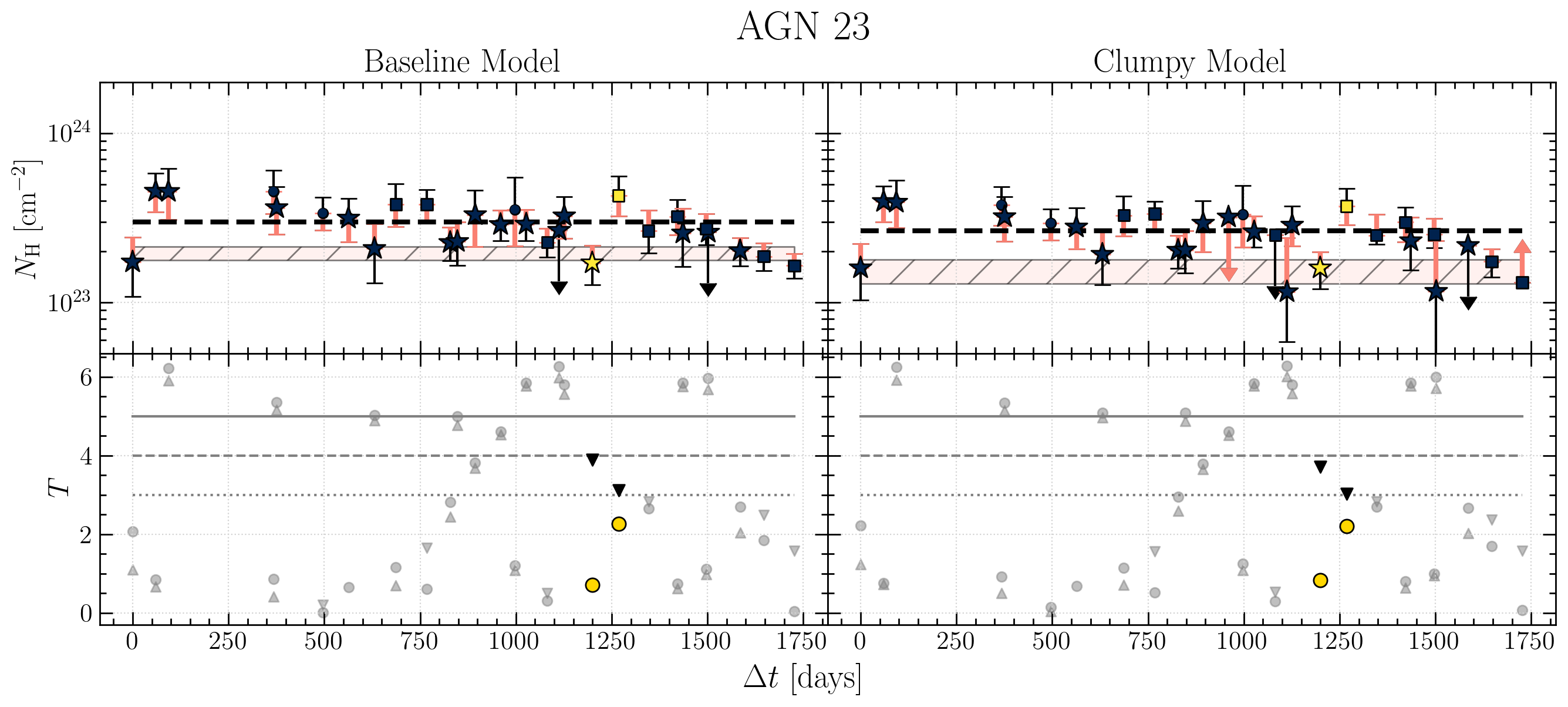}   
     \includegraphics[width=0.7\textwidth]{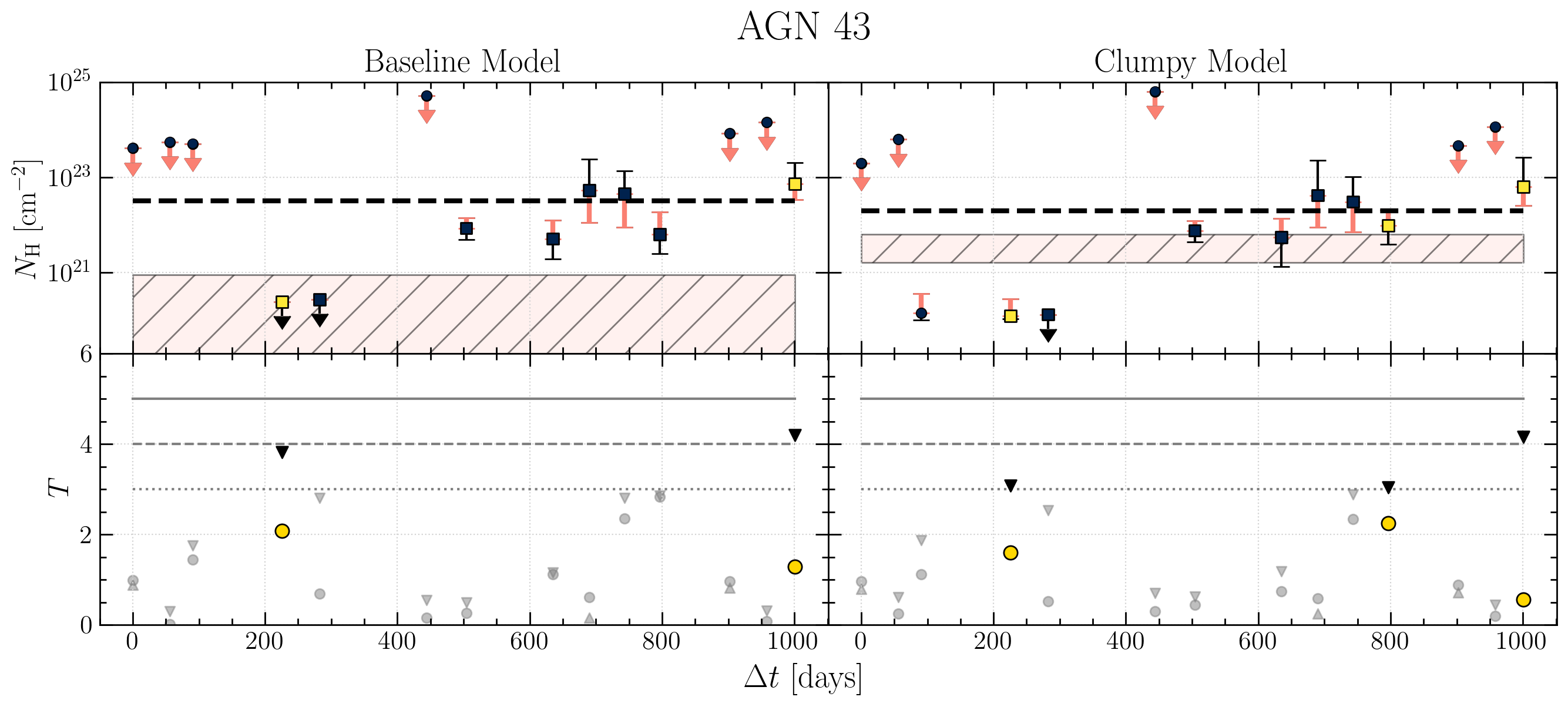} 
\end{figure*}%

\begin{figure*}
    \centering
     \includegraphics[width=0.7\textwidth]{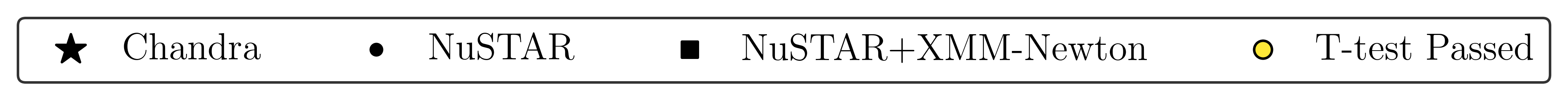} 
     \includegraphics[width=0.7\textwidth]{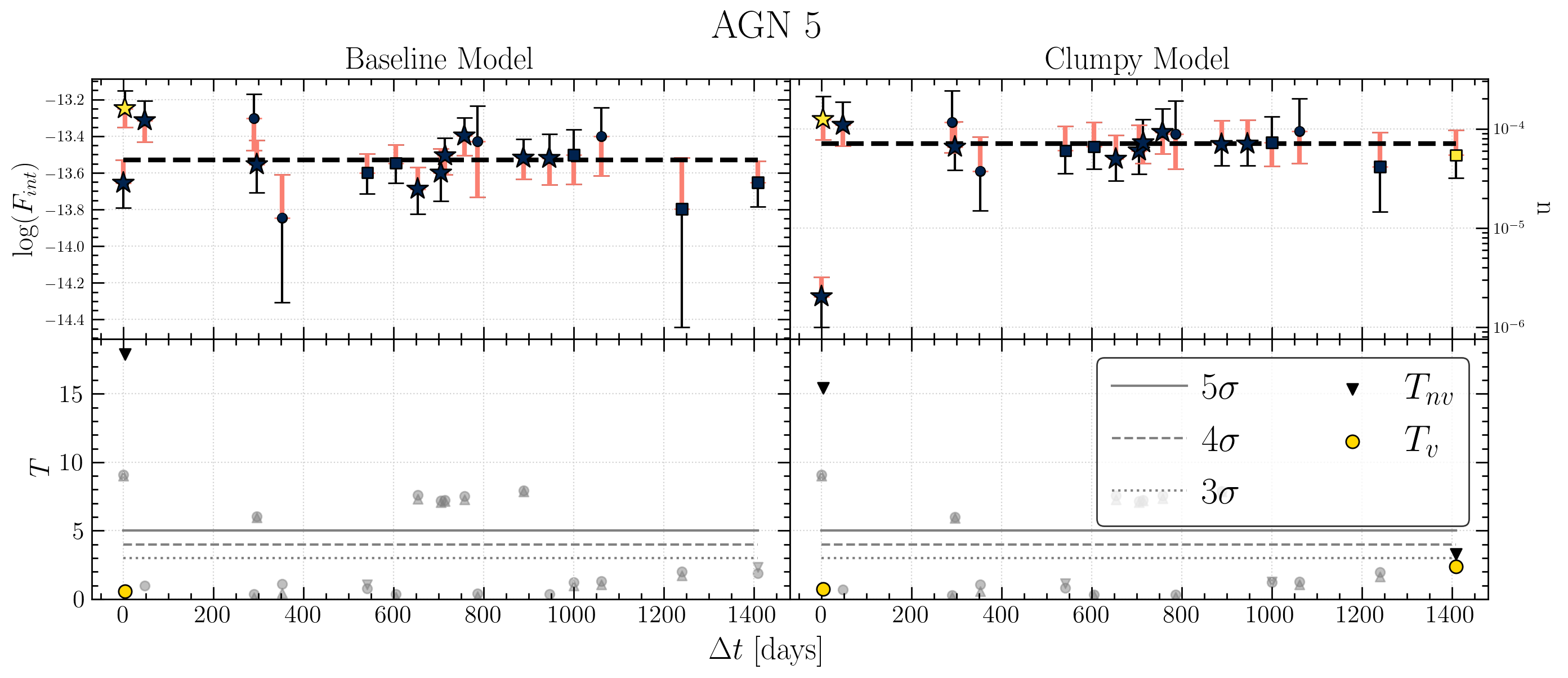} 
     \includegraphics[width=0.7\textwidth]{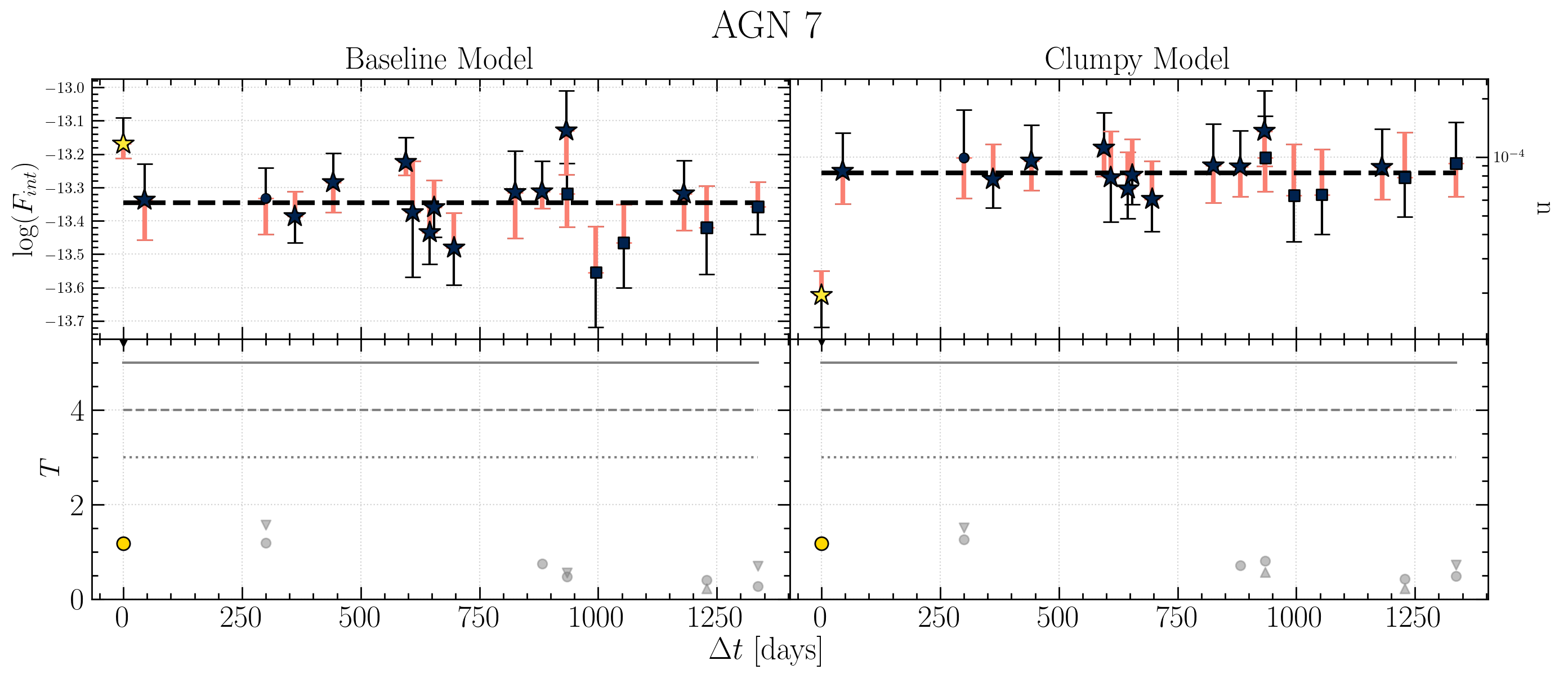}    
     \includegraphics[width=0.7\textwidth]{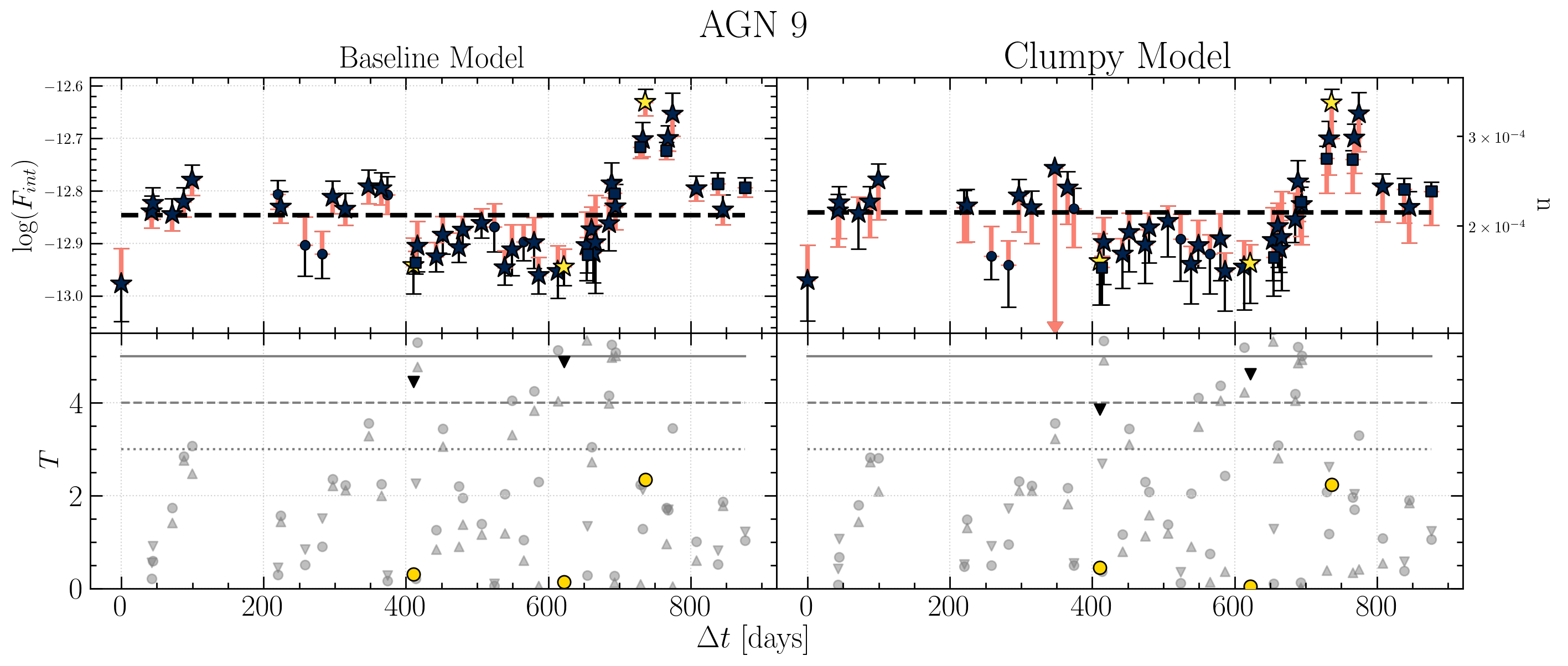} 
     \includegraphics[width=0.7\textwidth]{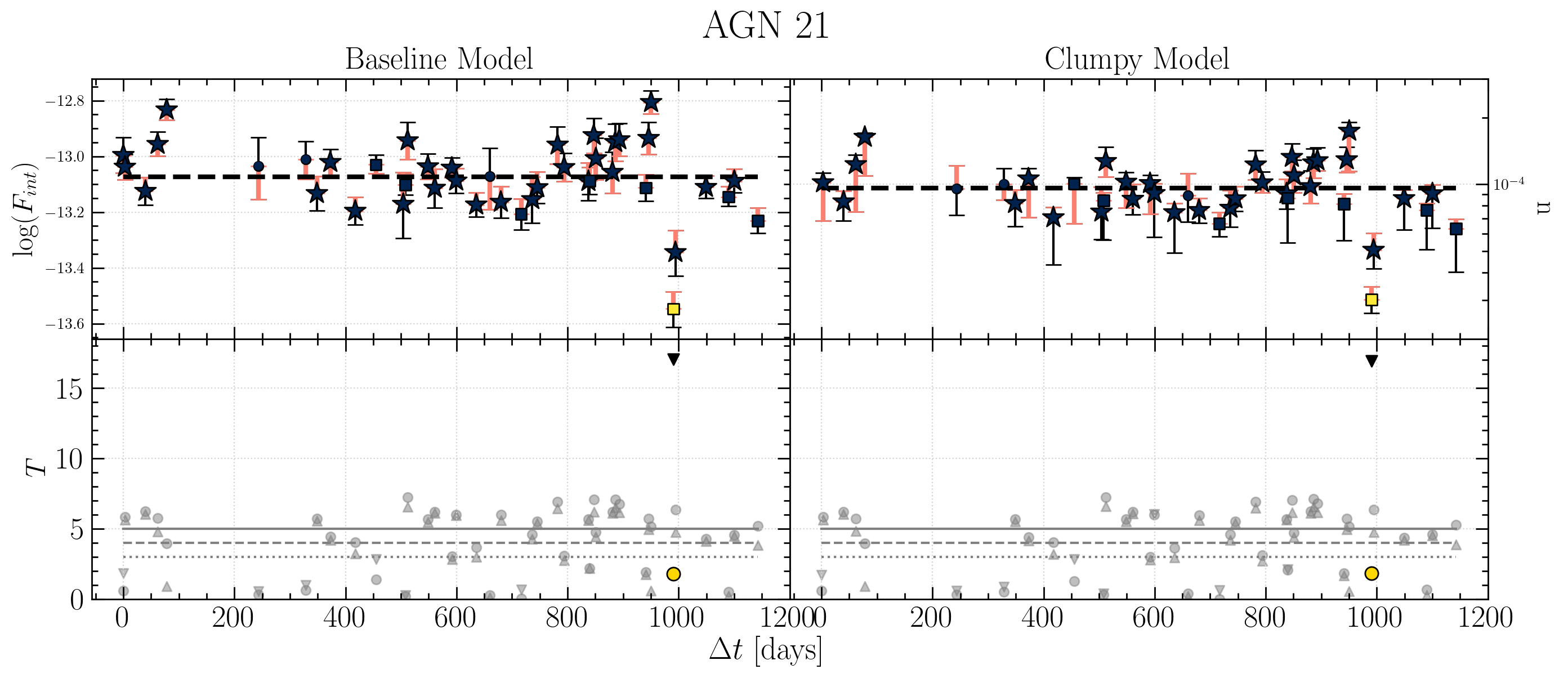} 
\end{figure*}%

\end{document}